\documentclass[a4paper,10pt]{article}

\usepackage{color}
\usepackage{graphicx}
\usepackage{amsmath}
\usepackage{graphics}
\usepackage{amsthm}
\usepackage{amsfonts}
\usepackage{url}
\usepackage{multirow}
\usepackage{rotating}
\usepackage{wrapfig}
\usepackage{authblk}
\usepackage[bookmarks=false]{hyperref} 
\usepackage{pifont}
\usepackage{array}
\usepackage{subfig}
\usepackage{float}
\usepackage{array}
\usepackage[margin=0.75in]{geometry}

\usepackage{amssymb}
\usepackage{fancyhdr}

\newcolumntype{@}{>{\global\let\currentrowstyle\relax}}
\newcolumntype{^}{>{\currentrowstyle}}
\newcommand{\rowstyle}[1]{\gdef\currentrowstyle{#1}%
  #1\ignorespaces
}

\newcommand{\tick}{\ding{51}}
\newcommand{\cross}{\ding{55}}

\newtheorem{definition}{Definition}

\newcommand{\Keywords}[1]{\par\noindent
{\small{\em Keywords\/}: #1}}

\newcommand{\rl}{\raggedleft}
\newcommand{\tn}{\tabularnewline}

\newcommand{\LV}{Low Voltage }
\newcommand{\MV}{Medium Voltage }
\newcommand{\HV}{High Voltage }

\newcommand{\PG}{Power Grid }

\newcommand{\etal}{\textit{et al.}}
\newcommand{\G}{Grid }

\newcommand{\SG}{Smart Grid }
\newcommand{\sw}{small-world }
\newcommand{\MLV}{Medium and Low Voltage }
\newcommand{\CNA}{Complex Network Analysis }

\newcommand{\pl}{power-law }
\newcommand{\cpl}{characteristic path length }
\newcommand{\apl}{average path length }
\newcommand{\Pl}{Power-law }

\title{Evolving \PG Topologies to Support Local Energy Trading}
\title{\PG Network Evolutions for Local Energy Trading}

\author{Giuliano~Andrea~Pagani} 
\author{Marco~Aiello}

\affil{Distributed Systems Group\\Johann Bernoulli Institute for Mathematics and Computer Science
\\University of Groningen\\ Groningen, The Netherlands\\
\vspace{0.3cm}
email: \url{{g.a.pagani,m.aiello}@rug.nl}\\
\url{http://www.cs.rug.nl/ds/}
}

\begin{document}

\maketitle

\begin{abstract}  
The shift towards an energy Grid dominated by prosumers (consumers and producers of energy) will inevitably have repercussions on the distribution infrastructure. Today it is a hierarchical one designed to deliver energy from large scale facilities to end-users. Tomorrow it will be a capillary infrastructure at the Medium and \LV levels that will support local energy trading among prosumers. In~\cite{PaganiAielloTSG2011}, we analyzed the Dutch \PG and made an initial analysis of the economic impact topological properties have on decentralized energy trading. In this paper, we go one step further and investigate how different networks topologies and growth models facilitate the emergence of a decentralized market. In particular, we show how the connectivity plays an important role in improving the properties of reliability and path-cost reduction. From the economic point of view, we estimate how the topological evolutions facilitate local electricity distribution, taking into account the main cost ingredient required for increasing network connectivity, i.e., the price of cabling.
\end{abstract}

\Keywords{
Power Grid, Decentralized energy trading, Complex Network Analysis
}

\section{Introduction}


Something is changing both in the way energy is produced and distributed due to the combined effects of technological advancements and introduction of new policies. In the last decades a clear trend has invested the energy sector, that of {\em unbundling.} That is the process of dismantling monopolistic and oligarchic energy system, by allowing a greater number of parties to operate in a certain role of the energy sector and market. The goal of unbundling is that of reducing costs for the end-users and providing better services through competition (e.g.,~\cite{cossent09,joskow08}). At the same time from the technological perspective, new energy generation facilities (mainly based on renewable sources) are becoming more and more accessible. These are increasingly convenient and available at both the industrial and the residential scale~\cite{mar:mic06,lovins02}. The term \emph{Smart Grid}, which does not yet have a unique agreed definition~\cite{brown09,smartGrid09}, is sometimes used to define the new scenario of a Grid with a high degree of delocalization in the production and exchange of energy. The new actors, who are both producers and consumers of energy, also known as {\em prosumers}, are increasing in number and will most likely demand a market with total freedom for energy trading~\cite{vai:pow05}.  In this coming scenario, the main role of the \HV \G may change, while the Distribution \G (i.e., \MV and \LV end of the Power Grid) becomes more and more important, while requiring a major update. In fact, the energy interactions between prosumers will increase and most likely occur at a rather local level, therefore involving the Low and Medium Voltage Grids. This evolution of the energy sector will inevitably call for an upgrade of the enabling distribution infrastructure so to facilitate local energy exchanges. An infrastructure comparable more to a ``peer-to-peer'' system on the Internet, rather than the current strictly hierarchical system. But how will the infrastructure evolve to enable and follow this trend?

In~\cite{PaganiAielloTSG2011}\footnote{An extended version of this
  paper is available electronically as technical report~\cite{pag:preprint11}.} we laid the foundation for a statistical study of the Medium and Low Voltage Grid with the aim of identifying salient topological properties of the \PG that affect decentralized energy exchange. We based that study on real samples of the Dutch Grid and provided an initial economic analysis of the possible barriers to delocalized trades. In this follow up paper, we go one step beyond and consider growth models for network topologies providing an analysis of which models suit best the purpose of local energy exchange. The tool for our study is Complex Network Analysis (CNA)~\cite{bar:lin03}. In particular, in the present case we use CNA as a synthesis tool by synthesizing networks using topological models coming from the literature of modeling technological, infrastructural and social network evolutions.

In order to evaluate the adequacy of the generated networks, we develop a set of metrics that capture the various aspects that networks suited for small-scale energy exchange need to satisfy. It is then quite straightforward to compare the results of the synthetic models with the real samples analyzed in~\cite{PaganiAielloTSG2011} and on that ground propose network models that best suit a prosumer-based local energy exchange. Finally, a quantitative evaluation of how the improvement in the topology directly influences electricity prices is then possible. 

We remark the novelty of this proposal with respect to previous CNA studies of the Power Grid. In fact, CNA has been used only on the High Voltage networks to get information on resilience to failures, and the \MLV Grids have been mostly ignored. Another novelty is the use of \CNA not as a tool for pure analysis of the existing, but to exploit it as a design tool for an infrastructure.  Using Graph Theory in the design of distribution systems is not completely new, several studies have incorporated Graph Theory elements in operation research techniques for Grid planning~\cite{crawford75,wall79}, but never, to the best of our knowledge, has Graph Theory been combined with global statistical measures to design the Grid.  In addition, we ground the design methods to investments by taking into account costs of \G cabling based on the types of cables typically used in real Distribution Networks (i.e., Norther Netherlands \MLV network samples). In summary, the paper proposes which topologies according to CNA-based metrics are best suited in terms of performance and reliability of the infrastructure for a local energy exchange, give an estimation of the cabling cost for the realization of such topologies and assess the advantages from the electricity  distribution point of view of the proposed topologies compared to the actual ones.

The paper is organized as follows. We open by analyzing the motivations for a new energy landscape and the required changes to the current Grid in Section~\ref{sec:motivations}. The background of Graph Theory  necessary for the present study is presented in Section~\ref{sec:graphBasics}. Section~\ref{sec:models} describes the main properties of the graph models; while the metrics exploited to compare the properties of the various generated graphs are described in Section~\ref{sec:metrics}. The analysis and discussion of the results is presented in Section~\ref{sec:results}. An overall discussion and illustration taking into account benefits and costs of evolution of topologies are considered in Sections~\ref{sec:discussion}. Section~\ref{sec:relWks} reviews the main approaches to Electrical \G and System design and evolution, while Section~\ref{sec:conclusion} provides a conclusion of the paper. A series of appendixes is included to provide extended coverage of topics related to the core of the paper, in particular, \ref{sec:appA} describes statistical properties of power cables' price and resistance; \ref{sec:AppAlfaBeta} provides an overview about the relationship between network topology a electricity price; appendix section concludes with \ref{sec:AppEngineering} which describes an engineering process based on a \CNA that can guide Grid and energy operators to shape their networks for the new local energy exchange paradigm.

\section{The Need for Evolution of the Grid}\label{sec:motivations}

In the XIX century electrical energy generation was considered a \textit{natural monopoly}. The cheapest way to produce electricity was in big power plants and then transmitting it across a country through a pervasive network of cables operated by a monopolistic state owned company. The situation has changed and now more and more companies are present in the energy business from energy production, to energy transmission and distribution, to retail and service provided to the end-user. To enable and accelerate this process, governments in the western world have promoted policies to open the electricity business and facilitate competition with the final aim to both modernize the energy sector and provide a more convenient service for the end-user. Even more on this path of enabling everybody to be a producer of energy is the possibility (sometimes incentivized with governments' policies) to have small-scale energy generation units such as photovoltaic panels, small-wind turbines and micro combined heat-and-power systems (micro-CHP) which are now all widely available and affordable for the end-user market. Such small-scale approach is beneficial to the electricity system in many ways: from reduced losses since source and load are closer, to system modularity, to smaller investments compared to large-scale energy solutions~\cite{lovins02}. Local generation based on renewables  is a boost for the transition towards a renewable-based energy supply. In fact, end-users generate their own energy and the additional supply is likely to be provided by other end-users in the neighborhood that have energy surplus generated by their renewable-based generating equipment. In such a context with many small-scale producers and still without an efficient and cheap energy storage technology a local energy exchange at the neighborhood or municipal level between end-users is foreseeable and desirable. Micro-grids increased performance in terms of reduced losses and power quality have been successfully tested~\cite{lasseter11,barri10}, but little attention has been devoted to the network topology of these type of Grids.

In the evolution of the actual electricity system to the \textit{Internet of Energy}~\cite{friedman08,vai:pow05} the end-users are producers in addition to the normal consumer role they have always had. Energy is then something that the user does no more need to negotiate through yearly or lifetime contracts, but can be traded between prosumers and consumers on a fully electronic and automated market. These energy exchanges are likely to be local inside a neighborhood, a village or a city where users that have small-scale renewable-based producing facilities can sell their energy surplus. This solution represents a ``win-win'' solution first for the environment envisioning more energy generated with renewable sources, and second for the prosumer who sells his surplus energy on the market obtaining some profits out of it. This latter aspect helps accelerating the return on investment made in purchasing the generating equipment. A benefit is also for the end-user that has more flexibility in choosing his energy provider and takes advantage of the cheaper tariffs of the prosumers. The traditional energy providers and distributors still play important roles even in this paradigm: the former provide traditional supply where or when prosumers are not available, the latter has even a more critical role in monitoring and providing a \G at \MLV that is efficient, failure resistant and that satisfies the needs of this new energy exchange paradigm.

This future scenario might impact deeply on the actual Grid infrastructure especially the \MLV section where the prosumer and consumer exchanges will happen. The Grid infrastructure with the associated reliability, losses, quality, performances and related transmission costs might act as an enabler or repression of the local energy exchange. The \MLV \G is likely to face important changes in its infrastructure to support \SG~\cite{brown08} and even more in enabling a scenario where energy producers are many and the interactions are at local scale. Usually, the lower end of the \G have been considered of small importance and less critical than the \HV infrastructures, however the tendency is likely to be reversed in a prosumer-based energy paradigm.  

In our study we resort to Complex Network Analysis, a branch of Graph Theory taking its root in the early studies of Erd{\H{o}}s and R{\'{e}}nyi~\cite{Erdos1959} on random graphs and considering statistical structural properties of very large graphs.  Although taking its root in the past, \CNA (CNA) is a relatively young field of research. The first systematic studies appeared in the late 1990s~\cite{Watts98,Strogatz2001, Barabasi1999,Albert2000} having the goal of looking at the properties of large networks with a complex systems behavior. Afterwards, \CNA has been used in many different fields of knowledge, from biology~\cite{Jeong2000} to chemistry~\cite{Doye2002}, from linguistics to social sciences~\cite{Milgram69}, from telephone call patterns~\cite{Aiello2000} to computer networks~\cite{Faloutsos1999} and web~\cite{albert99,Donato2007} to virus spreading~\cite{kephart91,colizza07,Gautreau2008} to logistics~\cite{Latora2002,Guimer2004,Colizza2006} and also inter-banking systems~\cite{Nationalbank2003}. Men-made infrastructures are especially interesting to study under the \CNA lenses, especially when they are large scale and grow in a decentralized and independent fashion, thus not being the result of a global, but rather of many local autonomous designs. The \PG is a prominent example.
In this work we consider a novel approach both in considering \CNA tools as a design instrument (i.e., CNA-related metrics are used in finding the most suited \MLV Grid for local energy exchange) and in focusing on the \MLV layers of the Power Grid. In fact, traditionally, \CNA studies applied to the \PG only evaluate reliability issues and disruption behavior of the \G when nodes or edges of the \HV layer are compromised.

In summary, the requirements of the new \PG enabling decentralized trading are:
\begin{enumerate}
\item Realizing the small-scale network paradigm;
\item Improving local energy exchange;
\item Supporting renewable-based energy production;
\item Encouraging the end-user (technically, economically and politically) to buy/sell energy locally;
\item Realizing networks easy to repeat 
 at different scales (i.e., neighborhood, small village, city, metropolis)
\item Reducing losses in the \MLV end of the \G ; and 
\item Enabling smartness in the automation of energy exchanges and their accounting.
\end{enumerate}
In the above general requirements several are tightly connected with the topology of the network, while others are more related to the control and ICT-oriented aspects of the Smart Grid. For the former aspects we provided a first investigation in our previous work~\cite{PaganiAielloTSG2011} and in the present work; the latter aspects are out of the scope of the present work and can be traced to other investigations such as~\cite{cecati11,Bu11,shafiullah10,potter09}.

\section{Graph Theory Background}\label{sec:graphBasics}

The approach used in this work to model the Power Grid and its evolution is based on Graph Theory and Complex Networks. Here we recall the basic definitions that we use throughout the paper and refer to standard textbooks such as~\cite{Bollobas79,Bollobas1998} for a broader introduction. First we define a graph for the Power Grid~\cite{pag:preprint1102}.
\begin{definition}[Power Grid graph] A
{\em Power Grid graph} is a graph $G(V,E)$ such that each element $v_i \in V$ is either a substation, transformer, or consuming unit of a physical Power Grid. There is an edge $e_{i,j}=(v_i,v_j) \in E$ between two nodes if there is physical cable connecting directly the elements represented by $v_i$ and $v_j$.  \end{definition}
One can also associate weights to the edges representing physical cable properties (e.g., resistance, voltage, supported current flow).
\begin{definition}[Weighted Power Grid graph]\label{def:wpgg}
A {\em Weighted Power Grid graph} is a Power Grid graph $G_w(V,E)$ with an additional function $f:E \to \mathbb{R}$ associating a real number to an edge representing the physical property of the corresponding cable (e.g., the resistance, expressed in Ohm, of the physical cable).
\end{definition}
A first classification of graphs is expressed in terms of their size. 
\begin{definition}[Order and size of a  graph]
Given the graph $G$ the order is given by $N=|V|$, while the
size is given by $M=|E|$.
\end{definition}
From \emph{order} and \emph{size} it is possible to have a global value for the connectivity of the vertexes of the graph, known as {\em average node degree} . That is $<\!k\!>=\frac{2M}{N}$. To characterize the relationship between a node and the others it is connected to, the following properties provide an indication of the bond between them.
\begin{definition}[Adjacency, neighborhood and degree]
  If $e_{x,y} \in E$ is an edge in graph $G$, then $x$ and $y$ are {\em adjacent,} or {\em neighboring,} vertexes, and the vertexes $x$ and $y$ are \emph{incident} with the edge $e_{x,y}$.  The set of vertexes adjacent to a vertex $x \in V$, called the \emph{neighborhood} of $x$, is denoted by $\Gamma_x$. The number $d(x)=|\Gamma_x|$ is the \emph{degree} of $x$.  \end{definition}

A measure of the average `density' of the graph is given by the
clustering coefficient, characterizing the extent to which vertexes
adjacent to any vertex $v$ are adjacent to each other.
\begin{definition}[Clustering coefficient (CC)]
The {\em clustering coefficient} $\gamma_v$ of $\Gamma_v$ is
\[
\gamma_v=\frac{|E(\Gamma_v)|}{\binom{k_v}{2}} 
\]
where $|E(\Gamma_v)|$ is the number of edges in the neighborhood of
$v$ and $\binom{k_v}{2}$ is the total number of $possible$ edges in
$\Gamma_v$.
\end{definition}
This local property of a node can be extended to an entire graph by
averaging over all nodes.

Another important property is how much any two nodes are far apart from each other, in particular the minimal distance between them or shortest path.
The concepts of \emph{path} and \emph{path length} are crucial to understand the way two vertexes are connected.
\begin{definition}[Path and  path length]
A \emph{path} of G is a subgraph $P$ of the form:
\[V(P)=\{x_0,x_1,\ldots,x_l\}, \hspace{10mm}
  E(P)=\{(x_0,x_1),(x_1,x_2),\ldots,(x_{l-1},x_l)\}.
\]
such that $V(P)\subseteq V\textrm{ and }E(P)\subseteq E$. The vertexes
$x_0$ and $x_l$ are \emph{end-vertexes} of $P$ and $l=|E(P)|$ is the
\emph{length} of $P$. A graph is {\em connected} if for any two
distinct vertexes $v_i,v_j\in V$ there is a finite path from $v_i$ to $v_j$.\\
\end{definition}

\begin{definition}[Distance]
Given a graph $G$ and vertexes $v_i$ and $v_j$, their \emph{distance} $d(v_i,v_j)$ is
the minimal length of any $v_i-v_j$ path in the graph. If there is no
$v_i-v_j$ path then it is conventionally  set to $d(v_i,v_j)=\infty$.\\
\end{definition}
\begin{definition}[Shortest path]
Given a graph $G$ and vertexes $v_i$ and $v_j$ the {\em shortest path} is
the the path corresponding to the minimum of to the set $\{|P_1|,|P_2|,\ldots,|P_k|\}$ containing the
lengths of all paths for which $v_i$ and $v_j$ are the end-vertexes.
\end{definition}
A global measure for a graph is given by its average distance among
any two nodes.
\begin{definition}[Average path length (APL)]
Let $v_i \in V$ be a vertex in graph $G$. The \emph{average path length} for $G$ $L_{av}$ is:
\[
 L_{av} = \frac{1}{N \cdot (N-1)} \sum_{i\neq j}d(v_i,v_j)
\]
where $d(v_i,v_j)$ is the finite distance between $v_i$ and $v_j$ and $N$ is the order of $G$.
\end{definition}
\begin{definition}[Characteristic path length (CPL)]
Let $v_i \in V$ be a vertex in graph $G$, the \emph{characteristic path length} for $G$, $L_{cp}$ is defined as the median of ${d_{v_i}}$ where:
\[
 d_{v_i} = \frac{1}{(N-1)} \sum_{i\neq j}d(v_i,v_j)
\]
is the mean of the distances connecting $v_i$ to any other vertex $v_j$ in $G$ and $N$ is the order of $G$.\\
\end{definition}
To describe the importance of a node with respect to minimal paths in the graph, the concept of betweenness helps. Betweenness (sometimes also referred as {\em load}) for a given vertex is the number of shortest paths between any other nodes that traverse it.
\begin{definition}[Betweenness]
 The {\em betweenness} $b(v)$ of vertex $v \in V$ is
\[
 b(v)=\sum_{v\neq s \neq t}\frac{\sigma_{st}(v)}{\sigma_{st}}
\]
where $\sigma_{st}(v)$ is 1 if the shortest path between vertex s and vertex t goes through vertex v, 0 otherwise and $\sigma_{st}$ is the number of shortest paths between vertex s and vertex t.
\end{definition}

Looking at large graphs, one is usually interested in global statistical measures rather than the properties of a specific node. A typical example is the node degree, where one measures the node degree probability distribution.
\begin{definition}[Node degree distribution]\label{def:ndd}
Consider the degree $k$ of a node in a graph as a random variable. The
function 
\[
N_k=\{v\in G:\: d(v)=k\}
\]
is called {\em probability node degree distribution}.
\end{definition}
The shape of the distribution is a salient characteristic of the network. For the Power Grid, the shape is typically either exponential or a Power-law~\cite{Barabasi1999,Amaral2000,PaganiAielloTSG2011,casals07}.
More precisely, an exponential node degree ($k$) distribution has a fast decay in the probability of having nodes with relative high node degree. The relation:
\[
 P(k)=\alpha e^{\beta k}
\]
follows, where $\alpha$ and $\beta$ are parameters of the specific network considered. On the contrary, a \Pl distribution has a slower decay with higher probability of having nodes with high node degree. It is expressed by the relation:
\[
 P(k)=\alpha k^{-\gamma}
\]
where $\alpha$ and $\gamma$ are parameters of the specific network considered. We remark that the graphs considered in the \PG domain are usually large, although finite, in terms of \textit{order} and \textit{size} thus providing limited and finite probability distributions.

A Graph can also be represented as a matrix, typically an adjacency matrix.  
\begin{definition}[Adjacency matrix] 
The adjacency matrix $A=A(G)=(a_{i,j})$ of a graph $G$ of order $N$ is the $N \times N$ matrix given by \[
 a_{ij} =
  \begin{cases}
   1      & \text{if } (v_i,v_j) \in E, \\
   0       & \text{otherwise.}
  \end{cases}
\]
\end{definition} 
We have now provided the basic definitions needed to present the modeling tools for the \PG evolutions.

\section{Modeling the \PG}\label{sec:models}

To address the question of what are the best suited topologies to characterize the \MLV Grids, we study models for graph generation proposed for technological complex networks. For each model we evaluate the properties of the network for several values of the \textit{order} of the graph. Following our analysis of the Northern Dutch \MLV \cite{PaganiAielloTSG2011}, we categorize networks as \textit{Small}, \textit{Medium} and \textit{Large}, see Table~\ref{tab:sizes}. We then analyze the properties of the networks coming from the generated models by applying relevant \CNA metrics and combine them appropriately. In this way, \CNA is not only a tool for analysis, but it becomes a design tool for the future electrical Grid.

\begin{table*}[htb]
\begin{center}
    \begin{tabular}{ | l | l | r | }
    \hline
    Network layer & Category & \textit{Order} \\ \hline
    \LV & Small & $\approx$20 \\ \hline
    \LV & Medium & $\approx$90 \\ \hline
    \LV & Large & $\approx$200 \\ \hline
    \MV & Small & $\approx$250 \\ \hline
    \MV & Medium & $\approx$500 \\ \hline
    \MV & Large & $\approx$1000 \\ \hline
    \end{tabular}
\end{center}
\caption{Categories of \MLV network and their \textit{order} based on~\cite{PaganiAielloTSG2011}.\label{tab:sizes}}
\end{table*}


Most studies using \CNA focus on extracting properties of networks arising from natural phenomena (e.g., food webs~\cite{dunne08}, protein interactions~\cite{Jeong2000}, neural networks of microorganism~\cite{Watts03}), and human generated networks (computer networks~\cite{Faloutsos1999}, the web~\cite{albert99}, transport systems~\cite{Guimer2004}) to try to understand which underlying rules characterize them. Here we look at network models that have proven successful in showing salient characteristics of technological networks (i.e, preferential attachment, Copying Model, \pl networks), social networks (i.e., small-world, Kronecker graph, recursive matrix) and natural phenomena as well (e.g., Random Graph, small-world, Forest Fire) to investigate which one is best suited for supporting local-scale energy exchange form a topological point of view. Next we provide a brief introduction to all the models used in the present study, while a more in-depth presentation is available for instance in~\cite{Chakrabarti2006} or~\cite{Newman2003a}.

\subsubsection*{Random Graph}

\begin{figure}[htbp!]
   \centering
   \includegraphics[height=0.35\textheight]{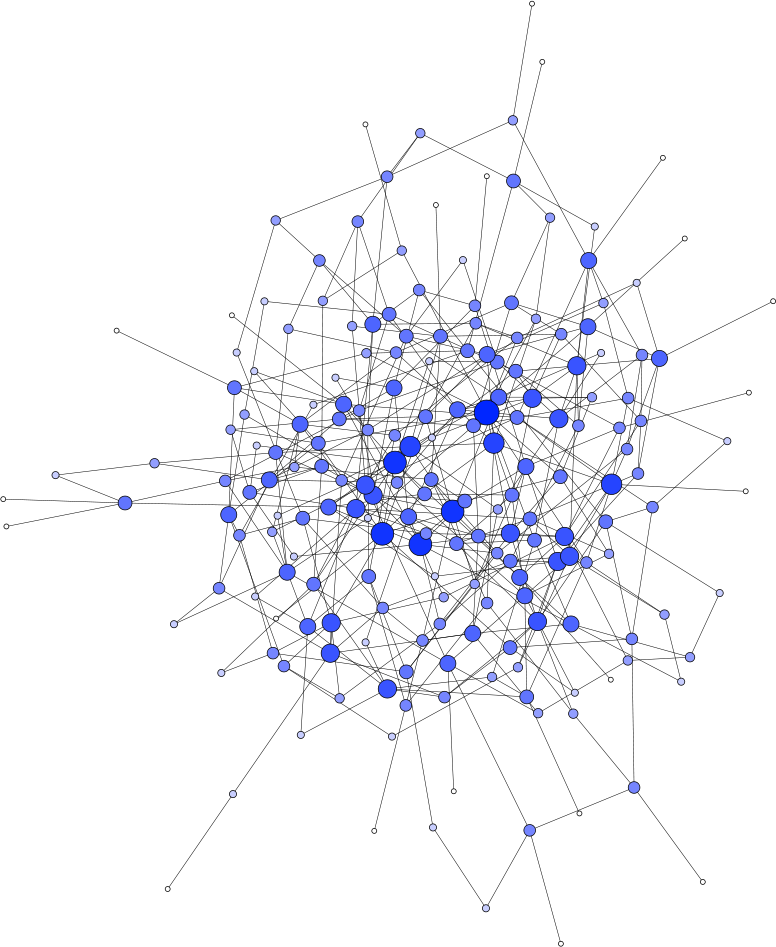}
   \caption{A Random Graph.}
\label{fig:rgraph}
\end{figure}

A Random Graph is a graph built by picking nodes under some random probability distribution and connecting them by edges. It is due to the pioneering studies of Erd{\H{o}}s and R{\'{e}}nyi~\cite{Erdos1959,Erdos1960}. More precisely, there are two ways to built a Random Graph, (a) the $G_{N,p}$ model proposed by Erd{\H{o}}s and R{\'{e}}nyi considers a set of $N$ nodes and for each pair of nodes an edge is added with a certain probability $p$; (b) the $G_{N,M}$ model considers with equal probability all the graphs having $N$ vertexes and exactly $M$ edges randomly selected among all the possible pairs of edges. The models have the same asymptotic properties. We use the $G_{N,M}$ model since we are interested in setting both the number of nodes and edges for the networks to generate. A Random Graph with \textit{order} 199 and \textit{size} 400 is shown in Figure~\ref{fig:rgraph}.

\subsubsection*{Small-world Graph}

\begin{figure}[htbp!]
   \centering
   \includegraphics[height=0.35\textheight]{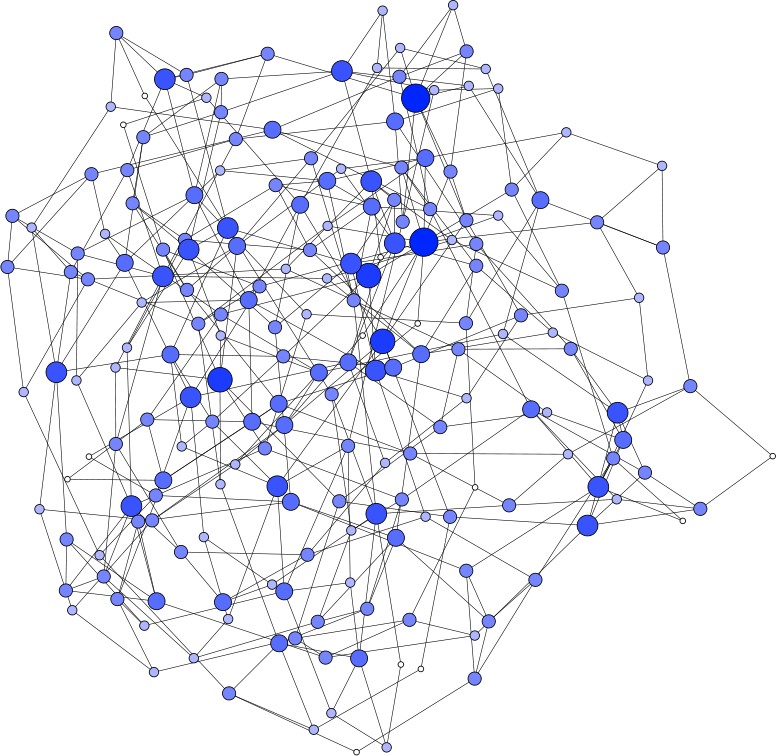}
   \caption{A \sw graph.}
\label{fig:swgraph}
\end{figure}

The \sw phenomenon became famous after the works of Milgram in the sociological context~\cite{milgram67,Milgram69} who found short chains of acquaintances connecting random people in the USA. More recently, the \sw characterization of graphs has been investigated by Watts and Strogatz~\cite{Watts03,Watts98} who showed the presence of the \sw property in many types of networks such as actor acquaintances, the \PG infrastructure and neural networks in worms. It is obtained from a regular lattice that connects the nodes followed by a process of rewiring the edges with a certain probability $p \in [0,1]$. The resulting graph has intermediate properties between the extreme situations of a regular lattice ($p=0$) and a random graph ($p=1$). In particular, \sw networks hold interesting properties: the characteristic path length is comparable to the one of a corresponding random graph ($L_{sw} \gtrsim L_{random}$), while the clustering coefficient has a value bigger than a random graph and closer to the one of a regular lattice ($CC_{sw}\gg CC_{random}$). A \sw Graph with \textit{order} 200 and \textit{size} 399 is shown in Figure~\ref{fig:swgraph}.

\subsubsection*{Preferential Attachment}

\begin{figure}[htbp!]
   \centering
   \includegraphics[height=0.35\textheight]{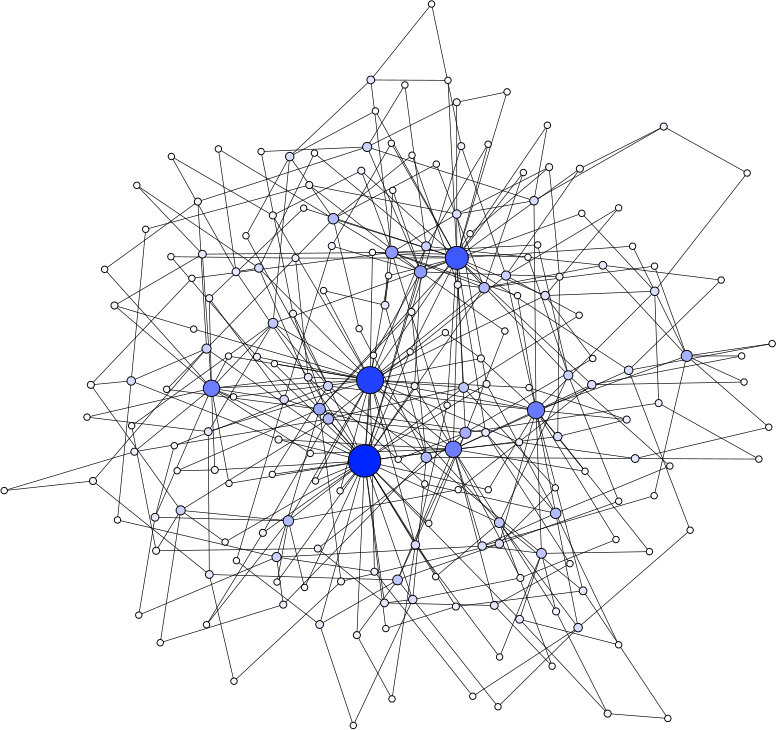}
   \caption{A Preferential Attachment graph.}
\label{fig:prefAtt}
\end{figure}

The preferential attachment model represents the phenomenon happening in real networks, where a fraction of nodes has a high connectivity while the majority of nodes has small node degree. This model is built upon the observation by Barab\'{a}si and Albert~\cite{Barabasi1999} of a typical pattern characterizing several type of natural and artificial networks. The basic idea is that whenever a node is added to the network and connects (through edges) to $m$ other nodes, those with higher degree are preferred for connection. In other words, the probability to establish an edge with an existing node $i$ is given by $\Pi(k_i) = \frac{k_i}{\sum_j{k_j}}$ where $k_i$ is the node degree of node $i$. One can see then that the more connected nodes have higher chances to acquire more and more edges over time in a sort of ``rich gets richer'' fashion; a phenomenon studied by Pareto~\cite{Pareto:1971} in relation to land ownership. The preferential attachment model reaches a stationary solution for the node degree probability that follows a \pl with $P(k)=\frac{2m^2}{k^{3}}$. A graph based on preferential attachment with \textit{order} 200 and \textit{size} 397 is shown in Figure~\ref{fig:prefAtt}.

\subsubsection*{R-MAT}

\begin{figure}[htbp!]
   \centering
   \includegraphics[height=0.35\textheight]{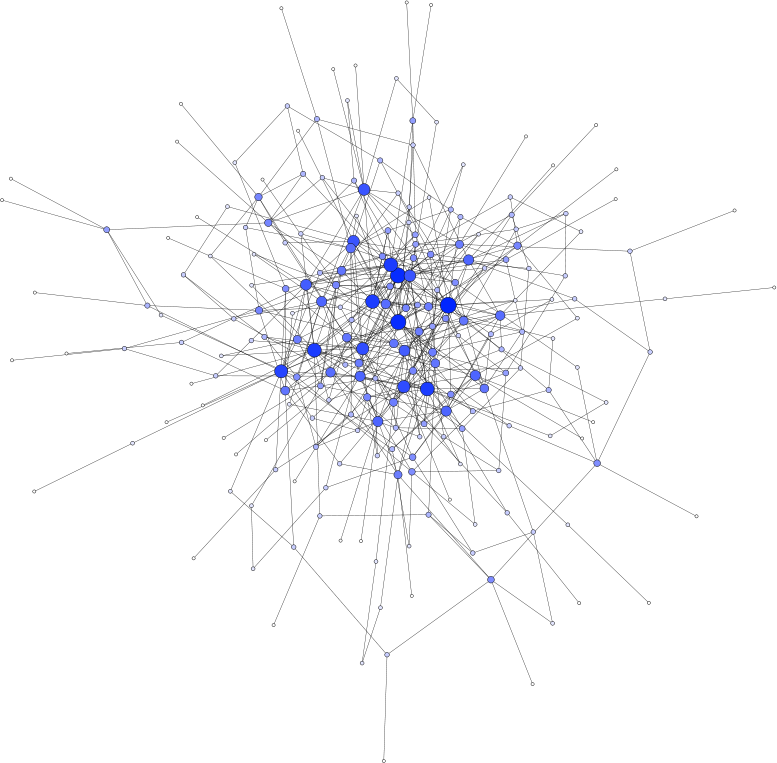}
   \caption{A R-MAT graph.}
\label{fig:rmatgraph}
\end{figure}

R-MAT (Recursive MATrix) is a model that exploits the representation of a graph through its adjacency matrix~\cite{Chakrabarti2004}. In particular, it applies a recursive method to create the adjacency matrix of the graph, thus obtaining a self-similar graph structure. This model captures the community-based pattern appearing in some real networks. Moreover, the generated  graph is characterized by a \pl node degree distribution while showing a small diameter. The idea is to start with an empty $N \times N$ matrix and then divide the square matrix into four partitions in which the nodes are present with a certain probability for each partition, specifically probabilities $a,b,c,d$ that sum to one. The procedure is then repeated dividing each partition again in four sub-partitions and associating the probabilities. The procedure stops when a $1 \times 1$ cell is reached in the iterative procedure. The $a,b,c,d$ partitions of the adjacency matrix have particular meaning: $a$ and $d$ represent the portions containing nodes belonging to different communities, while $b$ and $c$ represent the nodes that act as a link for the different communities (e.g., in a social network people with interests both in topics mostly popular in either $a$ or $d$ community). The recursive nature of this algorithm creates a sort of sub communities at each round. A graph based on R-MAT model with \textit{order} 222 and \textit{size} 499 is shown in Figure~\ref{fig:rmatgraph}.

\subsubsection*{Models Independent from the Average Node Degree}

When generating certain models there is no explicit dependence on the average node degree, these include Random Graph with power-law model, Copying Model, Forest Fire and Kronecker Graph which are presented next.

\paragraph*{Random Graph with Power-law}

A Random Graph with power-law model generates networks characterized by a \pl in the node degree distribution ($P(k)\sim k^{-\gamma}$) having the majority of nodes with a low degree and a small amount of nodes with a very high degree. Power-law distributions are very common in many real life networks both created by natural processes (e.g., food-webs, protein interactions) and by artificial ones (e.g., airline travel routes, Internet routing, telephone call graphs),~\cite{bar:lin03}. The types of networks that follow this property are also referred to as {\em Scale-free networks} (\cite{Barabasi2000,Barabasi2009,Albert2000}). From the dynamic point of view, these networks are modeled by a preferential attachment model.  In addition, reliability is a property of these graphs, that is, high degree of tolerance to random failures  and high sensitivity to targeted attacks towards high degree nodes or hubs~\cite{Albert2000,Moreno03,Crucitti2004a}.

\begin{figure}[htbp!]
   \centering
   \includegraphics[height=0.35\textheight]{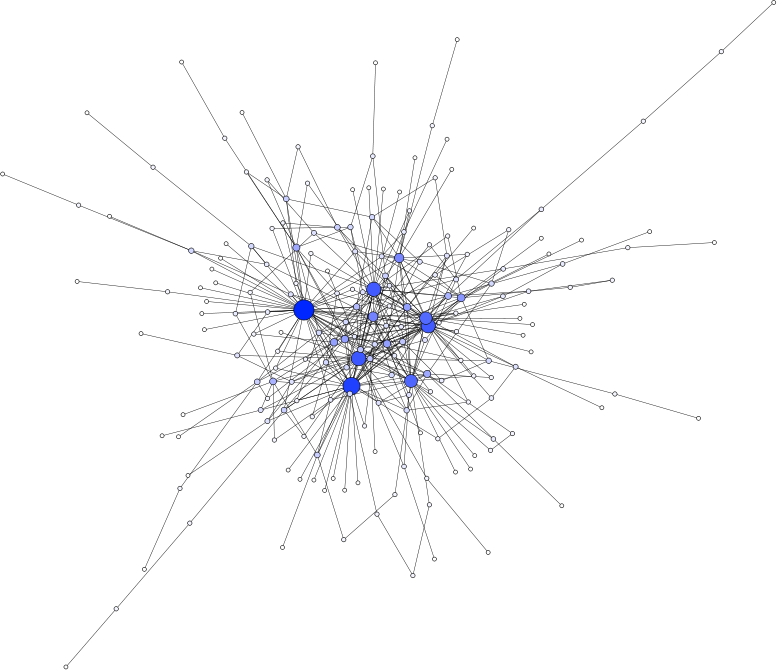}
   \caption{A Random Graph with \pl graph.}
\label{fig:rgraphpl}
\end{figure}

This model is characterized by the exponent of the \pl (i.e., $\gamma$) which governs the degree of each node. The edges between the nodes are then wired in a random fashion. As we have shown earlier, the other way of constructing a graph that is compliant with a \pl based node degree distribution is through the growth of the network and preferential attachment based on node degree. A Random Graph with power-law with \textit{order} 200 and \textit{size} 399 is shown in Figure~\ref{fig:rgraphpl}.

\paragraph*{Copying Model}

\begin{figure}[htbp]
   \centering
   \includegraphics[height=0.35\textheight]{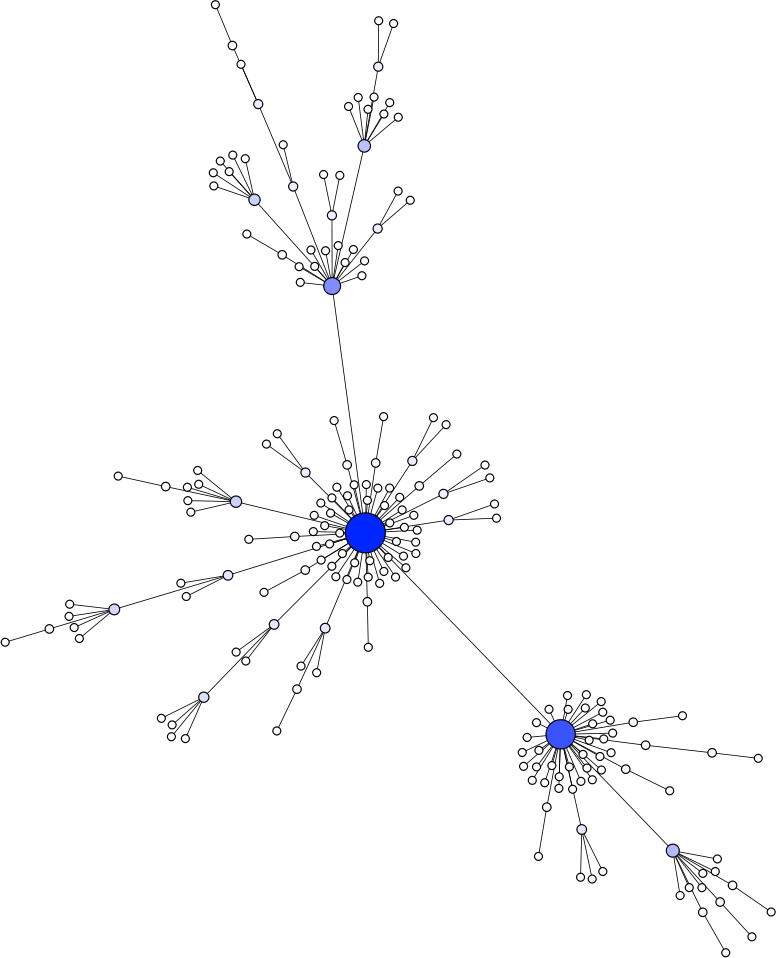}
   \caption{A Copying Model graph.}
\label{fig:copygraph}
\end{figure}

Replicating the structure underlying the links of WWW pages brought the development of the Copying Model~\cite{Kleinberg1999} capturing the tendency of members of communities with same interests to create pages with a similar structure of links. The basic intuition is to select a node and a number ($k$) of edges to add to the node. Then with a certain probability $\beta$, the edges are linked independently and uniformly at random to $k$ other nodes, while with probability $(1-\beta)$ the $k$ edges are copied from a randomly selected node $u$. If $u$ has more than $k$ edges, a subset is chosen, while if it has less than $k$ edges they are anyway copied and the remaining are copied from another randomly chosen node. It leads to a distribution for the incoming degree that follows a \pl with a characteristic parameter $\gamma_{in}=\frac{1}{1-\beta}$. A graph based on Copying Model with \textit{order} 200 and \textit{size} 199 is shown in Figure~\ref{fig:copygraph}.

\paragraph*{Forest Fire}

\begin{figure}[htbp]
   \centering
   \includegraphics[height=0.35\textheight]{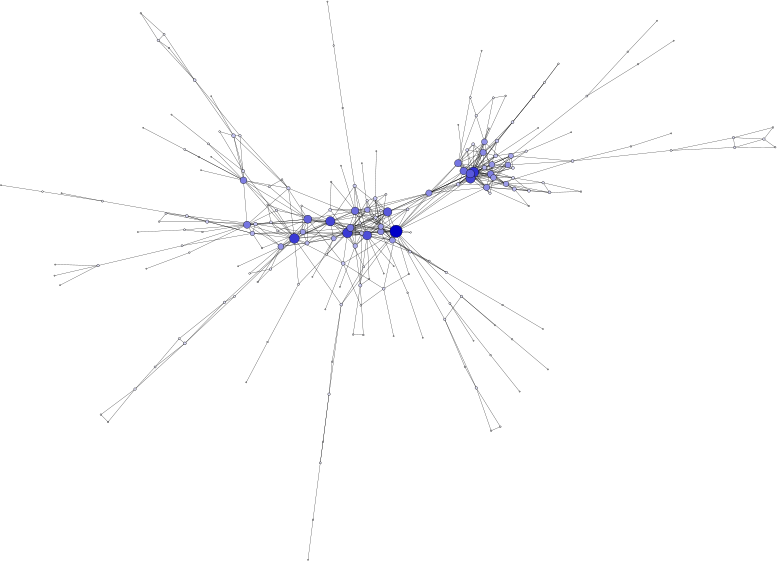}
   \caption{A Forest Fire graph.}
\label{fig:ffgraph}
\end{figure}

In order to capture dynamic aspects of the evolution of networks, Leskoveck \etal~\cite{Leskovec05} proposed the Forest Fire model. The intuition is that networks tend to densify in connectivity and shrink in diameter (i.e., the greatest shortest path in the network) during the growth process; technological, social and information networks show this phenomenon in their growth process. The model requires two parameters known as forward burning probability ($p$) and backward burning ratio ($r$).
The graph grows over time and at each discrete time step a node $v$ is added, then a node $w$, known as \textit{ambassador}, is chosen at random between the other nodes of the graph and a link between $v$ and $w$ is added. A random number $x$ (obtained from a binormal distribution with mean $(1-p)^{-1}$) is chosen and this is the number of out-links of node $w$ that are selected. Then a fraction $r$ times less than the out-links is chosen between the in-links and an edge is created with these as well. The process continues iterating choosing a new $x$ number for each of the nodes $v$ is now connected to. The idea, as the name of the model suggests, resembles the spreading of a fire in a forest that starts from the \textit{ambassador} node to a fraction (based on the probability parameters) of nodes it is connected to and goes on in a sort of chain reaction. This model leads to heavy tails both in the distribution of in-degree and out-degree node degree. In addition, a \pl is shown in the densification process: a new coming node tends to have most of his links in the community of his \textit{ambassador} and just few with other nodes. A graph based on Forest Fire model with \textit{order} 200 and \textit{size} 505 is shown in Figure~\ref{fig:ffgraph}.

\paragraph*{Kronecker Graph}

\begin{figure}[htbp]
   \centering
   \includegraphics[height=0.35\textheight]{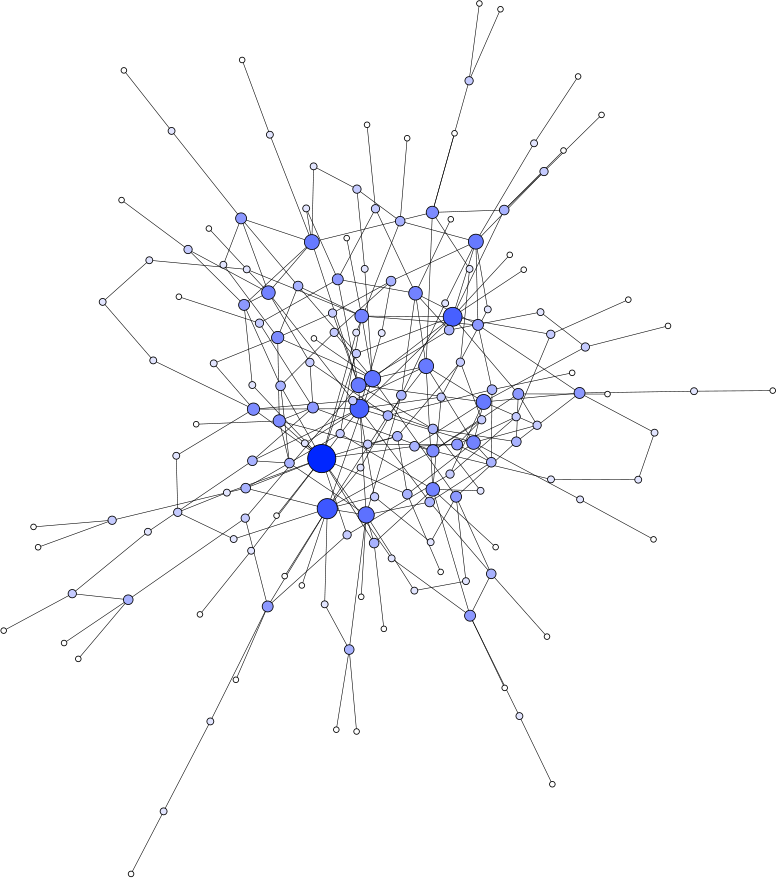}
   \caption{A Kronecker graph.}
\label{fig:krongraph}
\end{figure}

A generating model with a recursive flavor similar to R-MAT uses the Kronecker product applied to the adjacency matrix of a graph~\cite{Leskovec2010}. The Kroneker product is a non conventional way of multiplying two matrices.
\begin{definition}[Kronecker product]\label{def:kronProd}
 Given two matrices A and B with dimension ($n \times m$) and ($n' \times m'$) the Kronecker product between A and B is a matrix C with dimension ($n\cdot n' \times m \cdot m'$) with the following structure:

\[
  C=A \otimes B= \begin{pmatrix}
  a_{1,1}B & a_{1,2}B & \cdots & a_{1,m}B \\
  a_{2,1}B & a_{2,2}B & \cdots & a_{2,m}B \\
  \vdots  & \vdots  & \ddots & \vdots  \\
  a_{n,1}B & a_{n,2}B & \cdots & a_{n,m}B
 \end{pmatrix}
\]
\end{definition}
\begin{definition}[Kronecker Graph]\label{def:krongraph}
Given two graphs G and H with adjacency matrices A(G) and A(H), a Kronecker graph is a graph whose adjacency matrix is obtained by the Kronecker product between the adjacency matrices of G and H.
\end{definition}
If the Kronecker product is applied to the same matrix, therefore multiplying the matrix with itself in the Kronecker product fashion, a self similar structure arises in the graph. This situation can be seen as the increase of a community in a network and the further differentiation in sub-communities while the network grows.
This model creates networks that show a densification in the connectivity of its nodes, which provides a shrinking diameter over time. The idea is to apply the Kronecker product to the same matrix recursively. The procedure to create a graph based on the Kronecker product starts with a $N \times N$ matrix where each $x_{ij}$ element of the matrix represents a probability of having an edge between node $i$ and $j$. Thereafter, at each time step the network grows so that at step $k$ the network has $N^k$ nodes. By applying the Kronecker product to the same matrix leads to the emergence of self-similar fractal-like structures at different scales. This structure mimics a quite natural process that is the recursive growth of communities inside communities which are a miniature copy of a big community (i.e., the whole graph structure)~\cite{Leskovec2010}. A Kronecker Graph with \textit{order} 167 and \textit{size} 264 is shown in Figure~\ref{fig:krongraph}.

\section{Network metrics}\label{sec:metrics}

In~\cite{PaganiAielloTSG2011} we proposed a number of metrics useful for analyzing \PG topologies having in mind decentralized energy trading. We recall them here together with new ones, which we then apply to the evolution/growth models presented in Section~\ref{sec:models}.  We set two main categories of requirements: qualitative and quantitative desiderata the network should satisfy.

\subsection*{Qualitative requirements}

The main qualitative requirement we envision for the future Distribution Network relies on the modularity of the network topology. In the power system domain, the modularity is invoked as a solution that provides benefits reducing uncertainties in energy demand forecasting and costs for energy generation plants as well as risks of technological and regulatory obsolescence~\cite{lovins02,hoff97}. Modularity is usually required not only in the energy sector, but more generally in the design and creation of product or organizations~\cite{gershenson03}. It is also a principle that is promoted in innovation of complex systems~\cite{ethiraj04} for the benefits it provides in terms of reduced design and development time, adaptation and recombination.  We assess the modularity of a network as the ability of building the network using a self-similar recurrent approach and having a repetition of a kind of pattern in its structure.

\subsection*{Quantitative requirements}

As a global statistical tool, quantitative requirements are even more useful as they give a precise indication of network properties. 
Here are the relevant ones when considering efficiency, resilience and robustness of a power system.
\begin{itemize}
\item {\em Characteristic Path Length (CPL) lower or equal to the natural logarithm of order of graph:} $CPL \leq ln(N)$. This requirement represents having a general limited path when moving from one node to another. In the Grid this provides for a network with limited losses in the paths used to transfer energy from one node to another.
\item {\em Clustering Coefficient (CC) which is 5 times higher than a corresponding random graph with same order and size}: $CC \geq 5\times CC_{RG}$. Watts and Strogatz~\cite{Watts98} show that \sw networks have clustering coefficient such that $CC \gg CC_{RG}$. Here we require a similar condition, although less strong by putting a constant value of 5. This requirement is proposed in order to guarantee a local clustering among nodes since it is more likely that energy exchanges occur at a very local scale (e.g., neighborhood) when small-scale distributed energy resources are highly implemented. 
\item {\em Betweenness-related requirements:}
\begin{itemize}
\item {\em A low value for average betweenness in terms of order of the graph} $\overline{\upsilon}=\frac{\overline{\sigma}}{N}$, where $\overline{\sigma}$ is the average betweenness of the graph and $N$ is the \textit{order} of the graph. For the Internet V\'{a}zquez \etal~\cite{Vazquez2002} have found for this metric  $\overline{\upsilon}\approx 2.5$. Internet has proved successful to tolerate failures and attacks~\cite{Cohen2000,Albert2000},
therefore we require a similar value for this metric for the future Grid.
\item {\em A coefficient of variation for betweenness} $c_v=\frac{s}{\overline{x}} < 1$ where $s$ is the sample standard deviation and $\overline{x}$ is the sample mean of betweenness. Usually distribution with $c_v < 1$ are known as low-variance ones.
\end{itemize}
The above two requirements are generally considered to provide network resilience by limiting the number of critical nodes that have a high number of minimal paths traversing them. These properties provide distributions of shortest paths which are more uniform among all nodes.
\item {\em An index for robustness such that} $Rob_N \geq 0.45$. Robustness is evaluated with a random removal strategy and a node degree-based removal strategy by computing the average of the \textit{order} of the maximal connected component (MCC) of the graph between the two situations when the 20\% of the nodes of the original graph are removed~\cite{PaganiAielloTSG2011}. It can be written as $Rob_N = \frac{|MCC_{Random20\%}|+|MCC_{NodeDegree20\%}|}{2}$. Such a requirement is about double the value observed for current \MV and 33\% more for \LV samples~\cite{PaganiAielloTSG2011}.
\item {\em A measure of the cost related to the redundancy of paths available in the network:} $APL_{10^{th}} \leq 2\times CPL$. With this metric we consider the cost of having redundant paths available between nodes. In particular, we evaluate the 10$^{th}$ shortest path (i.e., the shortest path when the nine best ones are not considered) by covering a random sample of the nodes in the network (40\% of the nodes whose half represents source nodes and the other half represents destination nodes). The values for the paths considered are then averaged. In the case where there are less than ten paths available, the worst case path between the two nodes is considered. This last condition gives not completely significant values when applied to  networks with small connectivity (i.e., absence of redundant paths).
\end{itemize}
\begin{center}
\begin{table}[htbp]
\begin{center}
\begin{footnotesize}
    \begin{tabular}{ | l || c | c | c |}
    \hline
    \textbf{Metric} & \textbf{Efficiency} & \textbf{Resilience} & \textbf{Robustness} \\ \hline\hline
    CPL & \checkmark &  &  \\ \hline
    CC & \checkmark & &\\ \hline
    Avg. Betweenness &  & \checkmark & \\ \hline
    Betw. Coeff. of Variation &  & \checkmark & \\ \hline
    $Rob_N$&&& \checkmark \\ \hline
    $APL_{10^{th}}$ & \checkmark &\checkmark & \\
    \hline
    \end{tabular}
\caption{Metrics classification related to properties delivered to the network.}\label{tab:metricsClass}
\end{footnotesize}
\end{center}
\end{table}
\end{center}
The above quantitative metrics can be categorized into three macro categories with respect to how they affect a Power Grid: efficiency in the transfer of energy, resilience in providing alternative path if part of the network is compromised/congested and robustness to failures for network connectivity. Table~\ref{tab:metricsClass} summarized the property each metric assesses.

\section{Generating Smart Grids}\label{sec:results}

Having presented topological models and relevant \PG metrics to evaluate them, it is now time to perform the networks generation for the purpose of assessing their quality. The baseline network for metric analysis must be the real current \PG network. For this purpose, we use actual samples from the \MLV network of the Northern Netherlands (for a complete description of the data we refer to~\cite{PaganiAielloTSG2011,pag:preprint11}).

\begin{center}
\begin{table}[htbp]
\centering
\begin{footnotesize}
\begin{tabular}{|@p{1.7cm}|^p{1.5cm}|^p{0.9cm}|^p{0.6cm}|^p{1cm}|^p{1cm}|^p{1.2cm}|^p{1.5cm}|^p{2cm}|}
\hline
\rowstyle{\bfseries} Network sample & Model & \textit{Order} & \textit{Size} & {Avg. deg.} & {CPL} & {CC} & {Removal robustness ($Rob_N$)} & {Redundancy cost ($APL_{10^{th}}$)} \\ \hline
\hline
LV-Small &  Real data & \rl 21 & \rl 22 & \rl 2.095 & \rl 4.250 & \rl 0.00000 & \rl 0.338 & \rl \rl 12.364 \tn \hline
LV-Medium &  Real data & \rl 63 & \rl 62 & \rl 1.968 & \rl 5.403 & \rl 0.00000 & \rl 0.245 & \rl 5.607 \tn \hline
LV-Large & Real data & \rl 186 & \rl 189 & \rl 2.032 & \rl 17.930 & \rl 0.00000 & \rl 0.134 & \rl 56.733 \tn \hline
MV-Small & Real data & \rl 263 & \rl 288 & \rl 2.190 & \rl 12.672 & \rl 0.01117 & \rl 0.184 & \rl 20.905 \tn \hline
MV-Medium &  Real data & \rl 464 & \rl 499 & \rl 2.151 & \rl 13.107 & \rl 0.00035 & \rl 0.181 & \rl 18.399 \tn \hline
MV-Large & Real data & \rl 884 & \rl 1059 & \rl 2.396 & \rl 9.529 & \rl 0.00494 & \rl 0.298 & \rl 12.809 \tn \hline
\end{tabular}
\caption{Metrics for Dutch \MV and \LV samples.}\label{tab:samples}
\end{footnotesize}
\end{table}
\end{center}
Table~\ref{tab:samples} summarizes the values for the network metrics applied on the Dutch network samples. We notice that the average degree of the \MLV samples scores almost constantly around $<k>\approx2$ independently of the \textit{order} of the network. In the \LV networks we see a tendency towards the increase of characteristic path length, with a value about 18 when the \textit{order} and \textit{size} are about 200 nodes and edges, respectively. The same metric does not have the same clear tendency for the \MV samples. Considering the clustering coefficient there is a general tendency: a null value for the \LV samples and small, but at least significant, values for the \MV samples. These differences in both \cpl and clustering coefficient come from the difference in topology of the two networks. \LV is almost a non-mashed network which resembles for certain samples trees or closed chains with longer paths on average, especially when the network grows. On the other hand, the \MV network is more meshed (despite the same average node degree) with more connections that act as ``shortcuts''. It also has to some extent some redundancy in the connections between the neighborhood of a node, which implies a more significant clustering coefficient compared to the \LV network. The analysis of the robustness metric shows generally poor scores that decrease while the sample increase, at least for the \LV networks, while the tendency is not clear for the \MV samples considered. A common behavior for the \MV samples is the problem they experience in the biggest component connectivity, when the 20\% of the nodes with the highest degree are removed from the network: the robustness falls to 0.0456, 0.0366 and 0.0396 respectively for the \textit{Small}, \textit{Medium} and \textit{Large} sample. Considering the additional effort required when the first nine shortest paths are not available, we see a general increase especially for the \LV samples, where the $10^{th}$ \apl (redundancy cost column in Table~\ref{tab:samples}) increases three times for the \textit{Large} sample analyzed; the increase is still present in Medium Voltage, but it is limited compared to the \LV samples. This is again an indication that the \MV provides more efficient alternative paths to connect nodes. An exception in the results is the \LV \textit{Medium} size sample: here the $10^{th}$ path \apl is really close to the traditional characteristic path length. This is basically due to the absence of alternative paths, therefore the only paths between nodes are at the same time the best and worst case too. This reinforces once again the idea of a \LV network with a fixed structure (sort of chain or tree like) and a limited redundancy.

\begin{center}
\begin{table}[htbp]
\centering
\begin{footnotesize}
\begin{tabular}{|@p{1.7cm}|^p{1.5cm}|^p{0.9cm}|^p{0.6cm}|^p{1.5cm}|^p{1.5cm}|^p{1.2cm}|}
\hline
\rowstyle{\bfseries}
Network sample & Model & \textit{Order} & \textit{Size} & {Avg. betweenness} & {Avg. betw/order} & {Coeff. variation} \\ \hline
\hline
LV-Small & \rl Real data & \rl 21 & \rl 22 & \rl 70.286 & \rl 3.347 & \rl 0.643 \tn \hline
LV-Medium & \rl Real data & \rl 63 & \rl 62 & \rl 255.016 & \rl 4.048 & \rl 2.091 \tn \hline
LV-Large & \rl Real data & \rl 186 & \rl 189 & \rl 2928.227 & \rl 15.743 & \rl 1.207 \tn \hline
MV-Small & \rl Real data & \rl 263 & \rl 288 & \rl 1237.711 & \rl 4.706 & \rl 1.517 \tn \hline
MV-Medium & \rl Real data & \rl 464 & \rl 499 & \rl 3424.602 & \rl 7.381 & \rl 1.687 \tn \hline
MV-Large & \rl Real data & \rl 884 & \rl 1059 & \rl 7755.542 & \rl 8.773 & \rl 2.875 \tn \hline
\end{tabular}
\caption{Betweenness for Dutch \MV and \LV samples.}\label{tab:samplesBet}
\end{footnotesize}
\end{table}
\end{center}
Considering the betweenness-related metrics shown in Table~\ref{tab:samplesBet}, one notices an increase in the average betweenness as the samples become more numerous in the two segments of the network (i.e., \MV and Low Voltage). This same tendency is present also in the average betweenness to \textit{order} ratio: the biggest samples in terms of \textit{order} both of \LV and \MV   score highest. In particular, the \textit{Large} sample belonging to the \LV is almost twice the value of the biggest sample of the Medium Voltage. This again can be justified by the similar-to-tree structure of the \LV sample for which nodes responsible for the paths that enable sub-trees or sub-chains to be connected are the most high scoring for betweenness. This highly increases the average betweenness (while the mode is usually null). The coefficient of variation is above one for all the big samples and reaches almost three for the biggest sample belonging to the \MV network. Such a high value implies a high standard deviation in the betweenness of the nodes, an indication for an heavy-tail distribution.

\subsection*{Model Parameters}\label{sec:modelParams}

To model the future \PG we compare network topologies that quantitatively evolve in \textit{size} and \textit{order}. In particular, we consider the increase of average node degree ($<k> = \frac{2M}{N}$). This evolution implies new cables and costs. For the Random Graph, small-world, preferential attachment and R-MAT models, we consider an evolution in the magnitude of average node degree of $\approx 2$ then $\approx 4$ and $\approx 6$. The idea behind these values for average node degree is to study how the properties and metrics of networks change when increasing the connectivity of the network. 


Each of the models introduced in Section~\ref{sec:models} is defined by a set of parameters. Let us now consider some  meaningful values for each one of them.
\begin{itemize}
\item \textbf{Random Graph}. For the $G_{N,M}$ model, the only parameters needed are the \textit{order} and \textit{size} of the graph to be generated. We use the values shown in Table~\ref{tab:sizes} for the \textit{order}, and the \textit{size} is chosen accordingly to obtain an average node degree of two, four and six, respectively.
\item \textbf{Small-world Graph}. In addition to \textit{order}, the \sw model requires the specification of the average out-degree and the edge rewiring  probability. For the first parameter, we simply provide a value to obtain the desired average node degree (i.e., $<k>\approx 2$, $<k>\approx 4$ and $<k>\approx 6$). The latter parameter represents the probability of rewiring an edge connecting a source node to a different destination node chosen at random. We choose an intermediate approach between the regular lattice (i.e., rewiring probability $p=0$) and random graph extremes (i.e., rewiring probability $p=1$). In fact, we choose a rewiring probability $p=0.4$. This is to give slightly more emphasis to the regular structure of lattice than to the rewiring, since we expect the future \G to have more emphasis on a regular structure than random cabling. This last aspect also helps to satisfy the qualitative requirement of modularity.
\item \textbf{Preferential Attachment}. For the creation of a graph based on growth and preferential attachment model of Barab\'{a}si-Albert~\cite{Barabasi1999}, the only parameters needed are the \textit{order} and \textit{size} of the graph to be generated. We use the values shown in Table~\ref{tab:sizes} for \textit{order} parameter, while \textit{size} parameter is chosen accordingly to obtain an average node degree of two, four and six, respectively.
\item \textbf{R-MAT}. The R-MAT model requires several parameters. First of all, \textit{order} and \textit{size} of the network, then the $a,b,c,d$ parameters which represent the probabilities of the presence of an edge in a certain partition of the adjacency matrix. The order of the graph is chosen so that the nodes are a power of two, in particular, $2^n$ where usually $n=\lceil log_2 N\rceil$. Therefore, we consider for this model the following values for the \textit{order}: \{32,128,256\} for comparison with the Low Voltage, and \{256,512,1024\} for comparison with the Medium Voltage Grids. For the probability parameters, since we have an undirected graph, we have $b=c$, in addition the ratio found between $a$ and $b$, as in many real scenarios according to~\cite{Chakrabarti2004}, is about 3:1. We assume a more highly connected community ($a=0.46$) and a less connected community ($d=0.22$) and a relative smaller connectivity between the two communities ($b=c=0.16$).
\item \textbf{Copying Model}. The copying model requires, in addition to the \textit{order} of the graph, a value for the probability of copying (or not) edges from existing nodes. $(1-\beta)$ is the probability of copying nodes from another node. In the present study, we fix $\beta=0.2$ so as to have a high probability of having a direct (just one-hop since with probability 0.8 each new node copies the connections of another node and attaches directly to them) connection to what might be considered the most reliable energy sources present in the city or villages (at \MV level), while it represents single users or small aggregation of users with high energy capacity at \LV level.
\item \textbf{Forest Fire}. The Forest Fire model requires, in addition to the \textit{order} of the graph, two values representing the probability of forward and backward spread of the ``burning fire''. We choose the same value for both probabilities since our graph is not directed. To avoid a flooding of edges, we choose few small values to assign to forward and backward probability ($p_{fwd}=p_{bwd}=0.2$; $p_{fwd}=p_{bwd}=0.3$; $p_{fwd}=p_{bwd}=0.35$) that give realistic amounts of edges incident to a node on average and that can be compared with the models for which one is able to directly set \textit{order} and \textit{size}.
\item \textbf{Random Graph with Power-law}. For the model representing Random Graph with \pl in node degree distribution, the parameters required are essentially the \textit{order} of the network and the characteristic parameter of the \pl (known as the $\gamma$ coefficient). For the first parameter, we use the usual dimensions (see Table~\ref{tab:sizes}), while for the latter some additional considerations are necessary. We test different types of \pl coefficients characterizing real technological networks. For the non-electrical technological networks (i.e., technological networks which are not Power Grids) we average the values of the \pl characteristic parameter described in~\cite{Clauset07}; the details of the parameters are shown in Table~\ref{tab:plawClaus}. For the \PG networks the $\gamma$ values represent:
\begin{itemize}
\item the findings for the Western and Eastern \HV U.S. Power \G in~\cite{Chassin05}; the values are averaged to have a single $\gamma$, the details are shown in Table~\ref{tab:plawChassin};
\item the findings for the \HV U.S. Western \PG in~\cite{Barabasi1999} which reports a value $\gamma=4$;
\item the findings for the \MLV Dutch \G that follow a \pl in~\cite{PaganiAielloTSG2011}; the values are averaged to have a single $\gamma$, the details are shown in Table~\ref{tab:plawPag}.
\end{itemize}

\begin{table}[htbp]
\centering
\begin{minipage}[b]{0.8\linewidth}
\centering
\begin{center}
\begin{tabular}{|@l|^r|}
\hline
\rowstyle{\bfseries}Type of network & $\gamma$ \\ \hline
Internet degree  & 2.12  \\ \hline
Telephone calls received  & 2.09 \\ \hline
Blackouts &  2.3 \\ \hline
Email address book size &  3.5  \\ \hline
Hits to web-sites &  1.81  \\ \hline
Links to web sites  & 2.336  \\ \hline
 \hline 
Average & 2.359  \\ \hline
\end{tabular}
\caption{Power-law $\gamma$ parameters for technological networks~\cite{Clauset07}.}\label{tab:plawClaus}
\end{center}
\end{minipage}

\hspace{0.5cm}
\begin{minipage}[b]{0.5\linewidth}
\centering
\begin{center}
\centering
\begin{tabular}{|@l|^r|}
\hline
\rowstyle{\bfseries}Type of network & $\gamma$ \\ \hline
Eastern Interconnection  & 3.04  \\ \hline
Western System  & 3.09  \\ \hline 
\hline
Average & 3.065  \\ \hline
\end{tabular}
\caption{Power-law $\gamma$ parameters for \HV U.S. \PG~\cite{Chassin05}.}\label{tab:plawChassin}
\end{center}
\end{minipage}
\end{table}

\begin{table}
\begin{center}
\centering
\begin{tabular}{|@l|^r|}
\hline
\rowstyle{\bfseries}Type of network & $\gamma$ \\ \hline
LV\#5  & 2.402  \\ \hline
LV\#10  & 1.494 \\ \hline
MV\#2 &  1.977 \\ \hline
MV\#3 &  2.282  \\ \hline
 \hline 
Average & 2.039  \\ \hline
\end{tabular}
\caption{Power-law $\gamma$ parameters for Dutch \MLV \PG~\cite{PaganiAielloTSG2011}.}\label{tab:plawPag}
\end{center}
\end{table}

\item \textbf{Kronecker Graph}. For the Kronecker model, the required parameter is the initial dimension of the square matrix to apply the Kronecker product: a $2\times 2$ initiation matrix is a good starting model~\cite{Leskovec2010}. Once the structure of the matrix is defined the initial parameters for the generation matrix need to be evaluated. With a $2\times2$ adjacency matrix for the initial graph $G$:
\[
 A(G) =
 \begin{pmatrix}
  a & b  \\
   c & d 
 \end{pmatrix}
\]
the parameters can be interpreted in a similar fashion as in R-MAT:
\begin{itemize}
\item  $a$ models the ``core'' part of the network and the tightness of its connectivity.
\item $d$ models the ``perifery'' part of the network and the connectivity inside it.
\item $b, c$ model the relationships and interconnections between the core and the periphery.
\end{itemize}
The findings of Leskovec \etal~\cite{Leskovec2010} applying the Kronecker modeling to many different networks report a common recurrent structure for the parameters of the $2\times2$ Kronecker matrix initiator. In particular, the parameters tend to follow the empirical rule $a\gg b\geq c\gg d$ and are usually $a\approx1$, $b\approx c\approx0.6$ and $d\approx0.2$. 
In this work, we consider two sets of parameters characterizing the Kronecker initiator matrix. The first set is extracted and averaged from the technological and social networks parameters extracted from real sample data~\cite{Leskovec2010}. The second set of parameters is obtained applying the fitting procedure to a Kronecker graph to the UCTE \HV \PG data set used in~\cite{casals07,Corominas-murtra2008}, the \HV U.S. Western \PG data set used in~\cite{Watts98}, and the \MLV samples data set used in~\cite{PaganiAielloTSG2011}. All these values have been averaged to obtain just one $2\times 2$ Kronecker generation matrix. A summary of the values for the parameters of the Kronecker matrix used is given in Table~\ref{tab:kronPar}. One notices a very different structure in the matrix parameters between the social and other diverse technological networks and the Power Grid.
\begin{table}
\centering
\begin{tabular}{|@l|^p{2cm}|^p{2cm}|^p{2cm}|^p{2cm}|}
\hline
\rowstyle{\bfseries}Type of network & \multicolumn{4}{|c|}{\textbf{$2\times 2$ Kronecker generator parameters}} \\ \hline
 & a&c&b&d \\ \hline
Social-technological & \rl 0.9578 & \rl 0.4617& \rl  0.4623& \rl  0.3162\tn \hline
\PG & \rl  0.4547& \rl 0.8276& \rl 0.8504& \rl  0.0186\tn \hline
\end{tabular}
\caption{Probability parameters for the $2\times 2$ Kronecker matrix.}\label{tab:kronPar}
\end{table}
\end{itemize}

\begin{table}[htbp]
\begin{footnotesize}
\begin{center}
\begin{tabular}{|@p{1.7cm}|^p{1.58cm}|^p{0.9cm}|^p{0.6cm}|^p{1cm}|^p{1cm}|^p{1.2cm}|^p{1.5cm}|^p{2cm}|}
\hline
\rowstyle{\bfseries} Network type & Model & \textit{Order} & \textit{Size} & {Avg. deg.} & {CPL} & {CC} & {Removal robustness ($Rob_N$)} & {Redundancy cost ($APL_{10^{th}}$)} \\ \hline
\hline
LV-Small &  Small-world & \rl 20 & \rl 20 & \rl 2.000 & \rl 4.053 & \rl 0.00000 & \rl 0.330 & \rl 7.580 \tn \hline
LV-Medium &  Small-world & \rl 90 & \rl 90 & \rl 2.000 & \rl 11.820 & \rl 0.01593 & \rl 0.167 & \rl 12.932 \tn \hline
LV-Large & Small-world & \rl 200 & \rl 201 & \rl 2.010 & \rl 17.397 & \rl 0.01083 & \rl 0.109 & \rl 21.544 \tn \hline
MV-Small &  Small-world & \rl 250 & \rl 250 & \rl 2.000 & \rl 24.237 & \rl 0.00000 & \rl 0.087 & \rl 24.534 \tn \hline
MV-Medium &  Small-world & \rl 500 & \rl 501 & \rl 2.004 & \rl 28.084 & \rl 0.00000 & \rl 0.057 & \rl 35.413 \tn \hline
MV-Large &  Small-world & \rl 1000 & \rl 1001 & \rl 2.002 & \rl 47.077 & \rl 0.00000 & \rl 0.040 & \rl 60.074 \tn \hline
 \hline
LV-Small &  Preferential attachment & \rl 20 & \rl 19 & \rl 1.900 & \rl 2.579 & \rl 0.00000 & \rl 0.349 & \rl 2.800 \tn \hline
LV-Medium &  Preferential attachment & \rl 90 & \rl 89 & \rl 1.978 & \rl 4.315 & \rl 0.00000 & \rl 0.263 & \rl 4.471 \tn \hline
LV-Large &  Preferential attachment & \rl 200 & \rl 199 & \rl 1.990 & \rl 6.523 & \rl 0.00000 & \rl 0.206 & \rl 6.375 \tn \hline
MV-Small &  Preferential attachment & \rl 250 & \rl 249 & \rl 1.992 & \rl 5.426 & \rl 0.00000 & \rl 0.245 & \rl 5.570 \tn \hline
MV-Medium &  Preferential attachment & \rl 500 & \rl 499 & \rl 1.996 & \rl 5.705 & \rl 0.00000 & \rl 0.231 & \rl 5.745 \tn \hline
MV-Large &  Preferential attachment & \rl 1000 & \rl 999 & \rl 1.998 & \rl 6.976 & \rl 0.00000 & \rl 0.187 & \rl 6.908 \tn \hline
\hline
LV-Small &  Random Graph & \rl 17 & \rl 21 & \rl 2.471 & \rl 2.938 & \rl 0.07451 & \rl 0.390 & \rl 7.472 \tn \hline
LV-Medium &  Random Graph & \rl 78 & \rl 92 & \rl 2.359 & \rl 5.987 & \rl 0.03547 & \rl 0.418 & \rl 10.974 \tn \hline
LV-Large &  Random Graph & \rl 172 & \rl 207 & \rl 2.407 & \rl 6.254 & \rl 0.00736 & \rl 0.354 & \rl 10.796 \tn \hline
MV-Small &  Random Graph & \rl 224 & \rl 259 & \rl 2.313 & \rl 7.269 & \rl 0.00000 & \rl 0.322 & \rl 12.002 \tn \hline
MV-Medium &  Random Graph & \rl 435 & \rl 516 & \rl 2.372 & \rl 8.380 & \rl 0.00138 & \rl 0.321 & \rl 12.818 \tn \hline
MV-Large &  Random Graph & \rl 863 & \rl 1026 & \rl 2.378 & \rl 9.061 & \rl 0.00070 & \rl 0.328 & \rl 13.446 \tn \hline
 \hline
LV-Small &  R-MAT & \rl 27 & \rl 31 & \rl 2.296 & \rl 3.615 & \rl 0.00000 & \rl 0.356 & \rl 7.830 \tn \hline
LV-Medium &  R-MAT & \rl 88 & \rl 125 & \rl 2.841 & \rl 4.115 & \rl 0.05688 & \rl 0.369 & \rl 6.418 \tn \hline
LV-Large &  R-MAT & \rl 199 & \rl 261 & \rl 2.623 & \rl 5.495 & \rl 0.00737 & \rl 0.364 & \rl 8.774 \tn \hline
MV-Small &  R-MAT & \rl 195 & \rl 263 & \rl 2.697 & \rl 5.629 & \rl 0.00865 & \rl 0.378 & \rl 8.642 \tn \hline
MV-Medium &  R-MAT & \rl 365 & \rl 523 & \rl 2.866 & \rl 5.470 & \rl 0.01360 & \rl 0.396 & \rl 7.646 \tn \hline
MV-Large &  R-MAT & \rl 728 & \rl 1056 & \rl 2.901 & \rl 5.726 & \rl 0.00589 & \rl 0.363 & \rl 7.887 \tn \hline
\end{tabular}
\caption{Metrics for small-world, preferential attachment, Random Graph and R-MAT models with average node degree $\approx 2$.}\label{tab:gen2}
\end{center}
\end{footnotesize}
\end{table}

\begin{table}[htbp]
\begin{footnotesize}
\begin{center}
\begin{tabular}{|@p{1.7cm}|^p{1.58cm}|^p{0.9cm}|^p{0.6cm}|^p{1.5cm}|^p{1.5cm}|^p{1.2cm}|}
\hline
\rowstyle{\bfseries}
Network type & Model & \textit{Order} & \textit{Size} & {Avg. betweenness} & {Avg. betw/order} & {Coeff. variation} \\ \hline
\hline
LV-Small &  Small-world & \rl 20 & \rl 20 & \rl 62.300 & \rl 3.115 & \rl 0.804 \tn \hline
LV-Medium &  Small-world & \rl 90 & \rl 90 & \rl 985.956 & \rl 10.955 & \rl 1.307 \tn \hline
LV-Large &  Small-world & \rl 200 & \rl 201 & \rl 3429.720 & \rl 17.149 & \rl 1.260 \tn \hline
MV-Small &  Small-world & \rl 250 & \rl 250 & \rl 5881.296 & \rl 23.525 & \rl 1.598 \tn \hline
MV-Medium &  Small-world & \rl 500 & \rl 501 & \rl 13980.228 & \rl 27.960 & \rl 1.745 \tn \hline
MV-Large &  Small-world & \rl 1000 & \rl 1001 & \rl 47919.616 & \rl 47.920 & \rl 2.279 \tn \hline
\hline
LV-Small &  Preferential attachment & \rl 20 & \rl 19 & \rl 31.400 & \rl 1.570 & \rl 2.344 \tn \hline
LV-Medium &  Preferential attachment & \rl 90 & \rl 89 & \rl 293.400 & \rl 3.260 & \rl 3.068 \tn \hline
LV-Large &  Preferential attachment & \rl 200 & \rl 199 & \rl 1089.260 & \rl 5.446 & \rl 3.288 \tn \hline
MV-Small &  Preferential attachment & \rl 250 & \rl 249 & \rl 1096.144 & \rl 4.385 & \rl 3.972 \tn \hline
MV-Medium &  Preferential attachment & \rl 500 & \rl 499 & \rl 2401.680 & \rl 4.803 & \rl 5.049 \tn \hline
MV-Large &  Preferential attachment & \rl 1000 & \rl 999 & \rl 6061.288 & \rl 6.061 & \rl 6.240 \tn \hline
\hline
LV-Small &  Random Graph & \rl 17 & \rl 21 & \rl 31.059 & \rl 1.827 & \rl 1.157 \tn \hline
LV-Medium &  Random Graph & \rl 78 & \rl 92 & \rl 408.308 & \rl 5.235 & \rl 1.126 \tn \hline
LV-Large &  Random Graph & \rl 172 & \rl 207 & \rl 938.512 & \rl 5.456 & \rl 1.276 \tn \hline
MV-Small &  Random Graph & \rl 224 & \rl 259 & \rl 1474.143 & \rl 6.581 & \rl 1.265 \tn \hline
MV-Medium &  Random Graph & \rl 435 & \rl 516 & \rl 3415.890 & \rl 7.853 & \rl 1.204 \tn \hline
MV-Large &  Random Graph & \rl 863 & \rl 1026 & \rl 7081.119 & \rl 8.205 & \rl 1.264 \tn \hline
\hline
LV-Small &  R-MAT & \rl 27 & \rl 31 & \rl 70.593 & \rl 2.615 & \rl 1.320 \tn \hline
LV-Medium &  R-MAT & \rl 88 & \rl 125 & \rl 282.500 & \rl 3.210 & \rl 1.540 \tn \hline
LV-Large &  R-MAT & \rl 199 & \rl 261 & \rl 937.578 & \rl 4.711 & \rl 1.297 \tn \hline
MV-Small &  R-MAT & \rl 195 & \rl 263 & \rl 959.118 & \rl 4.919 & \rl 1.395 \tn \hline
MV-Medium &  R-MAT & \rl 365 & \rl 523 & \rl 1692.910 & \rl 4.638 & \rl 1.581 \tn \hline
MV-Large &  R-MAT & \rl 728 & \rl 1056 & \rl 3633.473 & \rl 4.991 & \rl 2.004 \tn \hline
\end{tabular}
\caption{Betweenness metrics for small-world, preferential attachment, Random Graph and R-MAT models with average node degree $\approx 2$.}\label{tab:genBet2}
\end{center}
\end{footnotesize}
\end{table}

\begin{table}[htbp]
\begin{footnotesize}
\begin{center}
\begin{tabular}{|@p{1.7cm}|^p{1.58cm}|^p{0.9cm}|^p{0.6cm}|^p{1cm}|^p{1cm}|^p{1.2cm}|^p{1.5cm}|^p{2cm}|}
\hline
\rowstyle{\bfseries} Network type & Model & \textit{Order} & \textit{Size} & {Avg. deg.} & {CPL} & {CC} & {Removal robustness ($Rob_N$)} & {Redundancy cost ($APL_{10^{th}}$)} \\ \hline
\hline
LV-Small &  Small-world & \rl 20 & \rl 39 & \rl 3.900 & \rl 2.289 & \rl 0.26000 & \rl 0.721 & \rl 4.720 \tn \hline
LV-Medium &  Small-world & \rl 90 & \rl 177 & \rl 3.933 & \rl 3.652 & \rl 0.14646 & \rl 0.780 & \rl 6.032 \tn \hline
LV-Large &  Small-world & \rl 200 & \rl 399 & \rl 3.990 & \rl 4.407 & \rl 0.15367 & \rl 0.767 & \rl 6.631 \tn \hline
MV-Small &  Small-world & \rl 250 & \rl 498 & \rl 3.984 & \rl 4.566 & \rl 0.12581 & \rl 0.779 & \rl 6.836 \tn \hline
MV-Medium &  Small-world & \rl 500 & \rl 1000 & \rl 4.000 & \rl 5.067 & \rl 0.10681 & \rl 0.764 & \rl 7.231 \tn \hline
MV-Large &  Small-world & \rl 1000 & \rl 1998 & \rl 3.996 & \rl 5.749 & \rl 0.10879 & \rl 0.781 & \rl 7.910 \tn \hline
\hline
LV-Small &  Preferential attachment & \rl 20 & \rl 37 & \rl 3.700 & \rl 2.263 & \rl 0.47341 & \rl 0.554 & \rl 4.380 \tn \hline
LV-Medium &  Preferential attachment & \rl 90 & \rl 177 & \rl 3.933 & \rl 2.910 & \rl 0.11216 & \rl 0.426 & \rl 4.788 \tn \hline
LV-Large &  Preferential attachment & \rl 200 & \rl 397 & \rl 3.970 & \rl 3.322 & \rl 0.09566 & \rl 0.448 & \rl 5.047 \tn \hline
MV-Small &  Preferential attachment & \rl 250 & \rl 497 & \rl 3.976 & \rl 3.504 & \rl 0.08400 & \rl 0.419 & \rl 4.998 \tn \hline
MV-Medium &  Preferential attachment & \rl 500 & \rl 997 & \rl 3.988 & \rl 3.687 & \rl 0.03929 & \rl 0.401 & \rl 5.232 \tn \hline
MV-Large &  Preferential attachment & \rl 1000 & \rl 1997 & \rl 3.994 & \rl 4.211 & \rl 0.01536 & \rl 0.401 & \rl 5.678 \tn \hline
\hline
LV-Small &  Random Graph & \rl 20 & \rl 40 & \rl 4.000 & \rl 2.079 & \rl 0.17667 & \rl 0.733 & \rl 4.350 \tn \hline
LV-Medium &  Random Graph & \rl 87 & \rl 180 & \rl 4.138 & \rl 3.174 & \rl 0.03418 & \rl 0.735 & \rl 5.368 \tn \hline
LV-Large &  Random Graph & \rl 199 & \rl 400 & \rl 4.020 & \rl 3.869 & \rl 0.03064 & \rl 0.734 & \rl 6.107 \tn \hline
MV-Small &  Random Graph & \rl 247 & \rl 500 & \rl 4.049 & \rl 4.057 & \rl 0.01681 & \rl 0.740 & \rl 6.432 \tn \hline
MV-Medium &  Random Graph & \rl 494 & \rl 1000 & \rl 4.049 & \rl 4.495 & \rl 0.00823 & \rl 0.749 & \rl 6.670 \tn \hline
MV-Large &  Random Graph & \rl 987 & \rl 2001 & \rl 4.055 & \rl 5.062 & \rl 0.00359 & \rl 0.738 & \rl 7.150 \tn \hline
\hline
LV-Small &  R-MAT & \rl 30 & \rl 59 & \rl 3.933 & \rl 2.517 & \rl 0.27360 & \rl 0.579 & \rl 4.511 \tn \hline
LV-Medium &  R-MAT & \rl 105 & \rl 250 & \rl 4.762 & \rl 3.019 & \rl 0.13039 & \rl 0.581 & \rl 4.490 \tn \hline
LV-Large &  R-MAT & \rl 227 & \rl 504 & \rl 4.441 & \rl 3.619 & \rl 0.04683 & \rl 0.601 & \rl 5.302 \tn \hline
MV-Small &  R-MAT & \rl 230 & \rl 496 & \rl 4.313 & \rl 3.736 & \rl 0.02940 & \rl 0.626 & \rl 5.381 \tn \hline
MV-Medium &  R-MAT & \rl 420 & \rl 1004 & \rl 4.781 & \rl 3.915 & \rl 0.00450 & \rl 0.591 & \rl 5.249 \tn \hline
MV-Large &  R-MAT & \rl 932 & \rl 2039 & \rl 4.376 & \rl 4.562 & \rl 0.00875 & \rl 0.690 & \rl 6.251 \tn \hline
\end{tabular}
\caption{Metrics for small-world, preferential attachment, Random Graph and R-MAT models with average node degree $\approx 4$.}\label{tab:gen4}
\end{center}
\end{footnotesize}
\end{table}

\begin{table}[htbp]
\begin{footnotesize}
\begin{center}
\begin{tabular}{|@p{1.7cm}|^p{1.58cm}|^p{0.9cm}|^p{0.6cm}|^p{1.5cm}|^p{1.5cm}|^p{1.2cm}|}
\hline
\rowstyle{\bfseries}
Network type & Model & \textit{Order} & \textit{Size} & {Avg. betweenness} & {Avg. betw/order} & {Coeff. variation} \\ \hline
\hline
LV-Small &  Small-world & \rl 20 & \rl 39 & \rl 24.900 & \rl 1.245 & \rl 0.654 \tn \hline
LV-Medium &  Small-world & \rl 90 & \rl 177 & \rl 235.244 & \rl 2.614 & \rl 0.653 \tn \hline
LV-Large &  Small-world & \rl 200 & \rl 399 & \rl 683.780 & \rl 3.419 & \rl 0.703 \tn \hline
MV-Small &  Small-world & \rl 250& \rl 498 & \rl 897.568 & \rl 3.590 & \rl 0.653 \tn \hline
MV-Medium &  Small-world & \rl 500& \rl 1000 & \rl 2043.600 & \rl 4.087 & \rl 0.706 \tn \hline
MV-Large &  Small-world & \rl 1000 & \rl 1998 & \rl 4762.808 & \rl 4.763 & \rl 0.677 \tn \hline
  \hline
LV-Small &  Preferential attachment & \rl 20& \rl 37 & \rl 23.100 & \rl 1.155 & \rl 1.505 \tn \hline
LV-Medium &  Preferential attachment & \rl 90 & \rl 177 & \rl 170.644 & \rl 1.896 & \rl 2.219 \tn \hline
LV-Large &  Preferential attachment & \rl 200 & \rl 397 & \rl 463.060 & \rl 2.315 & \rl 2.733 \tn \hline
MV-Small &  Preferential attachment & \rl 250 & \rl 497 & \rl 611.520 & \rl 2.446 & \rl 3.017 \tn \hline
MV-Medium &  Preferential attachment & \rl 500 & \rl 997& \rl 1342.864 & \rl 2.686 & \rl 3.484 \tn \hline
MV-Large &  Preferential attachment & \rl 1000 & \rl 1997 & \rl 3179.750 & \rl 3.180 & \rl 3.450 \tn \hline
 \hline
LV-Small &  Random Graph & \rl 20 & \rl 40 & \rl 23.600 & \rl 1.180 & \rl 0.807 \tn \hline
LV-Medium &  Random Graph & \rl 87 & \rl 180 & \rl 196.345 & \rl 2.257 & \rl 0.850 \tn \hline
LV-Large &  Random Graph & \rl 199 & \rl 400 & \rl 589.849 & \rl 2.964 & \rl 0.889 \tn \hline
MV-Small & Random Graph & \rl 247 & \rl 500 & \rl 766.389 & \rl 3.103 & \rl 0.857 \tn \hline
MV-Medium &  Random Graph & \rl 494 & \rl 1000 & \rl 1768.757 & \rl 3.580 & \rl 0.972 \tn \hline
MV-Large &  Random Graph & \rl 987 & \rl 2001 & \rl 4068.393 & \rl 4.122 & \rl 0.942 \tn \hline
  \hline
LV-Small &  R-MAT & \rl 30 & \rl 59 & \rl 44.000 & \rl 1.467 & \rl 1.342 \tn \hline
LV-Medium &  R-MAT & \rl 105 & \rl 250 & \rl 223.733 & \rl 2.131 & \rl 1.695 \tn \hline
LV-Large &  R-MAT & \rl 227 & \rl 504 & \rl 609.419 & \rl 2.685 & \rl 1.493 \tn \hline
MV-Small & R-MAT & \rl 230 & \rl 496 & \rl 650.374 & \rl 2.828 & \rl 1.468 \tn \hline
MV-Medium &  R-MAT & \rl 420 & \rl 1004 & \rl 1285.786 & \rl 3.061 & \rl 1.652 \tn \hline
MV-Large &  R-MAT & \rl 932 & \rl 2039 & \rl 3422.348 & \rl 3.672 & \rl 1.506 \tn \hline
\end{tabular}
\caption{Betweenness metrics for small-world, preferential attachment, Random Graph and R-MAT models with average node degree $\approx 4$.}\label{tab:genBet4}
\end{center}
\end{footnotesize}
\end{table}

\begin{table}[htbp]
\begin{footnotesize}
\begin{center}
\begin{tabular}{|@p{1.7cm}|^p{1.58cm}|^p{0.9cm}|^p{0.6cm}|^p{1cm}|^p{1cm}|^p{1.2cm}|^p{1.5cm}|^p{2cm}|}
\hline
\rowstyle{\bfseries} Network type & Model & \textit{Order} & \textit{Size} & {Avg. deg.} & {CPL} & {CC} & {Removal robustness ($Rob_N$)} & {Redundancy cost ($APL_{10^{th}}$)} \\ \hline
\hline
LV-Small &  Small-world & \rl 20 & \rl 59 & \rl 5.900 & \rl 1.816 & \rl 0.33250 & \rl 0.775 & \rl 3.470 \tn \hline
LV-Medium &  Small-world & \rl 90 & \rl 266 & \rl 5.911 & \rl 2.809 & \rl 0.20131 & \rl 0.794 & \rl 4.508 \tn \hline
LV-Large &  Small-world & \rl 200 & \rl 598 & \rl 5.980 & \rl 3.324 & \rl 0.13596 & \rl 0.797 & \rl 4.895 \tn \hline
MV-Small &  Small-world & \rl 250 & \rl 747 & \rl 5.976 & \rl 3.486 & \rl 0.14477 & \rl 0.798 & \rl 5.039 \tn \hline
MV-Medium &  Small-world & \rl 500 & \rl 1494 & \rl 5.976 & \rl 3.968 & \rl 0.14477 & \rl 0.799 & \rl 5.518 \tn \hline
MV-Large & Small-world & \rl 1000 & \rl 2996 & \rl 5.992 & \rl 4.429 & \rl 0.14854 & \rl 0.797 & \rl 5.905 \tn \hline
\hline
LV-Small &  Preferential attachment & \rl 20 & \rl 54 & \rl 5.400 & \rl 1.868 & \rl 0.34839 & \rl 0.749 & \rl 3.460 \tn \hline
LV-Medium &  Preferential attachment & \rl 90 & \rl 264 & \rl 5.867 & \rl 2.466 & \rl 0.16601 & \rl 0.742 & \rl 3.933 \tn \hline
LV-Large &  Preferential attachment & \rl 200 & \rl 594 & \rl 5.940 & \rl 2.854 & \rl 0.08772 & \rl 0.671 & \rl 4.130 \tn \hline
MV-Small &  Preferential attachment & \rl 250 & \rl 744 & \rl 5.952 & \rl 2.926 & \rl 0.08676 & \rl 0.705 & \rl 4.257 \tn \hline
MV-Medium &  Preferential attachment & \rl 500 & \rl 1495 & \rl 5.980 & \rl 3.185 & \rl 0.05017 & \rl 0.667 & \rl 4.481 \tn \hline
MV-Large &  Preferential attachment & \rl 1000 & \rl 2994 & \rl 5.988 & \rl 3.487 & \rl 0.03335 & \rl 0.679 & \rl 4.664 \tn \hline
\hline
LV-Small &  Random Graph & \rl 20 & \rl 60 & \rl 6.000 & \rl 1.684 & \rl 0.29599 & \rl 0.775 & \rl 3.370 \tn \hline
LV-Medium &  Random Graph & \rl 90 & \rl 270 & \rl 6.000 & \rl 2.640 & \rl 0.06987 & \rl 0.791 & \rl 4.298 \tn \hline
LV-Large &  Random Graph & \rl 200 & \rl 600 & \rl 6.000 & \rl 3.141 & \rl 0.03991 & \rl 0.777 & \rl 4.693 \tn \hline
MV-Small &  Random Graph & \rl 249 & \rl 750 & \rl 6.024 & \rl 3.230 & \rl 0.01934 & \rl 0.793 & \rl 4.884 \tn \hline
MV-Medium &  Random Graph & \rl 499 & \rl 1500 & \rl 6.012 & \rl 3.620 & \rl 0.00976 & \rl 0.792 & \rl 5.284 \tn \hline
MV-Large &  Random Graph & \rl 998 & \rl 3000 & \rl 6.012 & \rl 4.022 & \rl 0.00544 & \rl 0.791 & \rl 5.662 \tn \hline
\hline
LV-Small &  R-MAT & \rl 32 & \rl 87 & \rl 5.438 & \rl 2.194 & \rl 0.21179 & \rl 0.760 & \rl 3.945 \tn \hline
LV-Medium &  R-MAT & \rl 123 & \rl 374 & \rl 6.081 & \rl 2.926 & \rl 0.08173 & \rl 0.717 & \rl 4.377 \tn \hline
LV-Large &  R-MAT & \rl 249 & \rl 759 & \rl 6.096 & \rl 3.165 & \rl 0.04444 & \rl 0.736 & \rl 4.622 \tn \hline
MV-Small &  R-MAT & \rl 236 & \rl 747 & \rl 6.331 & \rl 3.143 & \rl 0.04982 & \rl 0.746 & \rl 4.389 \tn \hline
MV-Medium &  R-MAT & \rl 466 & \rl 1512 & \rl 6.489 & \rl 3.427 & \rl 0.04365 & \rl 0.743 & \rl 4.805 \tn \hline
MV-Large &  R-MAT & \rl 925 & \rl 3035 & \rl 6.562 & \rl 3.742 & \rl 0.02560 & \rl 0.723 & \rl 4.925 \tn \hline
\end{tabular}
\caption{Metrics for small-world, preferential attachment, Random Graph and R-MAT models with average node degree $\approx 6$.}\label{tab:gen6}
\end{center}
\end{footnotesize}
\end{table}

\begin{table}[htbp]
\begin{footnotesize}
\begin{center}
\begin{tabular}{|@p{1.7cm}|^p{1.58cm}|^p{0.9cm}|^p{0.6cm}|^p{1.5cm}|^p{1.5cm}|^p{1.2cm}|}
\hline
\rowstyle{\bfseries}
Network type & Model & \textit{Order} & \textit{Size} & {Avg. betweenness} & {Avg. betw/order} & {Coeff. variation} \\ \hline
\hline
LV-Small & Small-world & \rl 20 & \rl 39 & \rl 15.800 & \rl 0.790 & \rl 0.581 \tn \hline
LV-Medium &  Small-world & \rl 90 & \rl 177 & \rl 163.778 & \rl 1.820 & \rl 0.555 \tn \hline
LV-Large &  Small-world & \rl 200 & \rl 399 & \rl 464.330 & \rl 2.322 & \rl 0.617 \tn \hline
MV-Small &  Small-world & \rl 250 & \rl 498 & \rl 621.488 & \rl 2.486 & \rl 0.609 \tn \hline
MV-Medium &  Small-world & \rl 500 & \rl 1000 & \rl 1479.404 & \rl 2.959 & \rl 0.565 \tn \hline
MV-Large &  Small-world & \rl 1000 & \rl 1998 & \rl 3441.742 & \rl 3.442 & \rl 0.564 \tn \hline
  \hline
LV-Small &  Preferential attachment & \rl 20 & \rl 37 & \rl 15.900 & \rl 0.795 & \rl 1.292 \tn \hline
LV-Medium &  Preferential attachment & \rl 90 & \rl 177 & \rl 133.378 & \rl 1.482 & \rl 2.640 \tn \hline
LV-Large &  Preferential attachment & \rl 200 & \rl 397 & \rl 374.970 & \rl 1.875 & \rl 2.401 \tn \hline
MV-Small &  Preferential attachment & \rl 250 & \rl 497 & \rl 485.352 & \rl 1.941 & \rl 2.514 \tn \hline
MV-Medium &  Preferential attachment & \rl 500 & \rl 997 & \rl 1095.116 & \rl 2.190 & \rl 2.894 \tn \hline
MV-Large & Preferential attachment & \rl 1000 & \rl 1997 & \rl 2447.594 & \rl 2.448 & \rl 3.283 \tn \hline
 \hline
LV-Small &  Random Graph & \rl 20 & \rl 40 & \rl 14.700 & \rl 0.735 & \rl 0.662 \tn \hline
LV-Medium &  Random Graph & \rl 87 & \rl 180 & \rl 151.489 & \rl 1.683 & \rl 0.809 \tn \hline
LV-Large & Random Graph & \rl 199 & \rl 400 & \rl 431.090 & \rl 2.155 & \rl 0.835 \tn \hline
MV-Small &  Random Graph & \rl 247 & \rl 500 & \rl 563.839 & \rl 2.264 & \rl 0.710 \tn \hline
MV-Medium &  Random Graph & \rl 494 & \rl 1000 & \rl 1328.405 & \rl 2.662 & \rl 0.745 \tn \hline
MV-Large &  Random Graph & \rl 987 & \rl 2001 & \rl 3051.922 & \rl 3.058 & \rl 0.771 \tn \hline
  \hline
LV-Small &  R-MAT & \rl 30 & \rl 59 & \rl 38.000 & \rl 1.188 & \rl 0.989 \tn \hline
LV-Medium &  R-MAT & \rl 105 & \rl 250 & \rl 247.008 & \rl 2.008 & \rl 1.351 \tn \hline
LV-Large &  R-MAT & \rl 227 & \rl 504 & \rl 550.538 & \rl 2.211 & \rl 1.352 \tn \hline
MV-Small &  R-MAT & \rl 230 & \rl 496 & \rl 530.093 & \rl 2.246 & \rl 1.357 \tn \hline
MV-Medium &  R-MAT & \rl 420 & \rl 1004 & \rl 1169.382 & \rl 2.509 & \rl 1.506 \tn \hline
MV-Large &  R-MAT & \rl 932 & \rl 2039 & \rl 2599.496 & \rl 2.810 & \rl 1.731 \tn \hline
\end{tabular}
\caption{Betweenness-related metrics for small-world, preferential attachment, Random Graph and R-MAT models with average node degree $\approx 6$.}\label{tab:genBet6}
\end{center}
\end{footnotesize}
\end{table}

\begin{sidewaystable}[htbp]
\centering
\begin{footnotesize}
\begin{tabular}{|@p{1.7cm}|^p{7.5cm}|^p{0.9cm}|^p{0.6cm}|^p{1cm}|^p{1cm}|^p{1.2cm}|^p{1.5cm}|^p{2cm}|}
\hline
\rowstyle{\bfseries} Network type & Model & \textit{Order} & \textit{Size} & {Avg. deg.} & {CPL} & {CC} & {Removal robustness ($Rob_N$)} & {Redundancy cost ($APL_{10^{th}}$)} \\ \hline
\hline
LV-Small &  Copying Model & \rl 20 & \rl 19 & \rl 1.900 & \rl 2.053 & \rl 0.00000 & \rl 0.312 & \rl 2.060 \tn \hline
LV-Medium &  Copying Model & \rl 90 & \rl 89 & \rl 1.978 & \rl 3.337 & \rl 0.00000 & \rl 0.287 & \rl 3.996 \tn \hline
LV-Large &  Copying Model & \rl 200 & \rl 199 & \rl 1.990 & \rl 3.588 & \rl 0.00000 & \rl 0.271 & \rl 3.875 \tn \hline
MV-Small &  Copying Model & \rl 250 & \rl 249 & \rl 1.992 & \rl 3.253 & \rl 0.00000 & \rl 0.287 & \rl 2.958 \tn \hline
MV-Medium & Copying Model & \rl 500 & \rl 499 & \rl 1.996 & \rl 3.762 & \rl 0.00000 & \rl 0.273 & \rl 3.775 \tn \hline
MV-Large &  Copying Model & \rl 1000 & \rl 999 & \rl 1.998 & \rl 3.898 & \rl 0.00000 & \rl 0.280 & \rl 3.782 \tn \hline
\hline
LV-Small & Forest Fire (fwdPb=bckPb=0.2) & \rl 20 & \rl 22 & \rl 2.200 & \rl 3.263 & \rl 0.11476 & \rl 0.352 & \rl 6.760 \tn \hline
LV-Medium &  Forest Fire (fwdPb=bckPb=0.2) & \rl 90 & \rl 131 & \rl 2.911 & \rl 5.691 & \rl 0.25303 & \rl 0.279 & \rl 7.681 \tn \hline
LV-Large & Forest Fire (fwdPb=bckPb=0.2) & \rl 200 & \rl 295 & \rl 2.950 & \rl 5.475 & \rl 0.27328 & \rl 0.278 & \rl 7.323 \tn \hline
MV-Small &  Forest Fire (fwdPb=bckPb=0.2) & \rl 250 & \rl 336 & \rl 2.688 & \rl 6.725 & \rl 0.26020 & \rl 0.212 & \rl 8.551 \tn \hline
MV-Medium &Forest Fire (fwdPb=bckPb=0.2) & \rl 500 & \rl 792 & \rl 3.168 & \rl 7.084 & \rl 0.34939 & \rl 0.236 & \rl 8.425 \tn \hline
MV-Large & Forest Fire (fwdPb=bckPb=0.2) & \rl 1000 & \rl 1468 & \rl 2.936 & \rl 10.341 & \rl 0.28623 & \rl 0.170 & \rl 11.393 \tn \hline
 \hline
LV-Small &  Forest Fire (fwdPb=bckPb=0.3) & \rl 20 & \rl 34 & \rl 3.400 & \rl 2.711 & \rl 0.44833 & \rl 0.535 & \rl 4.750 \tn \hline
LV-Medium &  Forest Fire (fwdPb=bckPb=0.3) & \rl 90 & \rl 163 & \rl 3.622 & \rl 5.264 & \rl 0.51427 & \rl 0.317 & \rl 6.648 \tn \hline
LV-Large & Forest Fire (fwdPb=bckPb=0.3) & \rl 200 & \rl 505 & \rl 5.050 & \rl 4.224 & \rl 0.43460 & \rl 0.370 & \rl 5.410 \tn \hline
MV-Small &  Forest Fire (fwdPb=bckPb=0.3) & \rl 250 & \rl 527 & \rl 4.216 & \rl 5.231 & \rl 0.39254 & \rl 0.323 & \rl 6.508 \tn \hline
MV-Medium &  Forest Fire (fwdPb=bckPb=0.3) & \rl 500 & \rl 1185 & \rl 4.740 & \rl 5.260 & \rl 0.39249 & \rl 0.350 & \rl 6.264 \tn \hline
MV-Large &  Forest Fire (fwdPb=bckPb=0.3) & \rl 1000 & \rl 2461 & \rl 4.922 & \rl 5.606 & \rl 0.38572 & \rl 0.346 & \rl 6.416 \tn \hline
\hline
LV-Small &  Forest Fire (fwdPb=bckPb=0.35) & \rl 20 & \rl 38 & \rl 3.800 & \rl 2.421 & \rl 0.64278 & \rl 0.421 & \rl 4.880 \tn \hline
LV-Medium & Forest Fire (fwdPb=bckPb=0.35) & \rl 90 & \rl 212 & \rl 4.711 & \rl 3.697 & \rl 0.49510 & \rl 0.431 & \rl 5.359 \tn \hline
LV-Large &  Forest Fire (fwdPb=bckPb=0.35) & \rl 200 & \rl 707 & \rl 7.070 & \rl 3.472 & \rl 0.41380 & \rl 0.401 & \rl 4.738 \tn \hline
MV-Small &  Forest Fire (fwdPb=bckPb=0.35) & \rl 250 & \rl 1013 & \rl 8.104 & \rl 3.671 & \rl 0.44863 & \rl 0.411 & \rl 4.870 \tn \hline
MV-Medium &  Forest Fire (fwdPb=bckPb=0.35) & \rl 500 & \rl 2369 & \rl 9.476 & \rl 3.587 & \rl 0.49452 & \rl 0.409 & \rl 4.402 \tn \hline
MV-Large &  Forest Fire (fwdPb=bckPb=0.35) & \rl 1000 & \rl 7777 & \rl 15.554 & \rl 3.311 & \rl 0.48042 & \rl 0.408 & \rl 4.382 \tn \hline
\hline
LV-Small &  Kronecker (\PG parameters) & \rl 25 & \rl 25 & \rl 2.000 & \rl 4.500 & \rl 0.00000 & \rl 0.332 & \rl 7.269 \tn \hline
LV-Medium &  Kronecker (\PG parameters) & \rl 100 & \rl 110 & \rl 2.200 & \rl 6.465 & \rl 0.00000 & \rl 0.318 & \rl 12.358 \tn \hline
LV-Large &  Kronecker (\PG parameters) & \rl 202 & \rl 233 & \rl 2.307 & \rl 6.973 & \rl 0.00000 & \rl 0.325 & \rl 11.926 \tn \hline
MV-Small &  Kronecker (\PG parameters) & \rl 202 & \rl 233 & \rl 2.307 & \rl 7.012 & \rl 0.00000 & \rl 0.314 & \rl 11.747 \tn \hline
MV-Medium &  Kronecker (\PG parameters) & \rl 397 & \rl 504 & \rl 2.539 & \rl 6.290 & \rl 0.00303 & \rl 0.359 & \rl 9.554 \tn \hline
MV-Large &  Kronecker (\PG parameters) & \rl 809 & \rl 1047 & \rl 2.588 & \rl 6.741 & \rl 0.00185 & \rl 0.352 & \rl 9.750 \tn \hline
LV-Small &  Kronecker (Social and technological networks parameters) & \rl 25 & \rl 27 & \rl 2.160 & \rl 3.958 & \rl 0.04800 & \rl 0.401 & \rl 7.577 \tn \hline
LV-Medium &  Kronecker (Social and technological networks parameters) & \rl 90 & \rl 124 & \rl 2.756 & \rl 4.393 & \rl 0.01638 & \rl 0.392 & \rl 7.222 \tn \hline
LV-Large & Kronecker (Social and technological networks parameters) & \rl 176 & \rl 273 & \rl 3.102 & \rl 4.206 & \rl 0.01559 & \rl 0.375 & \rl 6.709 \tn \hline
MV-Small &  Kronecker (Social and technological networks parameters) & \rl 176 & \rl 273 & \rl 3.102 & \rl 4.206 & \rl 0.01933 & \rl 0.374 & \rl 6.701 \tn \hline
MV-Medium &  Kronecker (Social and technological networks parameters) & \rl 375 & \rl 603 & \rl 3.216 & \rl 4.628 & \rl 0.00457 & \rl 0.393 & \rl 6.653 \tn \hline
MV-Large & Kronecker (Social and technological networks parameters) & \rl 763 & \rl 1326 & \rl 3.476 & \rl 4.799 & \rl 0.00787 & \rl 0.391 & \rl 6.478 \tn \hline
\end{tabular}
\caption{Metrics for Copying, Forest Fire and Kronecker models.}\label{tab:other1-1}
\end{footnotesize}
\end{sidewaystable}

\begin{sidewaystable}[htbp]
\centering
\begin{footnotesize}
\begin{tabular}{|@p{1.7cm}|^p{7.5cm}|^p{0.9cm}|^p{0.6cm}|^p{1.5cm}|^p{1.5cm}|^p{1.2cm}|}
\hline
\rowstyle{\bfseries}
Network type & Model & \textit{Order} & \textit{Size} & {Avg. betweenness} & {Avg. betw/order} & {Coeff. variation} \\ \hline
\hline
LV-Small &  Copying Model & \rl 20 & \rl 19 & \rl 20.300 & \rl 1.015 & \rl 3.646 \tn \hline
LV-Medium &  Copying Model & \rl 90 & \rl 89 & \rl 254.978 & \rl 2.833 & \rl 3.524 \tn \hline
LV-Large &  Copying Model & \rl 200 & \rl 199 & \rl 555.330 & \rl 2.777 & \rl 5.369 \tn \hline
MV-Small &  Copying Model & \rl 250 & \rl 249 & \rl 523.216 & \rl 2.093 & \rl 7.697 \tn \hline
MV-Medium &  Copying Model & \rl 500 & \rl 499 & \rl 1393.884 & \rl 2.788 & \rl 8.300 \tn \hline
MV-Large &  Copying Model & \rl 1000 & \rl 999 & \rl 2857.788 & \rl 2.858 & \rl 11.396 \tn \hline
 \hline
LV-Small &  Forest Fire (fwdPb=bckPb=0.2) & \rl 20 & \rl 22 & \rl 46.300 & \rl 2.315 & \rl 1.315 \tn \hline
LV-Medium &  Forest Fire (fwdPb=bckPb=0.2) & \rl 90 & \rl 131 & \rl 410.556 & \rl 4.562 & \rl 2.280 \tn \hline
LV-Large &  Forest Fire (fwdPb=bckPb=0.2) & \rl 200 & \rl 295 & \rl 943.690 & \rl 4.718 & \rl 2.683 \tn \hline
MV-Small &  Forest Fire (fwdPb=bckPb=0.2) & \rl 250 & \rl 336 & \rl 1460.200 & \rl 5.841 & \rl 3.049 \tn \hline
MV-Medium &  Forest Fire (fwdPb=bckPb=0.2) & \rl 500 & \rl 792 & \rl 3147.128 & \rl 6.294 & \rl 3.569 \tn \hline
MV-Large &  Forest Fire (fwdPb=bckPb=0.2) & \rl 1000 & \rl 1468 & \rl 9416.916 & \rl 9.417 & \rl 4.324 \tn \hline
 \hline
LV-Small &  Forest Fire (fwdPb=bckPb=0.3) & \rl 20 & \rl 34 & \rl 35.500 & \rl 1.775 & \rl 1.363 \tn \hline
LV-Medium &  Forest Fire (fwdPb=bckPb=0.3) & \rl 90 & \rl 163 & \rl 371.200 & \rl 4.124 & \rl 2.301 \tn \hline
LV-Large &  Forest Fire (fwdPb=bckPb=0.3) & \rl 200 & \rl 505 & \rl 688.270 & \rl 3.441 & \rl 2.245 \tn \hline
MV-Small &  Forest Fire (fwdPb=bckPb=0.3) & \rl 250 & \rl 527 & \rl 1060.408 & \rl 4.242 & \rl 3.018 \tn \hline
MV-Medium &  Forest Fire (fwdPb=bckPb=0.3) & \rl 500 & \rl 1185 & \rl 2205.412 & \rl 4.411 & \rl 2.773 \tn \hline
MV-Large &  Forest Fire (fwdPb=bckPb=0.3) & \rl 1000 & \rl 2461 & \rl 4728.648 & \rl 4.729 & \rl 3.000 \tn \hline
  \hline
LV-Small &  Forest Fire (fwdPb=bckPb=0.35) & \rl 20 & \rl 38 & \rl 29.700 & \rl 1.485 & \rl 1.798 \tn \hline
LV-Medium &  Forest Fire (fwdPb=bckPb=0.35) & \rl 90 & \rl 212 & \rl 269.644 & \rl 2.996 & \rl 2.025 \tn \hline
LV-Large &  Forest Fire (fwdPb=bckPb=0.35) & \rl 200 & \rl 707 & \rl 565.740 & \rl 2.829 & \rl 1.673 \tn \hline
MV-Small &  Forest Fire (fwdPb=bckPb=0.35) & \rl 250 & \rl 1013 & \rl 755.272 & \rl 3.021 & \rl 2.302 \tn \hline
MV-Medium &  Forest Fire (fwdPb=bckPb=0.35) & \rl 500 & \rl 2369 & \rl 1446.308 & \rl 2.893 & \rl 2.507 \tn \hline
MV-Large &  Forest Fire (fwdPb=bckPb=0.35) & \rl 1000 & \rl 7777 & \rl 2598.518 & \rl 2.599 & \rl 3.020 \tn \hline
  \hline
LV-Small & Kronecker (\PG parameters) & \rl 25 & \rl 25 & \rl 86.560 & \rl 3.462 & \rl 1.198 \tn \hline
LV-Medium &  Kronecker (\PG parameters) & \rl 100 & \rl 110 & \rl 555.640 & \rl 5.556 & \rl 1.357 \tn \hline
LV-Large &  Kronecker (\PG parameters) & \rl 202 & \rl 233 & \rl 1224.347 & \rl 6.061 & \rl 1.499 \tn \hline
MV-Small &  Kronecker (\PG parameters) & \rl 202 & \rl 233 & \rl 1231.416 & \rl 6.096 & \rl 1.483 \tn \hline
MV-Medium &  Kronecker (\PG parameters) & \rl 397 & \rl 504 & \rl 2166.096 & \rl 5.456 & \rl 1.448 \tn \hline
MV-Large &  Kronecker (\PG parameters) & \rl 809 & \rl 1047 & \rl 4782.022 & \rl 5.911 & \rl 1.418 \tn \hline
LV-Small &  Kronecker (Social and technological networks parameters) & \rl 25 & \rl 27 & \rl 68.400 & \rl 2.736 & \rl 1.381 \tn \hline
LV-Medium &  Kronecker (Social and technological networks parameters) & \rl 90 & \rl 124 & \rl 316.844 & \rl 3.520 & \rl 1.340 \tn \hline
LV-Large &  Kronecker (Social and technological networks parameters) & \rl 176 & \rl 273 & \rl 584.193 & \rl 3.319 & \rl 1.601 \tn \hline
MV-Small &  Kronecker (Social and technological networks parameters) & \rl 176 & \rl 273 & \rl 590.193 & \rl 3.353 & \rl 1.557 \tn \hline
MV-Medium &  Kronecker (Social and technological networks parameters) & \rl 375 & \rl 603 & \rl 1386.560 & \rl 3.697 & \rl 1.816 \tn \hline
MV-Large &  Kronecker (Social and technological networks parameters) & \rl 763 & \rl 1326 & \rl 2992.993 & \rl 3.923 & \rl 1.894 \tn \hline
\end{tabular}
\caption{Betweenness metrics for Copying, Forest Fire and Kronecker models.}\label{tab:other1-1Bet}
\end{footnotesize}
\end{sidewaystable}

\begin{sidewaystable}[htbp]
\centering
\begin{footnotesize}
\begin{tabular}{|@p{1.7cm}|^p{9cm}|^p{0.9cm}|^p{0.6cm}|^p{1cm}|^p{1cm}|^p{1.2cm}|^p{1.5cm}|^p{2cm}|}
\hline
\rowstyle{\bfseries} Network type & Model & \textit{Order} & \textit{Size} & {Avg. deg.} & {CPL} & {CC} & {Removal robustness ($Rob_N$)} & {Redundancy cost ($APL_{10^{th}}$)} \\ \hline
\hline
LV-Small &  Power Law (Social and technological networks parameters) & \rl 20 & \rl 21 & \rl 2.100 & \rl 3.526 & \rl 0.00000 & \rl 0.409 & \rl 7.460 \tn \hline
LV-Small &  Power Law (East-West U.S. HV \PG networks parameters) & \rl 20 & \rl 19 & \rl 1.900 & \rl 4.105 & \rl 0.00000 & \rl 0.300 & \rl 4.120 \tn \hline
LV-Small &  Power Law (Western U.S. HV \PG networks parameters) & \rl 20 & \rl 19 & \rl 1.900 & \rl 4.632 & \rl 0.00000 & \rl 0.329 & \rl 4.800 \tn \hline
LV-Small &  Power Law (Dutch MV-LV \PG networks parameters) & \rl 20 & \rl 25 & \rl 2.500 & \rl 2.816 & \rl 0.00000 & \rl 0.400 & \rl 7.030 \tn \hline
LV-Medium &  Power Law (Social and technological networks parameters) & \rl 89 & \rl 118 & \rl 2.652 & \rl 4.091 & \rl 0.04839 & \rl 0.357 & \rl 6.626 \tn \hline
LV-Medium &  Power Law (East-West U.S. HV \PG networks parameters) & \rl 90 & \rl 95 & \rl 2.111 & \rl 5.169 & \rl 0.02386 & \rl 0.259 & \rl 10.241 \tn \hline
LV-Medium &  Power Law (Western U.S. HV \PG networks parameters) & \rl 90 & \rl 90 & \rl 2.000 & \rl 7.416 & \rl 0.00000 & \rl 0.221 & \rl 10.286 \tn \hline
LV-Medium &  Power Law (Dutch MV-LV \PG networks parameters) & \rl 90 & \rl 136 & \rl 3.022 & \rl 3.371 & \rl 0.14367 & \rl 0.359 & \rl 5.362 \tn \hline
LV-Large &  Power Law (Social and technological networks parameters) & \rl 200 & \rl 280 & \rl 2.800 & \rl 4.497 & \rl 0.01370 & \rl 0.351 & \rl 6.688 \tn \hline
LV-Large &  Power Law (East-West U.S. HV \PG networks parameters) & \rl 200 & \rl 210 & \rl 2.100 & \rl 6.975 & \rl 0.00000 & \rl 0.227 & \rl 12.646 \tn \hline
LV-Large &  Power Law (Western U.S. HV \PG networks parameters) & \rl 200 & \rl 199 & \rl 1.990 & \rl 11.256 & \rl 0.00000 & \rl 0.139 & \rl 12.774 \tn \hline
LV-Large &  Power Law (Dutch MV-LV \PG networks parameters) & \rl 200 & \rl 399 & \rl 3.990 & \rl 3.116 & \rl 0.13789 & \rl 0.372 & \rl 4.568 \tn \hline
MV-Small &  Power Law (Social and technological networks parameters) & \rl 250 & \rl 383 & \rl 3.064 & \rl 3.831 & \rl 0.05247 & \rl 0.357 & \rl 5.498 \tn \hline
MV-Small &  Power Law (East-West U.S. HV \PG networks parameters) & \rl 250 & \rl 257 & \rl 2.056 & \rl 9.127 & \rl 0.00269 & \rl 0.210 & \rl 15.052 \tn \hline
MV-Small &  Power Law (Western U.S. HV \PG networks parameters) & \rl 250 & \rl 249 & \rl 1.992 & \rl 13.277 & \rl 0.00000 & \rl 0.129 & \rl 14.047 \tn \hline
MV-Small &  Power Law (Dutch MV-LV \PG networks parameters) & \rl 250 & \rl 456 & \rl 3.648 & \rl 3.424 & \rl 0.08564 & \rl 0.354 & \rl 4.893 \tn \hline
MV-Medium &  Power Law (Social and technological networks parameters) & \rl 499 & \rl 721 & \rl 2.890 & \rl 4.026 & \rl 0.04681 & \rl 0.336 & \rl 5.530 \tn \hline
MV-Medium & Power Law (East-West U.S. HV \PG networks parameters) & \rl 500 & \rl 533 & \rl 2.132 & \rl 8.179 & \rl 0.00202 & \rl 0.246 & \rl 11.922 \tn \hline
MV-Medium &  Power Law (Western U.S. HV \PG networks parameters) & \rl 500 & \rl 500 & \rl 2.000 & \rl 17.385 & \rl 0.00271 & \rl 0.090 & \rl 18.189 \tn \hline
MV-Medium &  Power Law (Dutch MV-LV \PG networks parameters) & \rl 500 & \rl 1122 & \rl 4.488 & \rl 3.252 & \rl 0.13604 & \rl 0.364 & \rl 4.474 \tn \hline
MV-Large &  Power Law (Social and technological networks parameters) & \rl 1000 & \rl 1717 & \rl 3.434 & \rl 4.178 & \rl 0.04216 & \rl 0.345 & \rl 5.567 \tn \hline
MV-Large &  Power Law (East-West U.S. HV \PG networks parameters) & \rl 1000 & \rl 1086 & \rl 2.172 & \rl 8.194 & \rl 0.00232 & \rl 0.254 & \rl 11.099 \tn \hline
MV-Large &  Power Law (Western U.S. HV \PG networks parameters) & \rl 1000 & \rl 999 & \rl 1.998 & \rl 16.409 & \rl 0.00000 & \rl 0.085 & \rl 16.009 \tn \hline
MV-Large &  Power Law (Dutch MV-LV \PG networks parameters) & \rl 1000 & \rl 2404 & \rl 4.808 & \rl 3.224 & \rl 0.15003 & \rl 0.364 & \rl 4.446 \tn \hline
\end{tabular}
\caption{Metrics for random connected graphs showing a  power-law in node degree distribution.}\label{tab:other1-2}
\end{footnotesize}
\end{sidewaystable}

\begin{sidewaystable}[htbp]
\centering
\begin{footnotesize}
\begin{tabular}{|@p{1.7cm}|^p{9cm}|^p{0.9cm}|^p{0.6cm}|^p{1.5cm}|^p{1.5cm}|^p{1.2cm}|}
\hline
\rowstyle{\bfseries}
Network type & Model & \textit{Order} & \textit{Size} & {Avg. betweenness} & {Avg. betw/order} & {Coeff. variation} \\ \hline
\hline
LV-Small &  Power Law (Social and technological networks parameters) & \rl 20 & \rl 21 & \rl 52.600 & \rl 2.630 & \rl 1.112 \tn \hline
LV-Medium &  Power Law (Social and technological networks parameters) & \rl 89 & \rl 118 & \rl 278.382 & \rl 3.128 & \rl 1.755 \tn \hline
LV-Large &  Power Law (Social and technological networks parameters) & \rl 200 & \rl 280 & \rl 715.980 & \rl 3.580 & \rl 2.021 \tn \hline
MV-Small &  Power Law (Social and technological networks parameters) & \rl 250 & \rl 383 & \rl 760.968 & \rl 3.044 & \rl 2.946 \tn \hline
MV-Medium &  Power Law (Social and technological networks parameters) & \rl 499 & \rl 721 & \rl 1577.323 & \rl 3.161 & \rl 4.760 \tn \hline
MV-Large &  Power Law (Social and technological networks parameters) & \rl 1000 & \rl 1717 & \rl 3323.160 & \rl 3.323 & \rl 3.883 \tn \hline
 \hline
LV-Small &  Power Law (East-West U.S. HV \PG networks parameters) & \rl 20 & \rl 19 & \rl 59.200 & \rl 2.960 & \rl 1.213 \tn \hline
LV-Medium &  Power Law (East-West U.S. HV \PG networks parameters) & \rl 90 & \rl 95 & \rl 415.600 & \rl 4.618 & \rl 2.082 \tn \hline
LV-Large &  Power Law (East-West U.S. HV \PG networks parameters) & \rl 200 & \rl 210 & \rl 1230.020 & \rl 6.150 & \rl 2.271 \tn \hline
MV-Small &  Power Law (East-West U.S. HV \PG networks parameters) & \rl 250 & \rl 257 & \rl 2098.936 & \rl 8.396 & \rl 2.061 \tn \hline
MV-Medium &  Power Law (East-West U.S. HV \PG networks parameters) & \rl 500 & \rl 533 & \rl 3792.020 & \rl 7.584 & \rl 2.293 \tn \hline
MV-Large &  Power Law (East-West U.S. HV \PG networks parameters) & \rl 1000 & \rl 1086 & \rl 7716.384 & \rl 7.716 & \rl 2.864 \tn \hline
  \hline
LV-Small &  Power Law (Western U.S. HV \PG networks parameters) & \rl 20 & \rl 19 & \rl 71.200 & \rl 3.560 & \rl 0.906 \tn \hline
LV-Medium &  Power Law (Western U.S. HV \PG networks parameters) & \rl 90 & \rl 90 & \rl 584.756 & \rl 6.497 & \rl 1.521 \tn \hline
LV-Large &  Power Law (Western U.S. HV \PG networks parameters) & \rl 200 & \rl 199 & \rl 2122.370 & \rl 10.612 & \rl 1.983 \tn \hline
MV-Small &  Power Law (Western U.S. HV \PG networks parameters) & \rl 250 & \rl 249 & \rl 3136.032 & \rl 12.544 & \rl 2.207 \tn \hline
MV-Medium &  Power Law (Western U.S. HV \PG networks parameters) & \rl 500 & \rl 500 & \rl 8524.172 & \rl 17.048 & \rl 2.553 \tn \hline
MV-Large &  Power Law (Western U.S. HV \PG networks parameters) & \rl 1000 & \rl 999 & \rl 15956.820 & \rl 15.957 & \rl 3.103 \tn \hline
 \hline
LV-Small &  Power Law (Dutch MV-LV \PG networks parameters) & \rl 20 & \rl 25 & \rl 31.800 & \rl 1.590 & \rl 1.400 \tn \hline
LV-Medium &  Power Law (Dutch MV-LV \PG networks parameters) & \rl 90 & \rl 136 & \rl 221.133 & \rl 2.457 & \rl 2.284 \tn \hline
LV-Large &  Power Law (Dutch MV-LV \PG networks parameters) & \rl 200 & \rl 399 & \rl 455.410 & \rl 2.277 & \rl 3.007 \tn \hline
MV-Small &  Power Law (Dutch MV-LV \PG networks parameters) & \rl 250 & \rl 456 & \rl 647.864 & \rl 2.591 & \rl 2.836 \tn \hline
MV-Medium &  Power Law (Dutch MV-LV \PG networks parameters) & \rl 500 & \rl 1122 & \rl 1209.380 & \rl 2.419 & \rl 4.066 \tn \hline
MV-Large &  Power Law (Dutch MV-LV \PG networks parameters) & \rl 1000 & \rl 2404 & \rl 2356.038 & \rl 2.356 & \rl 5.758 \tn \hline
\end{tabular}
\caption{Betweenness metrics for random connected graphs showing a  power-law in node degree distribution.}\label{tab:other1-2Bet}
\end{footnotesize}
\end{sidewaystable}

\subsection*{Model Generation}

Given the parameters presented above, we generate the graphs with respect to the different models and analyze them according to the significant \PG metrics described in Section~\ref{sec:metrics}. We begin with the models for which it is possible to explicitly assign \textit{order} and \textit{size} (or one of these quantities and the average node degree); we then proceed analyzing the other models that do not explicitly allow to set the average node degree parameter.

\subsubsection*{Model generation implementation and metrics computation}

The values and graphs of generated topologies are obtained using software applications for network generation and analysis. In particular, for the model generation we developed C++ programs using the Stanford Network Analysis Project (SNAP) (\url{http://snap.stanford.edu/}) library that enables the generation of the network topologies described in Section~\ref{sec:models} and the assignment of the parameters described earlier in this section. The analysis of the generated graphs according to the metrics described in Section~\ref{sec:metrics} is performed with had-hoc created software based on the JAVA graph library JGraphT (\url{http://www.jgrapht.org/}). The only metric computed with the SNAP software is `betweenness' whose computation is based on the algorithm developed by Brandes~\cite{Brandes2001}. To perform the generation and computation of the metrics we used a PC with Intel Core2 Quad CPU Q9400 2.66GHz with 4GB RAM. The Operating system is based on the Linux kernel 2.6.32 with a 4.4.3 GCC compiler and JAVA framework 1.6. The versions of SNAP and JGraphT software libraries used are respectively v10.10.01 and v0.8.1.

\subsubsection*{Comparison of models with average node degree $<k>\approx 2$}

\begin{figure}
 \captionsetup{type=figure}
    \begin{center}
    \subfloat[Characteristic path length.]{\label{fig:cplNd2}\includegraphics[scale=.3]{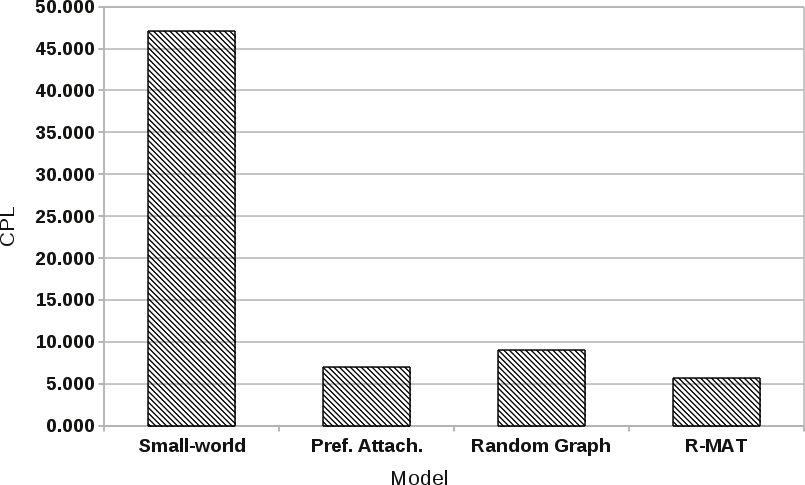}}
    \hspace*{4mm}
    \subfloat[Clustering coefficient.]{\label{fig:ccNd2}\includegraphics[scale=.3]{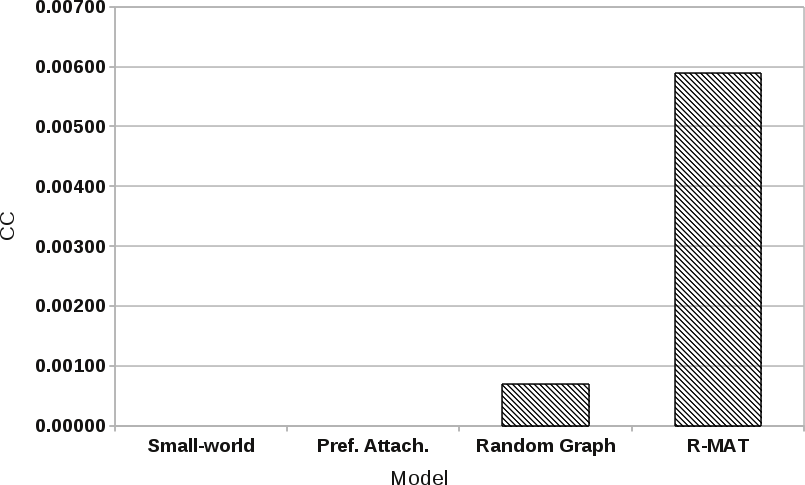}}\\    \hspace*{4mm}
    \subfloat[Removal robustness.]{\label{fig:robNd2}\includegraphics[scale=.3]{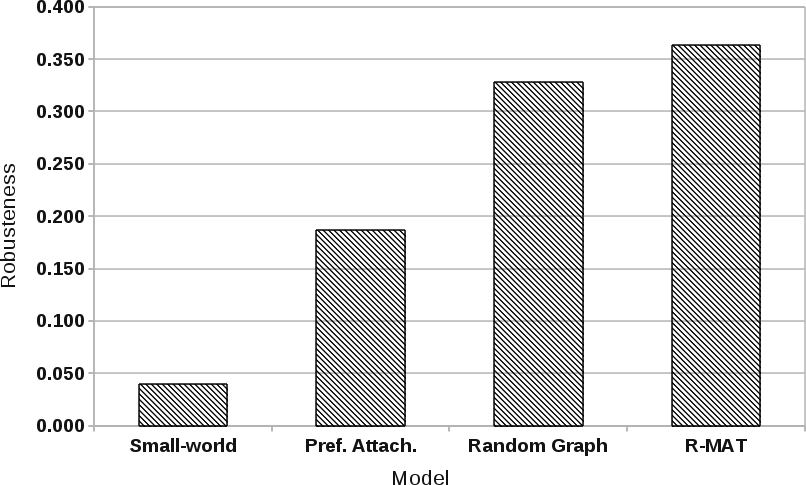}} 
   \end{center}
   \caption{Metrics for the \textit{Large} sample of \MV network type with average node degree $\approx 2$.}
    \label{fig:parNd2}
\end{figure}

The results for the metrics  with average degree $<k>\approx 2$ for the small-world, preferential attachment, Random Graph and R-MAT models score quite poorly, cf. Table~\ref{tab:gen2}. These low values are due to the small connectivity the networks show.  Especially, we highlight the poor results of the \sw model under these conditions: with such a small average degree, the \cpl tends to be very high particularly as the network grows, with a value that for the biggest network generated is higher than 45 and about 60 for the 10$^{th}$ path measure. In such a graph with small amount of edges, the clustering coefficient is also affected:  the neighbors of a node are not organized in tight clusters because the numbers of links available are limited, only the Random Graph and R-MAT have non zero values for some samples, although few. The robustness to failure is limited under these conditions, with the worst case corresponding to the \sw samples, while better results are shown by preferential attachment, Random Graph and R-MAT models. The last two score higher than 0.33 for this metric.
Considering the cost of redundancy, we generally see an increase in the \cpl as the \textit{order} of the graph grows; the best results are shown by the networks generated with the preferential attachment model that presents values close to the best ones of the average path length. This might be due to the absence of many redundant paths in such a loosely connected network (less than 10 shortest paths without cycles) between any two nodes. A graphical comparison for the results of the \textit{Large} sample for the \MV type considering characteristic path length, clustering coefficient and robustness metrics are given in Figure~\ref{fig:parNd2}.

\begin{figure}
 \captionsetup{type=figure}
    \centering
    \subfloat[Betweenness to \textit{order} ratio.]{\label{fig:cplNd2}\includegraphics[scale=.3]{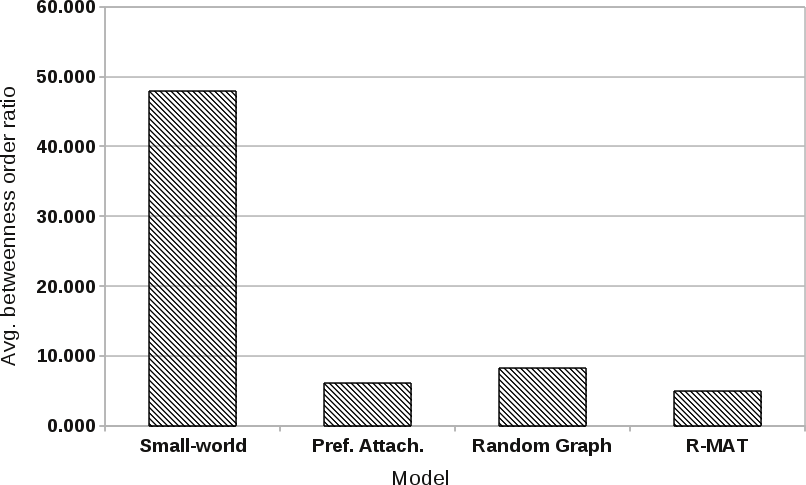}}
    \hspace*{4mm}
    \subfloat[Betweenness coefficient of variation.]{\label{fig:ccNd2}\includegraphics[scale=.3]{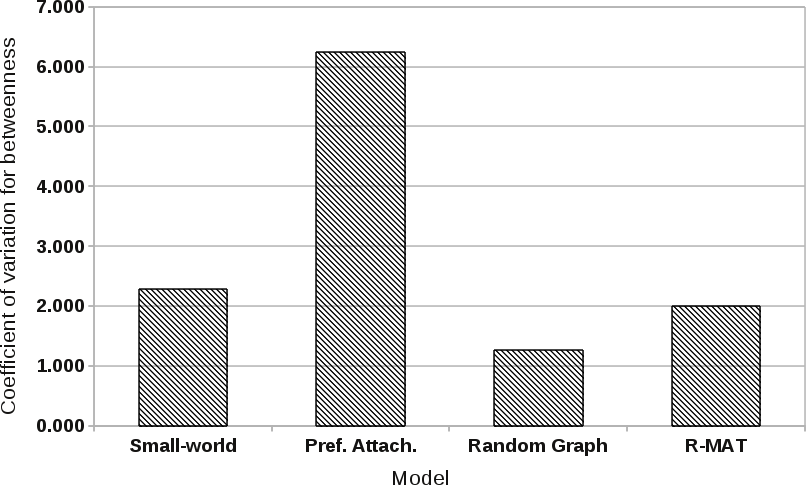}} 
    \caption{Metrics for the \textit{Large} sample of \MV network type with average node degree $\approx 2$.}
    \label{fig:parBetNd2}
\end{figure}

The betweenness analysis, whose results are presented in Table~\ref{tab:genBet2}, shows an average for each node that increases with the size of the graph. The difference is in the value of average betweenness for the \sw model compared to other models: for the largest networks (500 and 1000 nodes) the value is almost one order of magnitude higher. This is due to the lattice structure of \sw that with a $<k>\approx 2$ degenerates in a long ``closed-chain'' topology which involves many nodes. The amount of edges that provide a ``shortcut'' in the graph is limited. This is in line with the high \cpl just described. The R-MAT model scores well considering the desiderata we imposed for average betweenness \textit{order} ratio and coefficient of variation; the former is below 5 even for the biggest sample and the latter stays below 2. For the \sw sample, we experience a small coefficient of variation which reinforces the result indicating that almost all nodes have the same high betweenness close to the average. A graphical comparison for the results of the \textit{Large} sample for the \MV type considering average betweenness \textit{order} ratio and coefficient of variation metrics are shown in Figure~\ref{fig:parBetNd2}.

\subsubsection*{Comparison of models with average node degree $<k>\approx 4$}

\begin{figure}
 \captionsetup{type=figure}
    \centering
    \subfloat[Characteristic path length.]{\label{fig:cplNd4}\includegraphics[scale=.3]{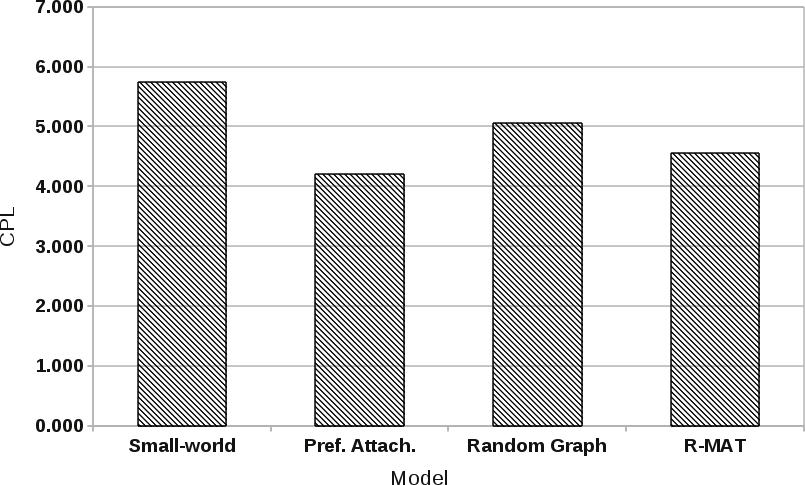}}
    \hspace*{4mm}
    \subfloat[Clustering coefficient.]{\label{fig:ccNd4}\includegraphics[scale=.3]{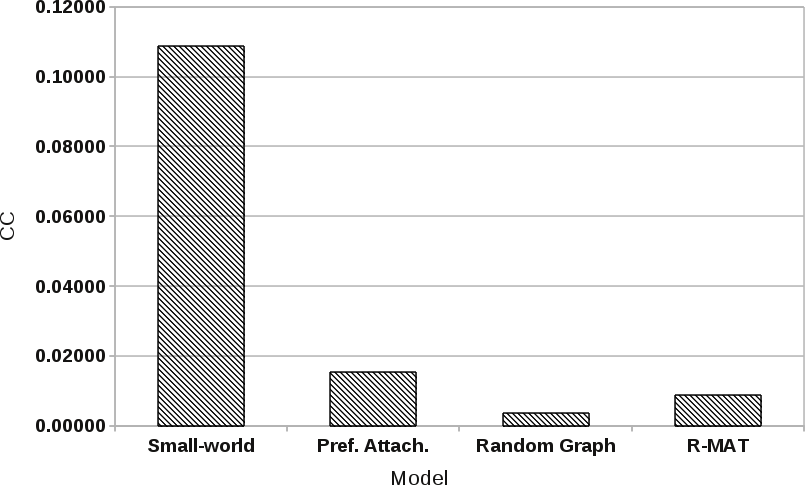}}\\
    \hspace*{4mm}
    \subfloat[Removal robustness.]{\label{fig:robNd4}\includegraphics[scale=.3]{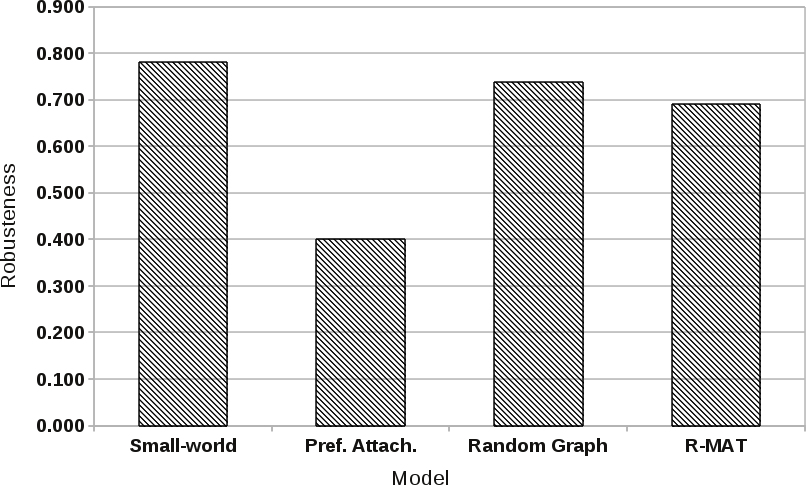}} 
    \caption{Results for metrics for the \textit{Large} sample of \MV network type with average node degree $\approx 4$.}
    \label{fig:parNd4}
\end{figure}

Table~\ref{tab:gen4} shows the results for small-world, preferential attachment, Random Graph and R-MAT models with an average degree $<k>\approx 4$. One notices the scores for the metrics improve compared to the $<k>\approx 2$ case. The average over the \cpl of all the samples reduces from around 10 to a value that is slightly less than 5. The clustering coefficient has values that are significant and all positive. The \sw model scores best in this specific metric since it relies on the lattice topology that with an average degree of 4 connects each node with four neighbors. In particular 3 triangle structures emerge in each neighborhood of a node. This provides a substantial contribution to the quite high clustering coefficient. Generally, all models score higher than the random graph with respect to the clustering coefficient (one of our desired properties). The addition of links provides enhanced robustness for the network too. Generally the \textit{order} of the biggest connected component is about 63\% of the initial \textit{order} of the network (averaging all the result for the models) while with a $<k>\approx2$ networks the value is just 27\%. Not surprisingly, the best scores for robustness are obtained by the Random Graph model since in this type of graph the nodes tend to have the same characteristics, and hubs are not present in the network. Quantitatively, quite similar results to Random Graph for robustness are shown by small-world graphs (for some samples the metric scores even higher) with a robustness that is close to 0.8. Preferential attachment and R-MAT models score lower than random and \sw models with values around 0.45 for the former and 0.6 for the latter (in both cases these values are almost double than those for the $<k>\approx2$ case). An explanation for this lower score compared to other models for preferential attachment and R-MAT models resides in their building properties: they admit the presence of hubs (the node degree distribution is characterized by a power-law)  that are highly sensitive for network robustness when targeted for removal. Considering the cost for the redundancy related to alternative paths, lower values appear for the preferential attachment model followed by the R-MAT and slightly higher for Random Graph and small-world. The worst case for this last model is a little smaller than 8 which is anyway only increased by 2 compared to the characteristic path length for the same sample. A graphical comparison for the results of the \textit{Large} sample for \MV type considering characteristic path length, clustering coefficient and robustness metrics are shown in Figure~\ref{fig:parNd4}.

\begin{figure}[h!]
 \captionsetup{type=figure}
    \centering
    \subfloat[Betweenness to \textit{order} ratio.]{\label{fig:cplNd4}\includegraphics[scale=.3]{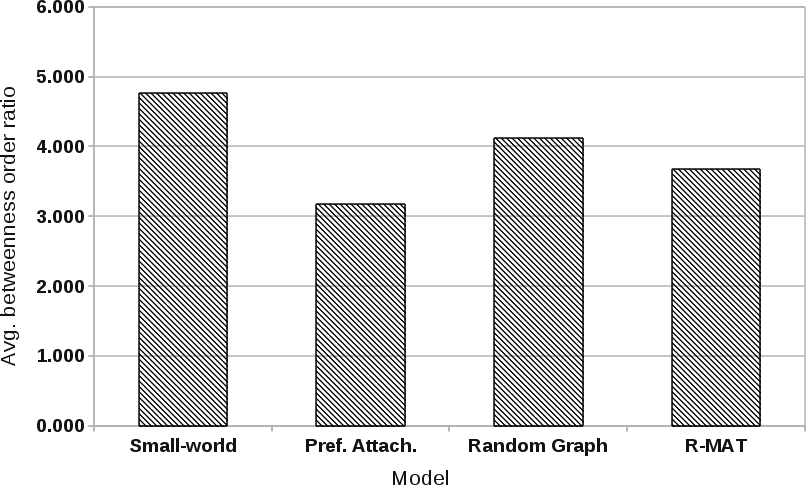}}
    \hspace*{4mm}
    \subfloat[Betweenness coefficient of variation.]{\label{fig:ccNd4}\includegraphics[scale=.3]{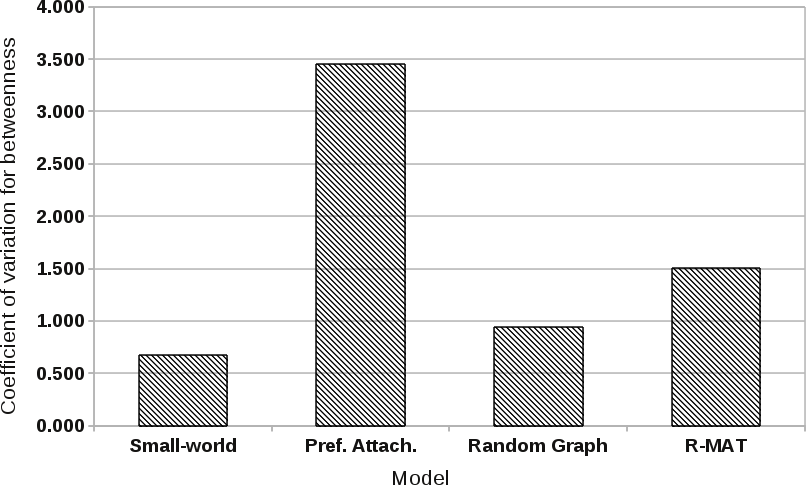}} 
    \caption{Results for metrics for the \textit{Large} sample of \MV network type with average node degree $\approx 4$.}
    \label{fig:parBetNd4}
\end{figure}

Analyzing the betweenness we see a general improvement in the metrics compared to the $<k>\approx2$ case, cf.\ Table~\ref{tab:genBet4}. The most important improvement is for the \sw model which, with approximately 4 connections per node, substantially reduces the average betweenness by a factor of 10 compared to the $<k>\approx 2$ case. Although the \sw model performs worse than other models for the average betweenness \textit{order} ratio, the coefficient of variation performs the best. It reinforces the idea that is in the model itself: nodes that do not differ much in their properties (the underlying lattice structure) have a small variation in the betweenness of nodes. The  preferential attachment and R-MAT models, which generate networks with a fraction of nodes that have a very high connectivity due to the \pl in the node degree distribution, reach a higher coefficient of variation for betweenness. A graphical comparison for the results of the \textit{Large} sample for \MV type considering average betweenness \textit{order} ratio and coefficient of variation metrics are shown in Figure~\ref{fig:parBetNd4}.

\subsubsection*{Comparison of models with average node degree $<k>\approx 6$}

\begin{figure}
 \captionsetup{type=figure}
    \centering
    \subfloat[Characteristic path length.]{\label{fig:cplNd6}\includegraphics[scale=.3]{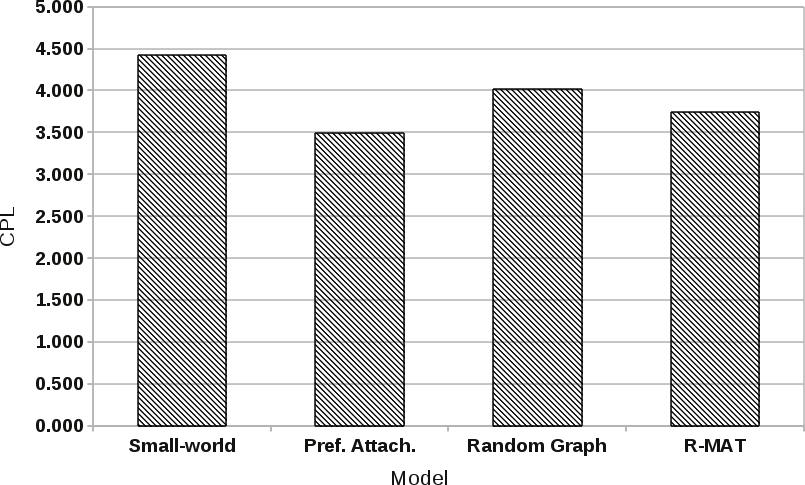}}
    \hspace*{4mm}
    \subfloat[Clustering coefficient.]{\label{fig:ccNd6}\includegraphics[scale=.3]{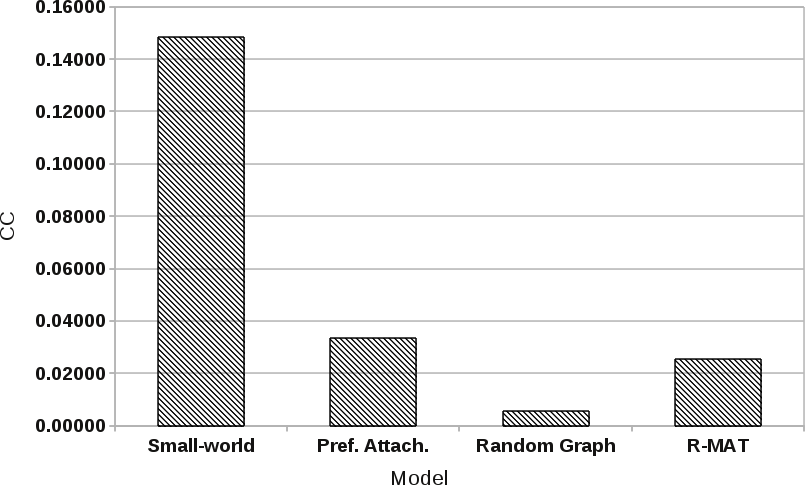}} \\
    \hspace*{4mm}
    \subfloat[Removal robustness.]{\label{fig:robNd6}\includegraphics[scale=.3]{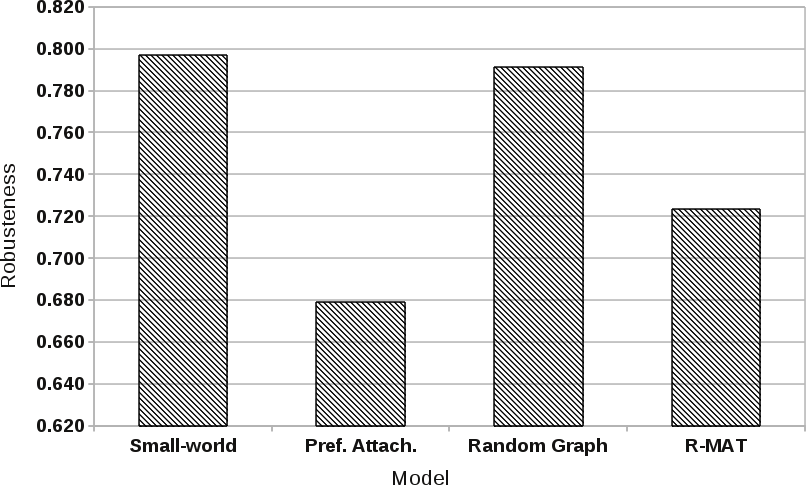}} 
    \caption{Metrics for the \textit{Large} sample of \MV network type with average node degree $\approx 6$.}
    \label{fig:parNd6}
\end{figure}

Table~\ref{tab:gen6} shows the results for small-world, preferential attachment, Random Graph and R-MAT models with an average degree $<k>\approx 6$. The scores for the metrics considered improve even more with respect to those of Tables~\ref{tab:gen2} and~\ref{tab:gen4}. The \cpl of all the samples has reduced to a value that, considering the average over all the samples with $<k>\approx6$, is about 3; yet 2 hops lower than the situation with $<k>\approx4$. The same tendency for clustering coefficient found for samples in Table~\ref{tab:gen4} applies to this situation too. The \sw model scores highest since the neighbors of a node have nine connections with each other, thus substantially contributing to a high coefficient. For the R-MAT and preferential attachment models the clustering coefficient decreases as the \textit{order} of the graph increases, but still for the biggest sample generated (1000 nodes) it is about one order of magnitude higher than a corresponding random graph. It is interesting to highlight how the clustering coefficient for the \sw model tends to stabilize over the 0.14 value for the biggest samples (250, 500 and 1000 nodes, respectively). Regarding robustness, on average it increases to a value higher than 0.75. However it worths to notice how the increment mainly involves the preferential attachment and the R-MAT models which improve respectively from 0.44 and 0.61 to 0.70 and 0.73, on average. Therefore, the additional connectivity is more beneficial to \pl distributions than the others which seem to have already hit the upper bound for this metric with the $<k>\approx4$ situation. The cost of the redundant paths with this enhanced connectivity is reduced even more and on average the 10$^{th}$ shortest path is just 1.5 hops higher than the \cpl for the same network.  A graphical comparison for the results of the \textit{Large} sample for \MV type considering characteristic path length, clustering coefficient and robustness metrics are shown in Figure~\ref{fig:parNd6}.

\begin{figure}
 \captionsetup{type=figure}
    \centering
    \subfloat[Betweenness to \textit{order} ratio.]{\label{fig:cplNd6}\includegraphics[scale=.3]{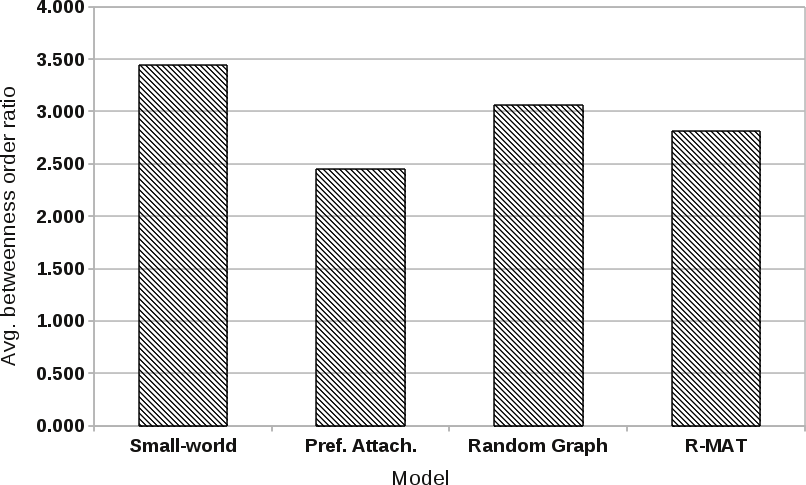}}
    \hspace*{4mm}
    \subfloat[Betweenness coefficient of variation.]{\label{fig:ccNd6}\includegraphics[scale=.3]{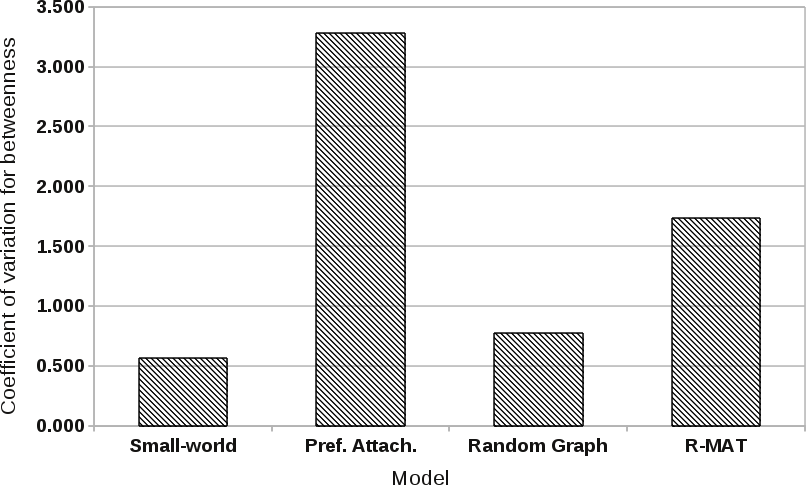}} 
    \caption{Metrics for the \textit{Large} sample of \MV network type with average node degree $\approx 6$.}
    \label{fig:parBetNd6}
\end{figure}

Having increased the average degree to 6 brings benefits to the betweenness statistics too, cf.\  Table~\ref{tab:genBet6}. The benefits on the average betweenness \textit{order} ratio are about 25\% higher compared to the $<k>\approx4$ situation; this ratio therefore is now very close to the experimental values that have been found for the Internet (i.e., $\approx 2.5$) which is one of our desiderata. The preferential attachment model, especially scores  lower than the Internet threshold value for all the categories of samples considered. As already mentioned for the samples with $<k>\approx4$, the coefficient of variation for betweenness, even in this $<k>\approx6$ situation, scores best for the non \pl topologies (i.e., \sw and Random Graph) that show a value below the unit for all the dimensions of samples considered. The improvement for this metric for preferential attachment and R-MAT models are present but limited, in fact, they score higher than 3 and 1.7, respectively, in the worst case. A graphical comparison for the results of the \textit{Large} sample for \MV type considering average betweenness \textit{order} ratio and coefficient of variation metrics are shown in Figure~\ref{fig:parBetNd6}.

\subsubsection*{Models Independent from the Average Node Degree}

The Copying, Forest Fire, and Kronecker models are not generated using explicitly the average node degree, cf.\ 
 Section~\ref{sec:modelParams}. Therefore, we consider the \PG metrics on them separately. We remark however that, though not explicitly used as input parameter, the average node degree of the generated graphs has similar values to those of Random Graphs, \sw and preferential attachment models generated with the same order. Tables~\ref{tab:other1-1} and~\ref{tab:other1-2} contain the results for the metrics analyzed, while  Tables~\ref{tab:other1-1Bet} and~\ref{tab:other1-2Bet} contain the results for betweenness. 

 The {\bf Copying Model} results are comparable with the values for the metrics analyzed in Table~\ref{tab:gen2}, since the way the model is created provides a constant average node degree $<k>\approx2$. One can see that the Copying Model, which leads to a \pl in the node degree distribution, scores better than the \sw and preferential attachment ones in \cpl and robustness. The small cost in the average 10$^{th}$ shortest path is due to the computation of the worst case path (in fact in these conditions of network connectivity there are not ten paths between two nodes) which due to the very small meshed structure that is created it admits in the majority of cases just one path. With the way the model is implemented in the simulation environment used (Stanford Network Analysis Platform - SNAP\footnote{\url{http://snap.stanford.edu/}}), a node just copies a link from another chosen node and therefore there is no possibility to generate ``triangle'' structures between nodes which are essential to have a non-null clustering coefficient; that is why this metrics has such score.
%
Considering the results for betweenness and comparing the values for Copying Model with the results obtained for models with imposing $<k>\approx2$, the Copying Model scores with an average smaller betweenness, this translates into a betweenness to \textit{order} ratio that is better than other samples. On the other hand, the coefficient of variation is quite high given the difference in betweenness: extremely high only for few nodes in the network that sustain the majority of the shortest paths, while the majority of the nodes participates only in the shortest paths for which they are end nodes for the path. The statistical mode for the betweenness values of each category for Copying Model is in fact null.

For the {\bf Forest Fire model}, we assign different forward and backward burning probabilities to obtain values for the average degree to some extent comparable with the other models. The model with $p_{fwd}=p_{bwd}=0.2$ can be compared to models with $<k>\approx2$. The Forest Fire scores definitely better than all the others in clustering coefficient. This is not surprising, if one recalls the algorithm behind the model: an ambassador node is chosen and with a certain probability a certain number of ambassador's neighbors nodes are chosen to establish link to. One can see how many triangle-like structures tend to appear from such a generating method. The same observations can be done for the Forest Fire with $p_{fwd}=p_{bwd}=0.3$ when compared to models with $<k>\approx4$: the \cpl scores almost like the other models, while this model suffers deeply in the robustness metric which for the biggest samples obtain a score which is half compared to the other generating models with $<k>\approx4$. This is due to the very high damages imposed to network connectivity when high degree nodes are removed: for the biggest sample (order of about 1000 nodes), when the 20\% of nodes with highest degree are removed, the biggest connected component is just 2\% of the original graph \textit{order}. This is typical of heavy-tailed distributions which Forest Fire models empirically~\cite{Leskovec05}. The metric that scores best is again the clustering coefficient that is three times higher (for the biggest sample) than the already quite high value of the \sw model. Even when we consider denser Forest Fire networks (i.e., $p_{fwd}=p_{bwd}=0.35$) the comparison with the model with $<k>\approx6$ brings to the same conclusions: far better clustering coefficient, but an important weakness to node removal.
Betweenness for the Forest Fire model shows a known trend when varying the average node degree, the more the networks becomes connected the better the metrics related to betweenness become. For the samples with a burning probability of $p_{fwd}=p_{bwd}=0.35$, the betweenness to \textit{order} ratio stays below 3. The same behavior applies to the coefficient of variation, although it generally scores worse than the samples already analyzed with similar average degree.

The results shown by the networks generated with the {\bf Kronecker model} using the parameters extracted from the \PG networks show metrics values similar to the ones computed from the physical samples with almost equal \textit{order}. Especially the parameters for the \PG create networks with an average node degree, \cpl and robustness that mimic what we found for the current Dutch \MLV samples. Even the very low clustering coefficient (very often down to zero) is something we already recorded in real \PG samples~\cite{PaganiAielloTSG2011}. When the networks generated with parameters extracted from \PG are compared with the networks generated from social and technological networks, one sees a general improvement in all the metrics under analysis: a reduction of a couple of hops in the characteristic path length, a higher clustering coefficient which is similar to the values obtained for random graphs. 
Generally the social-technological based Kronecker networks score more than 30\% better than the corresponding based on \PG parameters for characteristic path length. We also see how the networks based on the Kronecker model show an almost constant, or decreasing value, for the average 10$^{th}$ path and a \cpl that very slowly increases with the order of the network. To some extent this tendency is something that the Kronecker model aims to achieve: densification of the network over time, i.e., when more nodes become part of the network the effective diameter of the networks becomes smaller.
Considering betweenness, one sees a smaller average betweenness for the networks based on technological and social parameters than the ones generated with \PG parameters. Despite quite high values for both average betweenness and standard deviation, the samples produced with \PG parameters have a smaller coefficient of variation compared to the techno-social networks. The comparison of Kronecker models with networks generated with $<k>\approx2$ models shows better values of the former compared to \sw and preferential attachment models while the results are quite similar for Random Graph and R-MAT models.

Considering the results of {\bf Random Graph with Power-law models}, there is a difference for the networks generated with smaller $\gamma$ parameters (i.e., \MLV Dutch Grid $\gamma\approx 2$ and social and technological networks $\gamma\approx 2.3$) which score better than the ones with higher $\gamma$ (i.e., U.S. Eastern Interconnect and Western Grid $\gamma\approx 3$ and U.S. Western \PG $\gamma\approx 4$). The first two sets of samples show a denser network with higher average node degree, almost double compared to the other two sets, this results in a beneficial behavior for the metrics computed which present a smaller characteristic path length. This set of networks with small $\gamma$ is comparable for the \cpl property to the values obtained for networks generated with $<k>\approx4$. The second set of samples (i.e., higher $\gamma$ parameter) shows results that are similar to the ones obtained for samples generated with $<k>\approx2$. A general property that applies to all these \pl based samples is the problem they suffer, as already mentioned, from targeted attack involving the nodes with high degree, which justifies very poor scores for robustness metric.
The betweenness analysis for the \pl based models shows an average betweenness value that is smaller for the networks with a lower value for the $\gamma$ coefficient so that they score best in the betweenness to \textit{order} ratio. On the other hand, a lower $\gamma$ implies a higher probability in the presence of nodes that have higher node degree; usually there is quite a good positive correlation between the node degree and the betweenness the nodes have to sustain (high degree implies high betweenness for that node). It is therefore understandable why the coefficient of variation is higher for the networks characterized by a low $\gamma$  than the ones with higher \pl characteristic parameter. 

\begin{figure}
 \captionsetup{type=figure}
    \centering
    \hspace*{-1cm}
    \subfloat[Characteristic path length.]{\label{fig:cplOthers}\includegraphics[scale=.3]{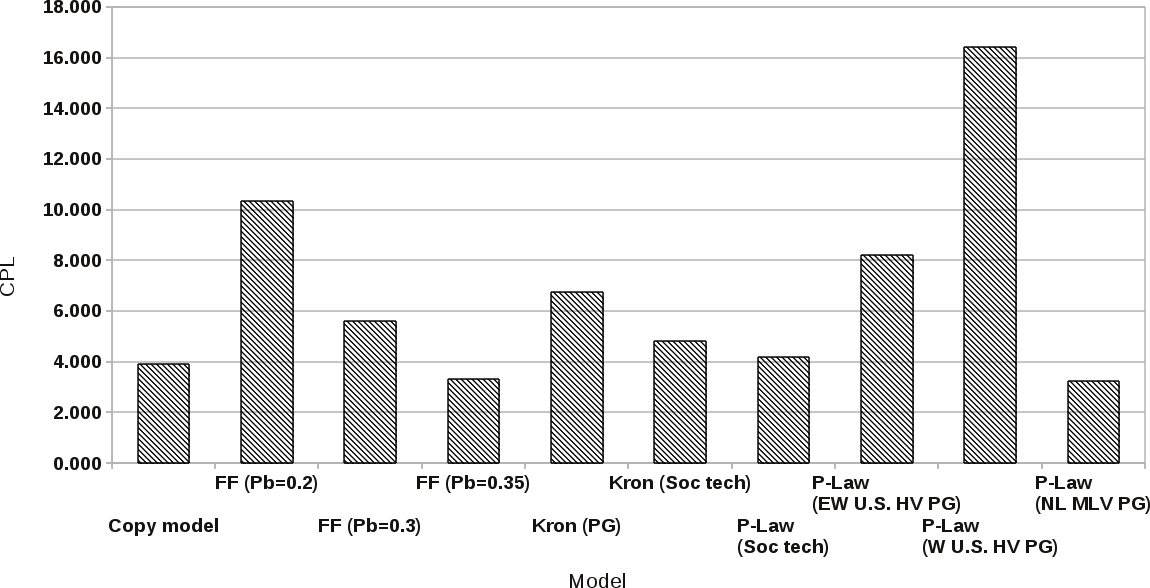}}
    \hspace*{4mm}
    \subfloat[Clustering coefficient.]{\label{fig:ccOthers}\includegraphics[scale=.3]{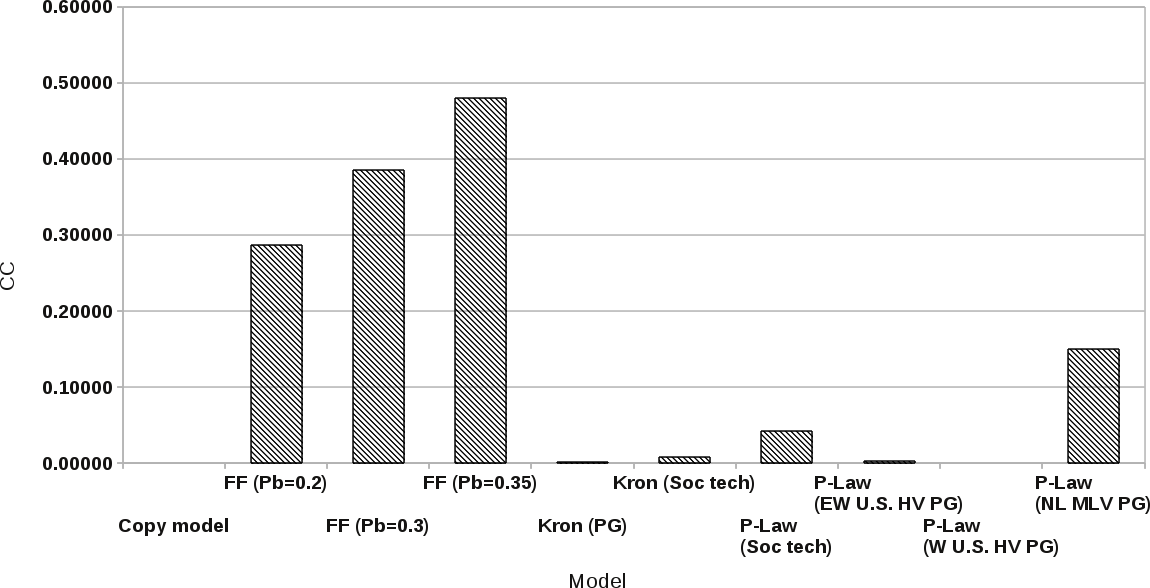}} 
    
    \subfloat[Removal robustness.]{\label{fig:robOthers}\includegraphics[scale=.3]{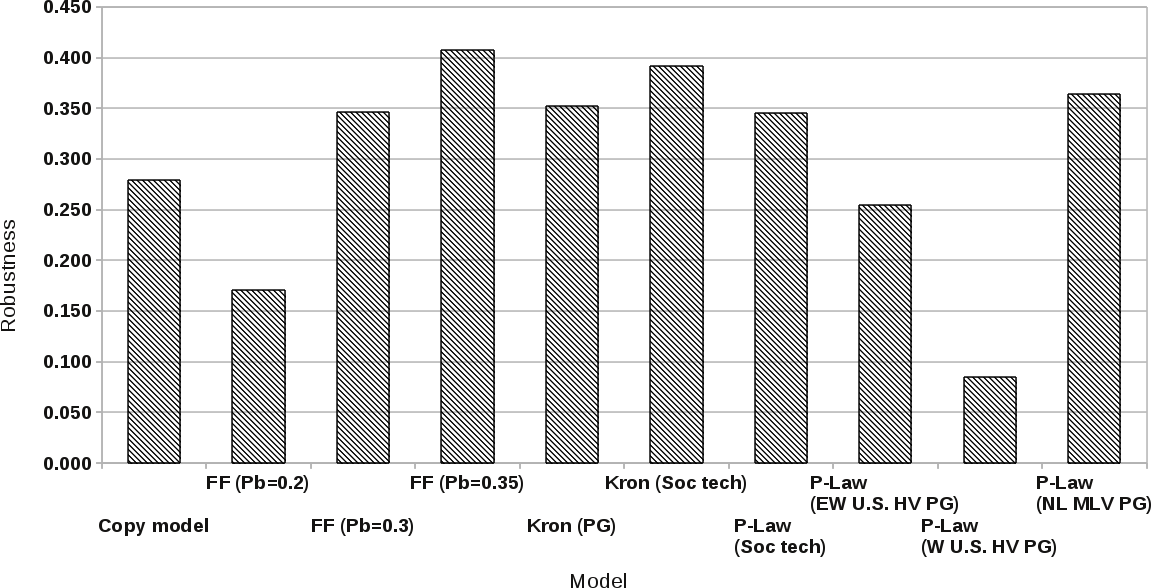}} 
    \caption{Metrics for the \textit{Large} sample of \MV network type for models independent from node degree.}
    \label{fig:parOthers}
\end{figure}

\begin{figure}
 \captionsetup{type=figure}
    \centering
    \hspace*{-0.85cm}
    \subfloat[Betweenness to \textit{order} ratio.]{\label{fig:betRadOthers}\includegraphics[scale=.3]{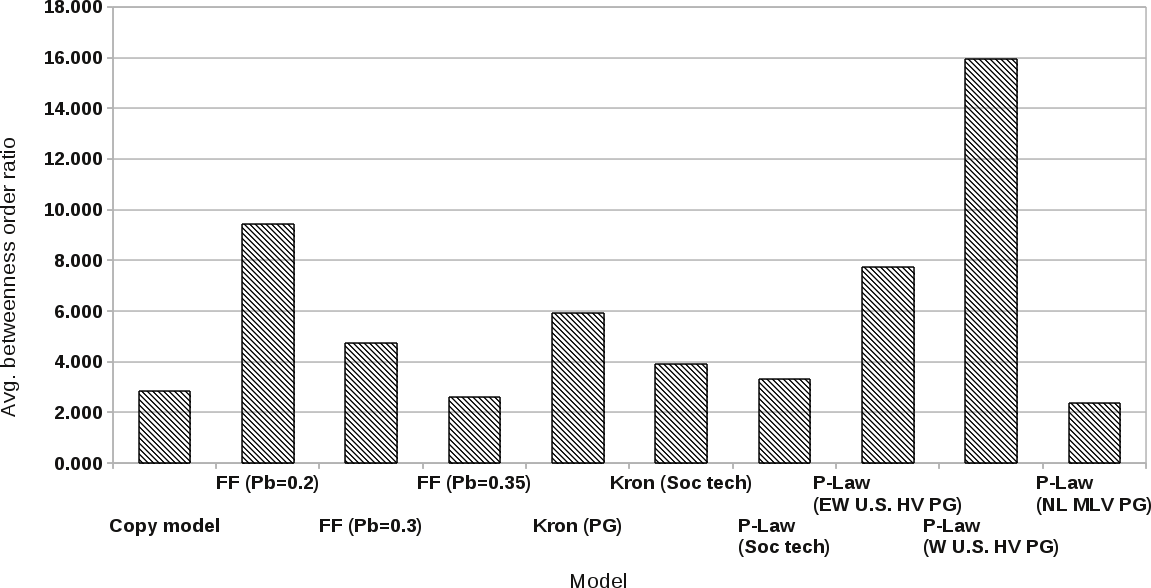}}
    \hspace*{4mm}
    \subfloat[Betweenness coefficient of variation.]{\label{fig:cVarOthers}\includegraphics[scale=.3]{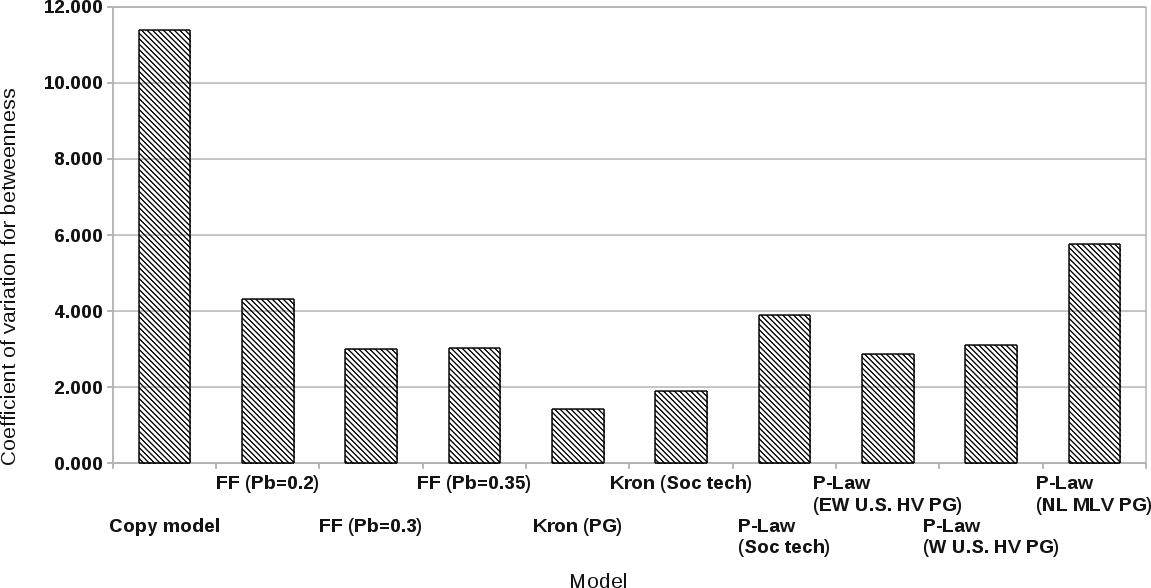}} 
    \caption{Metrics for the \textit{Large} sample of \MV network type for models independent from node degree.}
    \label{fig:parBetOthers}
\end{figure}

A graphical comparison of the results for networks without explicit dependence to average node degree for the \textit{Large} sample for \MV type considering characteristic path length, clustering coefficient and robustness metrics are shown in Figure~\ref{fig:parOthers}, while a summary of the results for betweenness for the same sample are illustrated in Figure~\ref{fig:parBetOthers}.

\subsubsection*{Comparing The Generated Topologies with the Physical Ones}

The analysis of the Northern Netherlands Grid shows an average degree almost constant of about $<k>\approx2$. Thus, it is fair to compare the generated models with similar average degree, the Copying Model ones and the Random Graphs with \pl in node degree distribution based on the data of Eastern and Western  \HV U.S. \PG and the U.S. Western \HV \PG since all generate networks with average node degree $<k>\approx2$. Generated models, except the model based on Random Graph with power-law, score better than the physical topologies for all the metrics considered; the \cpl scores half for the R-MAT and Copying Model cases in comparison to the real data. Also synthetic networks are more robust than the real data samples: R-MAT and Random Graph score constantly above 0.3 for robustness metric while real data hardly obtain this value. Clustering coefficients are quite similar since in this configuration with limited connectivity  having triangle structures in the network is rare, however we see that R-MAT model has almost always significant clustering coefficient values. An exception is the \sw model which scores almost always worse than the real data samples, in fact, under this situation of limited average node degree it is actually not fully correct to consider this synthetic topology a ``small-world''. The same sort of considerations can be done considering betweenness values: except the \sw model all the other synthetic ones score better for the average betweenness to \textit{order} ratio metric, while for the coefficient of variation the situation is similar.
If one considers the satisfaction of the desiderata for the actual samples of the Dutch Medium and Low Voltage Grid, summarized in Table ~\ref{tab:desiderataSatisfReal}, we notice that all parameters are not satisfied. However, networks generated according to the models with almost the same average node degree (networks with $<k>\approx2$ in Table~\ref{tab:desiderataSatisf} and networks based on Random Graph with \pl based on data from Eastern and Western \HV U.S. \PG and the U.S. Western \HV \PG in Table~\ref{tab:desiderataSatisfIndND}) do not satisfy all the desiderata as well. Therefore, this highlights that the first ingredient for the next generation of Grids to enable local energy exchange is an increase connectivity.

\begin{table}[htbp]
\centering
\begin{footnotesize}
\begin{tabular}{|l|c|}
  \hline
  \textbf{Desiderata}& Northern Netherlands \\ &\MLV samples \\
  \hline
  
  Modularity & \cross  \\  \hline
  $CPL\leq ln(N)$ & \cross \\  \hline
  $CC\geq 5\times CC_{RG}$ & \cross \\  \hline
  $\overline{\upsilon}=\frac{\overline{\sigma}}{N} \approx 2.5$& \cross  \\ \hline
  $c_v \leq 1$&\cross  \\ \hline
  $Rob_N \geq 0.45$&\cross  \\ \hline
  $APL_{10^{th}} \leq 2 \times CPL$&$\approx$ \\ \hline
\end{tabular}
\end{footnotesize}
\caption{Desiderata parameter compliance of real \MLV network samples of Northern Netherlands Grid.}\label{tab:desiderataSatisfReal}
\end{table}


Increasing the average node degree naturally provides for better values for the metrics, as shown in Table~\ref{tab:desiderataSatisf}. The case of the \sw model is emblematic. While the $<k>\approx2$ case scores extremely poor as there are not enough ``shortcuts'' in the networkso that they can not improve much the characteristic path length.
Actually under such small average degree the condition Watts and Strogatz impose for their model is not completely satisfied (i.e., $n\gg k\gg ln(n)\gg 1$, where $k$ is the average node degree and $n$ is the \textit{order} of the graph). When we move closer to satisfying the \sw condition by increasing the average node degree,  the value of the metrics suddenly changes and the models score extremely high. The \sw scores best for the clustering property and resilience to failures in $<k>\approx4$ situations. Under these conditions also the betweenness values are quite concentrated around the mean with a coefficient of variation that does not exceed the unit.

\begin{sidewaystable}[htbp]
\centering
\begin{footnotesize}
\begin{tabular}{|l|c|c|c|c|c|c|c|c|c|c|c|c|}
  \hline
  \textbf{Desiderata}&\multicolumn{4}{|c|}{\textbf{Average node degree $<k>\approx2$}}&\multicolumn{4}{|c|}{\textbf{Average node degree $<k>\approx4$}}&\multicolumn{4}{|c|}{\textbf{Average node degree $<k>\approx6$}} \\
  \hline
  & SW & Pref. Attach.&Rnd. Graph&R-MAT& SW & Pref. Attach.&Rnd. Graph&R-MAT& SW & Pref. Attach.&Rnd. Graph&R-MAT\\ \hline
  Modularity & $\approx$ &\cross &\cross &\tick & $\approx$ &\cross &\cross &\tick & $\approx$ &\cross &\cross &\tick \\  \hline
  $CPL\leq ln(N)$ & \cross &$\approx$ &$\approx$ &\tick & \tick &\tick &\tick &\tick & \tick &\tick &\tick &\tick \\  \hline
  $CC\geq 5\times CC_{RG}$ & \cross &\cross &N/A &$\approx$ & \tick &\tick &N/A &\cross & \tick &\tick &N/A &$\approx$ \\  \hline
  $\overline{\upsilon}=\frac{\overline{\sigma}}{N} \approx 2.5$& \cross &\cross &\cross &\cross &\cross &$\approx$ &\cross &\cross & $\approx$ &\tick &$\approx$ &\tick \\ \hline
  $c_v \leq 1$&\cross &\cross &\cross &\cross & \tick &\cross &\tick &\cross & \tick &\cross &\tick &\cross \\ \hline
  $Rob_N \geq 0.45$&\cross &\cross &\cross &\cross & \tick &\tick &\tick &\tick & \tick &\tick &\tick &\tick \\ \hline
  $APL_{10^{th}} \leq 2 \times CPL$&\tick &\tick &\tick &\tick & \tick &\tick &\tick &\tick & \tick &\tick &\tick  &\tick \\ \hline
\end{tabular}
\end{footnotesize}
\caption{Desiderata parameter compliance of the generated models with node degree $<k>\approx2,4,6$.}\label{tab:desiderataSatisf}
\end{sidewaystable}

Comparing the average values of the generated models for increasing node degree, one notices a natural improvement of the metrics, cf.\ Table~\ref{tab:metricsImpr}. In fact, we have a reduction in \cpl of about 60\% and an increase in the clustering coefficient of one order of magnitude, at the same time the robustness doubles. With $<k>\approx6$ the improvement compared to the metrics is less prominent, being between 10\% and 20\%.
From the comparison of the metric results in Table~\ref{tab:desiderataSatisf} one sees that the \sw model almost always satisfies the desiderata requirement from a quantitative point of view when the average node degree is at least 4. From a qualitative point of view, the \sw model shows to some extent certain characters of modularity being generated starting from a regular lattice and then rewiring a certain fraction of the edges.

\begin{table}
\centering
\begin{center}
\begin{tabular}{|l|p{2cm}|p{2cm}|p{2cm}|}
\hline
\textbf{Avg. node degree transition} & \multicolumn{3}{|c|}{\textbf{Average metric improvement (\%)}}\\ \hline
  & CPL &CC&Robustness \\ \hline
$<k>\approx2 \rightarrow <k>\approx4$ & \rl 61.7	& \rl 941.6& \rl 128.5 \tn \hline
$<k>\approx4 \rightarrow <k>\approx6$ & \rl 18.0	& \rl 11.8&	\rl 19.6\tn \hline
\end{tabular}
\caption{Comparison of generated topologies for varying average node degree.}\label{tab:metricsImpr}
\end{center}
\end{table}

\begin{sidewaystable}[htbp]
\centering
\begin{footnotesize}
\begin{tabular}{|l|c|c|c|c|c|c|c|c|c|c|c|c|}  \hline
  \textbf{Desiderata}& Copying & Forest Fire  & Forest Fire  & Forest Fire &Kronecker &Kronecker & RG with \pl &RG with \pl &RG with \pl &RG with \pl \\ 
  &Model & (pb=0.2)& (pb=0.3)& (pb=0.35)& (PG params) & (social & (social& (East-West US  & (Western US & (NL MLV  \\ 
  & &  & & & & net params)& net params)& HV PG params) &  HV PG params)& PG params)\\ \hline
  Modularity & $\approx$ &\cross &\cross &\cross & \tick &\tick &\cross &\cross & \cross &\cross \\  \hline
  $CPL\leq ln(N)$ & \tick &\cross &$\approx$ &\tick & \cross &\tick &\tick &\cross & \cross &\tick  \\  \hline
  $CC\geq 5\times CC_{RG}$ & \cross &\tick &\tick &\tick & \cross &\cross &\tick &$\approx$ & \cross &\tick \\  \hline
  $\overline{\upsilon}=\frac{\overline{\sigma}}{N} \approx 2.5$& $\approx$ &\cross &\cross &$\approx$ &\cross &\cross &\cross &\cross & \cross &\cross \\ \hline
  $c_v \leq 1$&\cross &\cross &\cross &\cross & \cross &\cross &\cross &\cross & \cross &\cross \\ \hline
  $Rob_N \geq 0.45$&\cross &\cross &\cross &$\approx$ & \cross &\cross &\cross &\cross & \cross &\cross  \\ \hline
  $APL_{10^{th}} \leq 2 \times CPL$&N/A &\cross &\tick &\tick & \tick &\tick &\tick &\tick & \tick &\tick  \\ \hline
\end{tabular}
\end{footnotesize}
\caption{Desiderata parameter compliance of the generated models.}\label{tab:desiderataSatisfIndND}
\end{sidewaystable}

The models independent form average node degree perform generally worse than the other models in satisfying the desiderata values for the \PG metrics. The adherence to the target values  are shown in Table~\ref{tab:desiderataSatisfIndND}; one sees the general prevalence of requirement dissatisfaction, especially parameters involving betweenness are never satisfied by these generated samples.

From the topological analysis one can see that between the models analyzed  when there is a minimal connectivity ($<k>\approx4$ or $<k>\approx6$) the \sw stands out, cf. Table~\ref{tab:desiderataSatisf}. In Table~\ref{tab:summarytable} the models with explicit dependence on node degree are once again compared by assigning a ``tick'' sign (\tick) for the fulfillment of each of the following properties: qualitative topological parameters (i.e., modularity), quantitative topological parameters (Table~\ref{tab:desiderataSatisf}) and the thrift in network realization (e.g., addition of cables which represent a cost). This last parameter given is just a rough estimation, a more detailed analysis of cost in realizing a network belonging to Medium or Low Voltage with a certain size (i.e., \textit{Small}, \textit{Medium} or \textit{Large}) and the economic benefits in electricity distribution arising from the enhanced connectivity is provided in Section~\ref{sec:discussion}. From Table~\ref{tab:summarytable} we conclude that networks generated with \sw model with average degree $<k>\approx4$ provide the the best balance to satisfy the desiderata of the future Power Grid.

\begin{table}[h!]
\centering
\begin{footnotesize}
\begin{tabular}{|l|c|c|c|}
  \hline
  \textbf{Network Model}&\textbf{Avg. node degree $<k>\approx2$}&\textbf{Avg. node degree $<k>\approx4$}&\textbf{Avg. node degree $<k>\approx6$} \\
  \hline
   \textbf{Small-world} &\tick\tick &\tick\tick\tick &\tick\tick\\ \hline
  \textbf{Preferential Attachment} & \tick &\tick\tick &\tick \\  \hline
  \textbf{Random Graph} & \tick &\tick\tick &\tick \\  \hline
  \textbf{R-MAT} & \tick\tick &\tick\tick &\tick \tick \\ \hline
\end{tabular}
\end{footnotesize}
\caption{Summary table considering satisfaction of modularity, performance and cabling cost for generated models with node degree $<k>\approx2,4,6$.}\label{tab:summarytable}
\end{table}

\section{Economic Considerations}~\label{sec:discussion}

Traditionally the problem of evaluating the expansion of an electrical system is a complex task that involves both the use of modeling, usually based on operation research optimization techniques and linear programming~\cite{garver70,lee74,belagari75}, and the experience and vision of experts in the field supported by computer systems. In this latter case computers acquire knowledge based on previous experts' decision and, based on the electrical physical constraints of the domain, are then able to support \PG evolution decision~\cite{teive98} finding the most suitable technical and economical solution. With more distributed generating facilities at local scale, traditional methods have limits and need to be modified or updated to take into account the new scenario the Smart Grid framework brings into play. The models that we have so far analyzed as being candidates for the vision of the future \SG need also to be evaluated from the economic point of view. How much will it cost to generate electrical infrastructures according to these models? What is the actual cost of adding a physical edge to the topology?

\subsection*{The Cost of Adding Edges}

One important difference that a physical infrastructure such as the \PG has compared to the WWW or social networks is the physical presence of cables that have to connect the \MV substations or \LV end-users generating units. If establishing a link from a Web page to another one is free, on the other hand, each increase in connectivity in the \PG implies costs in order to adequate the substation or end-user premise involved and the cables required for the connection. To assess these cost in the \MLV infrastructure, we consider a simple relation where the cost of cabling and cost of substations are added:
\begin{equation}
C_{imp}=\sum_{j=1}^{N} Ssc_{j}+\sum_{i=1}^{M}Cc_i
\end{equation}
where $C_{impl}$ stands for cost for implementation, $Ssc_j$ is the adaptation cost for the substation $j$ and $Cc_i$ is the cost for the cable $i$.  The cost of the cable can be expressed as a linear function of the distance the cable $i$ covers: $Cc_i= C_{uc}\cdot l$ where $C_{uc}$ is the cable cost per unit of length and $l$ is the length of the cable.  Several types of cables exist which are used for power transmission and distribution with varying physical characteristics and costs, in addition also the cost for installation can vary significantly~\cite{natGrid}. In the present work, though, we simply consider cabling costs and ignore substation ones. While the former are directly tied to the topology and length of the links, the latter pricing is too dependent on other factors. As a source of data for cable type and pricing, we have been provided (courtesy of Enexis B.V.) with cables characteristics and prices together with topological information for 11 network samples belonging to the \LV network and 13 samples belonging to the \MV of the Northern Netherlands.

\subsection*{Statistical consideration over cables' price}\label{subsec:priceStat}

\begin{figure}
   \centering
   \includegraphics[width=1\textwidth]{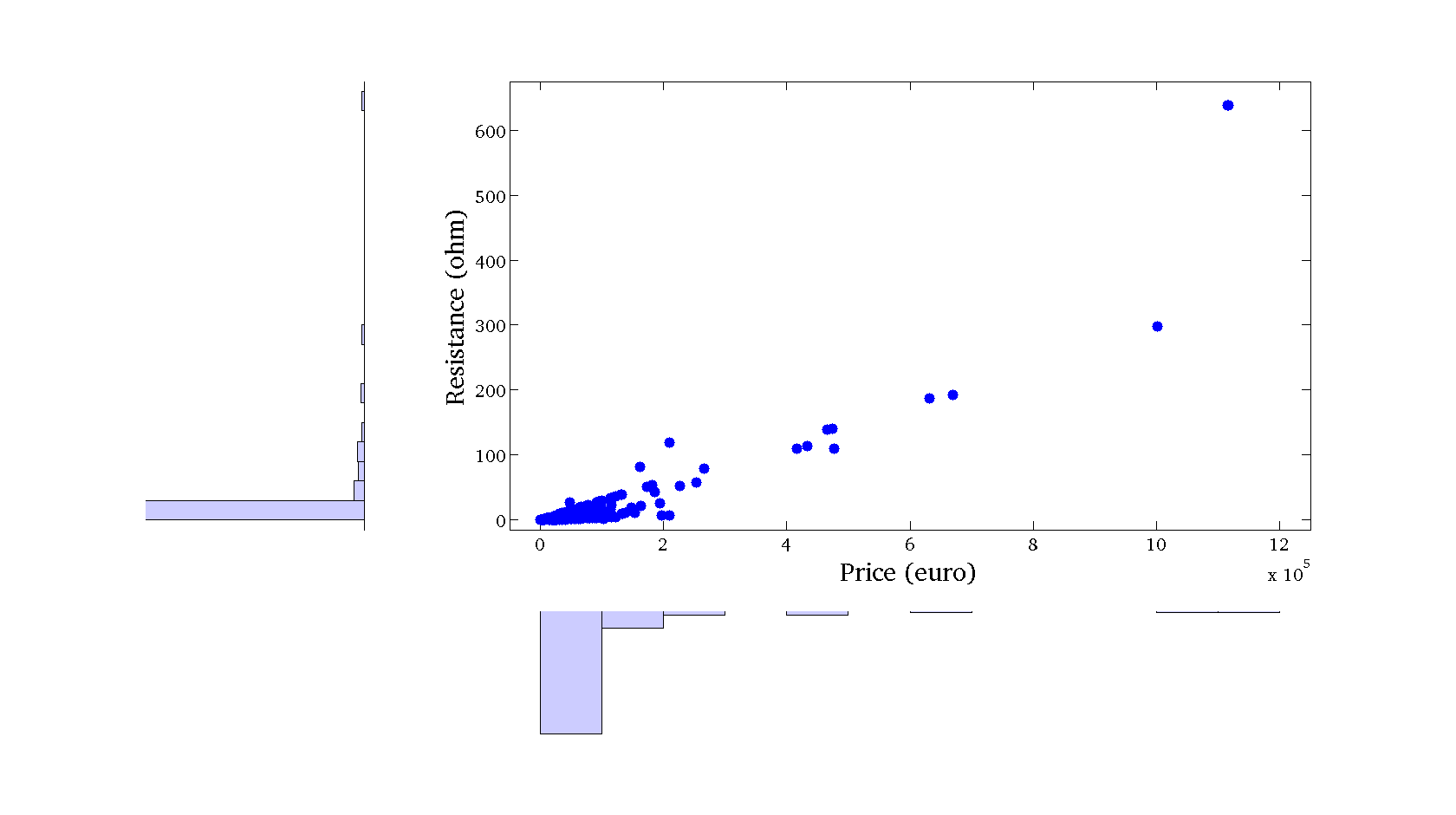}
   \caption{Price-Resistance pairs joint plot for the \MV \textit{Small} size sample.}
\label{fig:corrMvSmall}
\end{figure}

\begin{figure}
   \centering
   \includegraphics[width=1\textwidth]{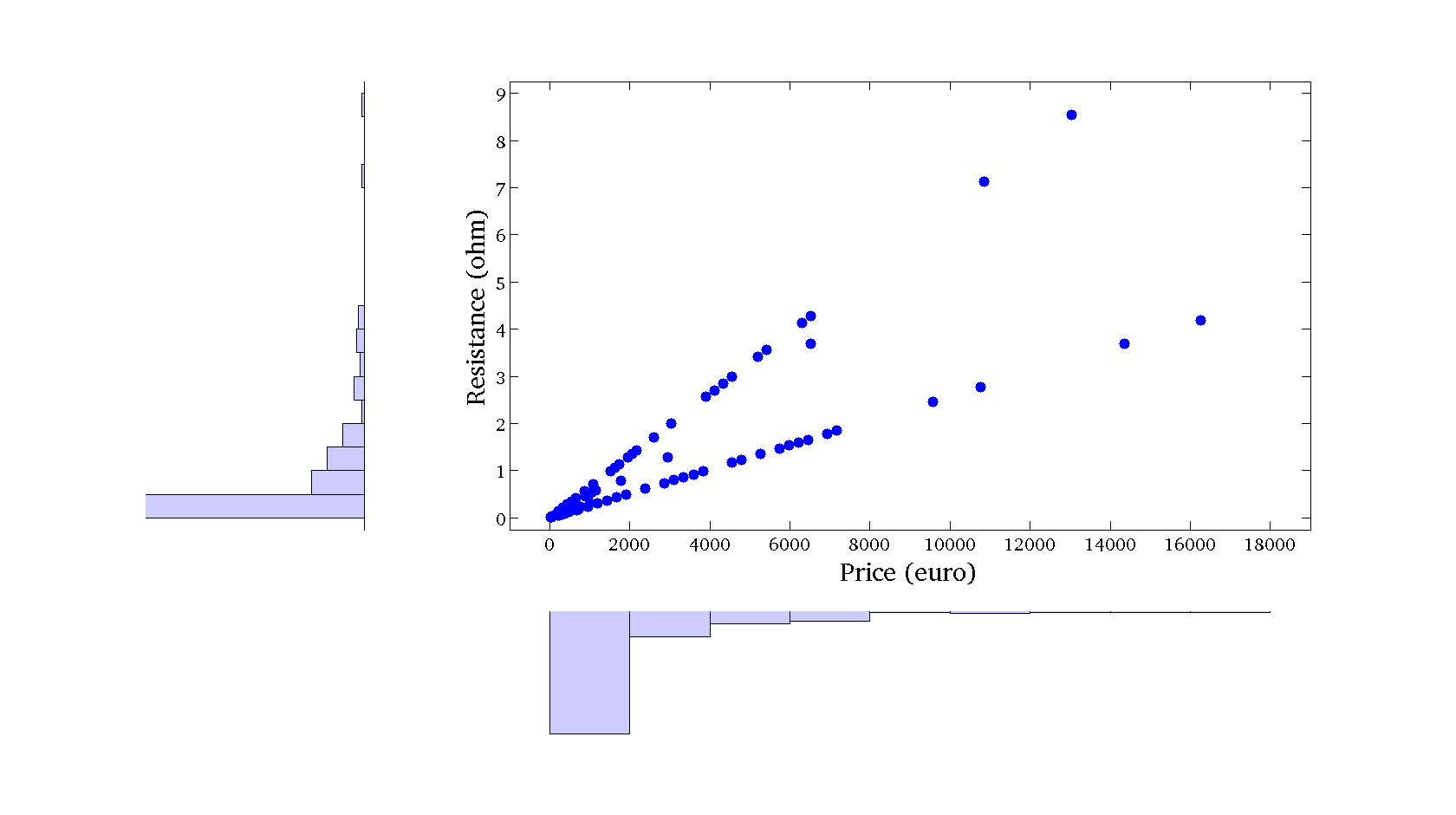}
   \caption{Price-Resistance pairs joint plot for the \LV large size sample.}
\label{fig:corrLvLarge}
\end{figure}

Extracting probability distributions of physical and price data out of North Netherlands data samples, shows interesting correlations. The length of the cables plays an important role for both total resistance and price. If one considers the correlation between the price and resistance, high values are found. Using Spearman's rank correlation coefficient~\cite{kendall48}, shown in Table~\ref{tab:rankCorr}, one can evaluate to what extent the variation tendency characterizing two variables can be described by a monotonic relationship. In other words, one has an indication of the correlation between price and resistance. Especially, for generating synthetic networks it is important to obtain values for both the properties of the cables that are similar to the ones actually used in practice. Plotting the two variables characterizing each cable one notices that the majority of the samples concentrates in the lower tails for the joint distribution. Figures~\ref{fig:corrMvSmall} and~\ref{fig:corrLvLarge} show the relation between the price and resistance where the values concentrate in the lower corner of the $price\times resistance$ space.
\begin{table}
\begin{center}
\centering
\begin{tabular}{|l|p{4.5cm}|}
\hline
\textbf{Sample type} & \textbf{Spearman's rank correlation Price-Resistance}\\ \hline
\LV- Small & \rl 0.962\tn \hline
\LV- Medium  & \rl 0.974\tn \hline
\LV- Large  & \rl 0.937\tn \hline
\MV- Small & \rl 0.787\tn \hline
\MV- Medium  & \rl 0.634\tn \hline
\MV- Large  & \rl 0.946\tn \hline
\end{tabular}
\caption{Spearman's rank correlation for \LV and \MV representative samples.}\label{tab:rankCorr}
\end{center}
\end{table}

In the chart in Figure~\ref{fig:corrLvLarge} one notices the two distinct lines that deviate from the low-left corner. They represent the two main types of cables that are used in that sample of the \LV network to cover different distances and that result in increasing in price and resistance when longer lines are realized.
This opens a new perspective: {\em evaluate for each type of cable used in a certain sample (Small, Medium and Large) how the length of the cables used are distributed.} In fact, given a certain type of cable and its length all other interesting properties for our analysis are then available (i.e., cable total resistance, cable total cost and cable current supported).

\begin{figure}
   \centering
   \includegraphics[width=1\textwidth]{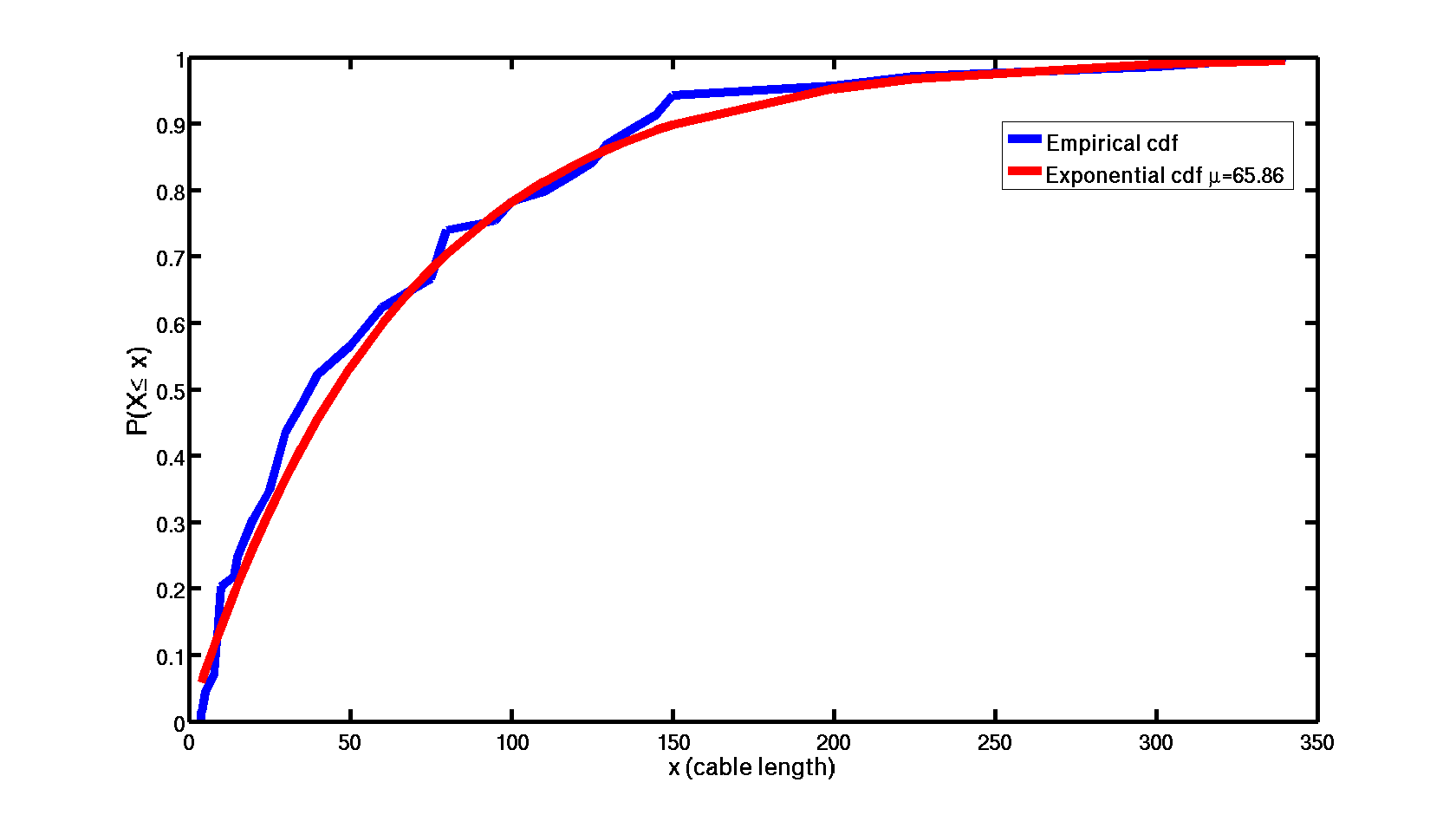}
   \caption{Cumulative distribution function for cable length (meters) for cable type ``VMvK(h)as 4x150 al'' in Northern Netherlands sample \LV size Large.}
\label{fig:cdfLV}
\end{figure}

\begin{figure}
   \centering
   \includegraphics[width=1\textwidth]{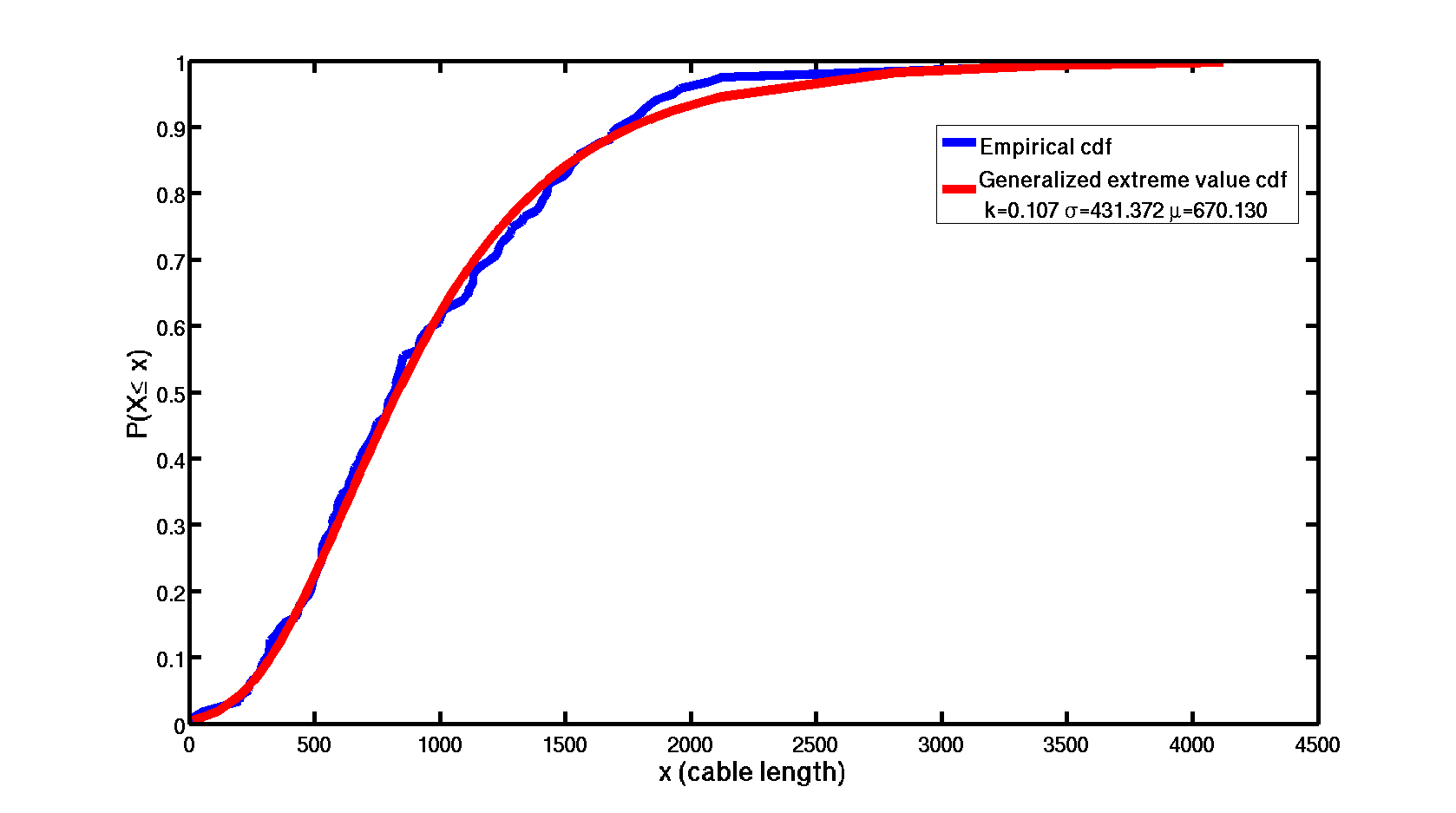}
   \caption{Cumulative distribution function for cable length (meters) for cable type ``3x1x70al'' in Northern Netherlands \MV sample size Medium.}
\label{fig:cdfMV}
\end{figure}

\begin{figure}
   \centering
   \includegraphics[width=1\textwidth]{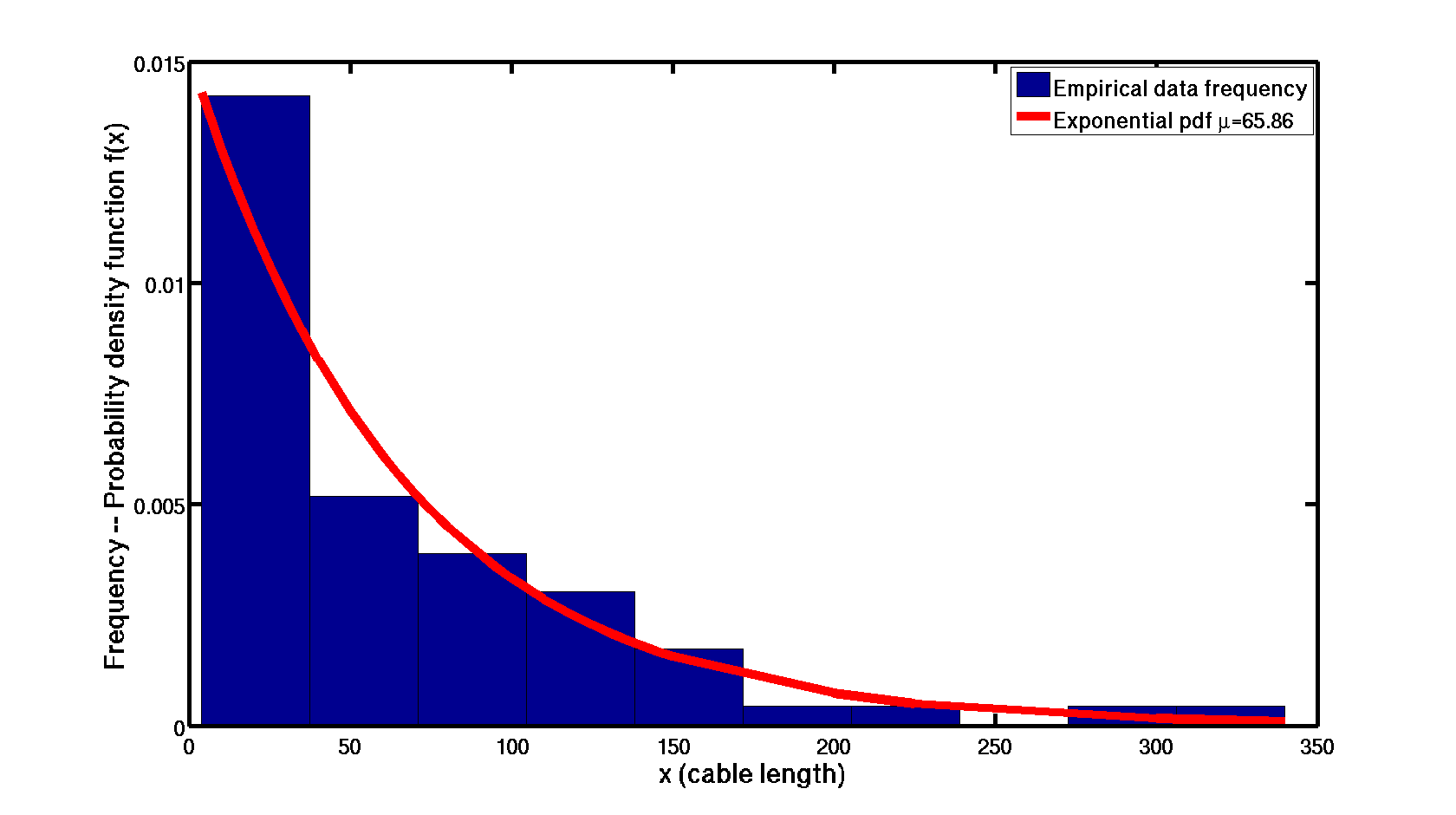}
   \caption{Probability density function for cable length (meters) for cable type ``VMvK(h)as 4x150 al'' in Northern Netherlands sample \LV size Large.}
\label{fig:pdfLV}
\end{figure}

\begin{figure}
   \centering
   \includegraphics[width=1\textwidth]{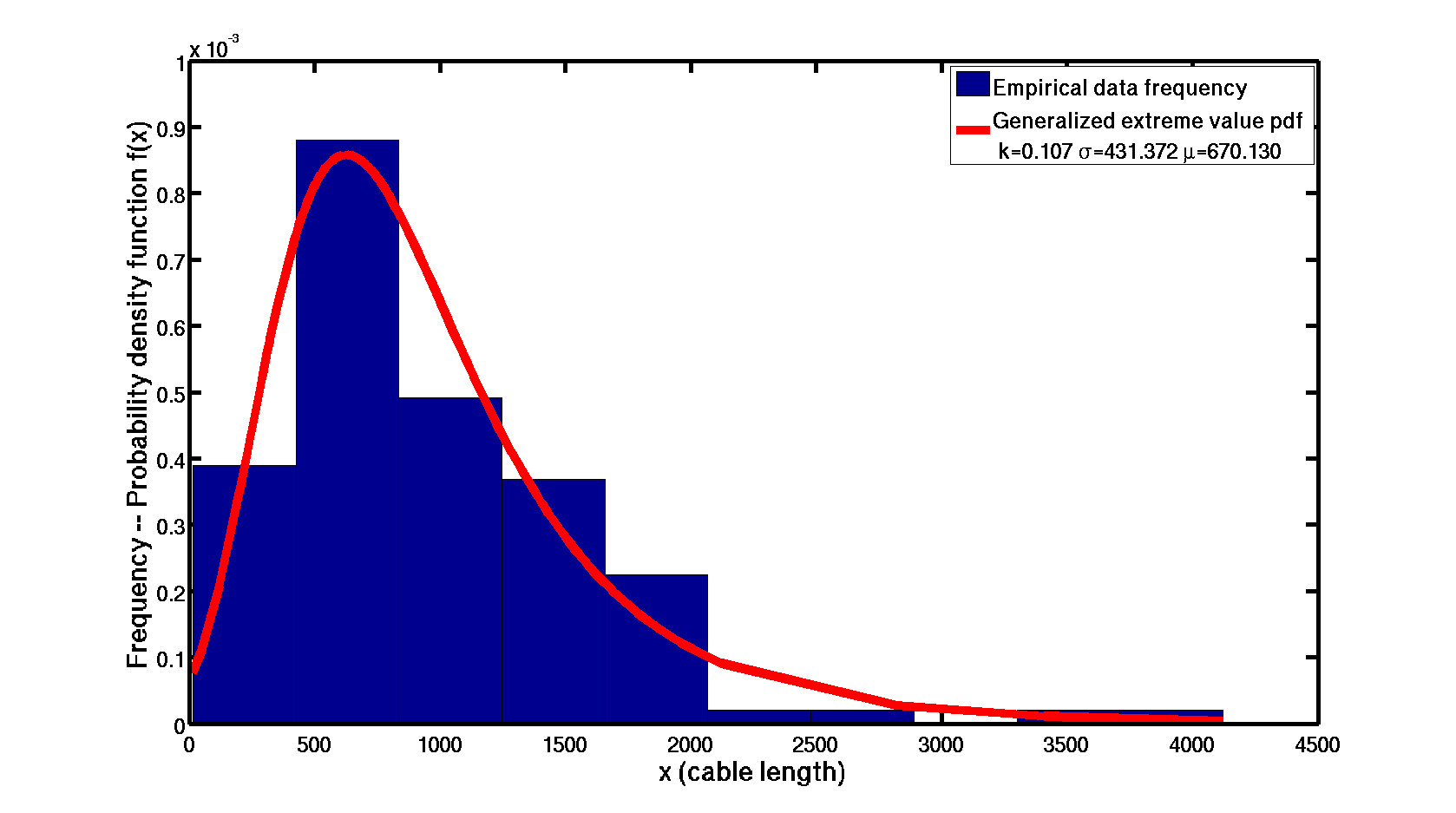}
   \caption{Probability density function for cable length (meters) for cable type ``3x1x70al'' in Northern Netherlands \MV sample size Medium.}
\label{fig:pdfMV}
\end{figure}

A general tendency appears when fitting the distribution of lengths to cable types belonging to \LV and Medium Voltage: a fast decay in lengths' probability distribution with the majority of lengths for the \LV cables types in the order of tens of meters, and \MV cables hundreds of meters. Fitting the length to a statistical probability distribution gives a good approximation for the \LV cable lengths as exponential distributions ($y=f_X(x;\mu)=\frac{1}{\mu}e^{\frac{-x}{\mu}}$). Figures~\ref{fig:cdfLV} and~\ref{fig:pdfLV} show respectively the cumulative distribution probability and the probability density functions for a certain type of cable belonging to the \LV network. The use of the Kolmogorov-Smirnov test~\cite{massey51} lets us accept the hypothesis in favor of this distribution. The situation is slightly different for the \MV cables where the distribution that generally fits best the data is the generalized extreme value distribution ($y=f_X(x;k,\mu,\sigma)=\frac{1}{\sigma}(1+k\frac{x-\mu}{\sigma})^{-1-\frac{1}{k}}$exp$\{-(1+k\frac{x-\mu}{\sigma})^{-\frac{1}{k}}\}$); even in this case the Kolmogorov-Smirnov test supports this hypothesis. A graphical representation of the probability cumulative distribution function and the probability density function per cable type of to the \MV network are shown in Figures~\ref{fig:cdfMV} and~\ref{fig:pdfMV}, respectively.

Assume that, statistically speaking, the distribution of the lengths for each type of cable in the synthetic networks are the same as in the real samples. Therefore, knowing the probability of using a certain type of cable $i$ ($p_{cable_i}=\frac{\#cable_i}{\sum_{k}\#cable_k}$ where $\#cable_i$ is the number of occurrences of cable type $i$ in a certain network sample) that has a certain cost and resistance per meter and a specific current supported, it is then possible to estimate the cables that are used in the synthetic samples together with their properties.

\subsection*{Economic Benefits of Highly Connected Topologies}

Once the information about cable prices is available, it is possible to estimate the cost for realizing a network with a certain connectivity and if such networks are able to lower the (economic) barrier towards decentralized trading. The results for \LV networks with an average node degree $<k>\approx2$ are shown in Table~\ref{tab:priceLVk2}. The results for $<k>\approx4$ and $<k>\approx6$ are about two and three times more expensive since there is an increase in the number of edges by the same quantity; for completeness the results are summarized in Tables~\ref{tab:priceLVk4} and~\ref{tab:priceLVk6}.

\begin{table}
\begin{center}
\centering
\begin{tabular}{|l|p{2cm}|p{4cm}|}
\hline
\textbf{Sample type} & \textbf{\textit{Size}}&\textbf{Cost (thousand euro)}\\ \hline
\LV- Small & \rl $\approx 20$	&\rl $\approx 30$\tn \hline
\LV- Medium  & \rl $\approx 90$	&\rl$\approx 78$\tn \hline
\LV- Large  & \rl $\approx 200$	&\rl $\approx 449$\tn \hline
\end{tabular}
\caption{Cabling cost for $<k>\approx2$ synthetic samples for \LV networks.}\label{tab:priceLVk2}
\end{center}
\end{table}

\begin{table}
\begin{center}
\centering
\begin{tabular}{|l|p{2cm}|p{4cm}|}
\hline
\textbf{Sample type} & \textbf{\textit{Size}}&\textbf{Cost (thousand euro)}\\ \hline
\LV- Small & \rl $\approx 40$	&\rl $\approx 51$\tn \hline
\LV- Medium  & \rl $\approx 180$	&\rl $\approx 174$\tn \hline
\LV- Large  & \rl $\approx 400$	&\rl $\approx 827$\tn \hline
\end{tabular}
\end{center}
\caption{Cabling cost for $<k>\approx4$ synthetic samples for \LV networks.}\label{tab:priceLVk4}
\end{table}

\begin{table}
\begin{center}
\centering
\begin{tabular}{|l|p{2cm}|p{4cm}|}
\hline
\textbf{Sample type} & \textbf{\textit{Size}}&\textbf{Cost (thousand euro)}\\ \hline
\LV- Small & \rl $\approx 60$	&\rl $\approx 76$\tn \hline
\LV- Medium  & \rl $\approx 270$	&\rl $\approx 254$\tn \hline
\LV- Large  & \rl $\approx 600$	&\rl $\approx 1239$\tn \hline
\end{tabular}
\end{center}
\caption{Cabling cost for $<k>\approx6$ synthetic samples for \LV networks.}\label{tab:priceLVk6}
\end{table}

For \MV networks, it is important to clarify that the information available for cables' prices in this study are only partial and limited to some technologies (only few cross sections of aluminum and copper cables). Anyway, in order to have a glimpse of costs for this type of the network, we fitted to the best interpolating curve the available prices as a function of the cross section. The relation between price and cross section for aluminum cables fits best to a cubic polynomial, while for the copper ones is linear; in this way we can have an estimation for the prices for all the types of cables involved knowing their cross section.
\begin{table}
\begin{center}
\centering
\begin{tabular}{|l|p{2cm}|p{4cm}|}
\hline
\textbf{Sample type} & \textbf{\textit{Size}}&\textbf{Cost (millions euro)}\\ \hline
\LV- Small & \rl $\approx 250$	&\rl $\approx 32$\tn \hline
\LV- Medium  & \rl $\approx 500$	&\rl $\approx 42$\tn \hline
\LV- Large  & \rl $\approx 1000$	&\rl $\approx 43$\tn \hline
\end{tabular}
\end{center}
\caption{Cabling cost for $<k>\approx2$ synthetic samples for \MV networks.}\label{tab:priceMVk2}
\end{table}
The results for the networks with an average node degree $<k>\approx2$ are shown in Table~\ref{tab:priceMVk2}. The results for $<k>\approx4$ and $<k>\approx6$ are just two and three times more expensive since there is an increase in the number of edges by these same factors; for completeness, the results are shown in Tables~\ref{tab:priceMVk4} and~\ref{tab:priceMVk6}. The small difference in costs between the \textit{Medium} and \textit{Large} types of networks is related mainly to the different technologies (i.e., cable types) in the cables that are used for these types of networks.

\begin{table}
\begin{center}
\centering
\begin{tabular}{|l|p{2cm}|p{4cm}|}
\hline
\textbf{Sample type} & \textbf{\textit{Size}}&\textbf{Cost (millions euro)}\\ \hline
\LV- Small & \rl$\approx 500$	&\rl $\approx 55$\tn \hline
\LV- Medium  & \rl$\approx 1000$	&\rl $\approx 88$\tn \hline
\LV- Large  & \rl$\approx 2000$	&\rl $\approx 86$\tn \hline
\end{tabular}
\end{center}
\caption{Cabling cost for $<k>\approx4$ synthetic samples for \MV networks.}\label{tab:priceMVk4}

\end{table}

\begin{table}
\begin{center}
\centering
\begin{tabular}{|l|p{2cm}|p{4cm}|}
\hline
\textbf{Sample type }&\textbf{ \textit{Size}}&\textbf{Cost (millions euro)}\\ \hline
\LV- Small & \rl $\approx 750$	&\rl $\approx 80$\tn \hline
\LV- Medium  & \rl $\approx 1500$	&\rl $\approx 132$\tn \hline
\LV- Large  & \rl $\approx 3000$	&\rl $\approx 131$\tn \hline
\end{tabular}
\end{center}
\caption{Cabling cost for $<k>\approx6$ synthetic samples for \MV networks.}\label{tab:priceMVk6}
\end{table}

Price alone is not enough to describe future scenarios. It is important to investigate how an enhanced connectivity is beneficial to the electricity distribution costs. We have provided the benefits for more connected networks at the beginning of this section, however those results consider only the topology without any parameter related to the properties of the cables (e.g., resistance and supported current). In order to consider the effects of topology in electricity distribution costs, we have developed a set of metrics that associate topological properties of \PG networks to costs in electricity distribution. We have applied these metrics in the analysis of the \MLV Grid of the Northern Netherlands in~\cite{PaganiAielloTSG2011}. In order to apply these metrics to \PG networks  \textit{weights} are essential, representing physical quantities such as resistance of the cable and maximal operating current supported by the cable. Once we have the statistical information about the types and the length of the cables used in a specific type of physical network, (i.e., Medium or Low Voltage and its \textit{Small}, \textit{Medium} or \textit{Large} size) it is possible to assign \textit{weights} to the edges of the generated graphs. This is done under the assumption that the same type of cables are used and that the distances covered in general (i.e., statistically) remain the same.
In~\cite{PaganiAielloTSG2011} we proposed to consider two types of measures that influence electricity price: one related to the dissipation and losses aspects on the Grid called $\alpha$, and a second one which takes into account the aspects of reliability in the network called $\beta$. Formally, these measures have the following expressions:
\begin{align} 
\alpha & = f(L_{line_N},L_{substation_N})\label{eq:alphaMainText};
 \end{align}
\begin{align}
 \beta & = f( Rob_{N},Red_{N},Cap_{N})\label{eq:betaMainText}.
 \end{align}
In equation~\ref{eq:alphaMainText} the factors influencing $\alpha$ are the losses happening in the electrical lines ($L_{line_N}$) and the losses arising at substations ($L_{substation_N}$). On the other hand, the parameters influencing $\beta$ consider the robustness of the network to failures ($Rob_{N}$), the loss experienced in following redundant paths between nodes ($Red_{N}$) and the available capacity in the lines connecting nodes of the network ($Cap_{N}$). The relationship between $\alpha$ and $\beta$ and the price of electricity is considered quadratic as other components (e.g., fuel price) influencing electricity price hold this relationship~\cite{Harris06}. Considering how $\alpha$ and $\beta$ parameters are computed~\cite{PaganiAielloTSG2011}, the smaller the values for each, the smaller the impact topology has on electricity prices (see~\ref{sec:AppAlfaBeta}). For completeness the essential information about topology and electricity cost-related metrics are more thoroughly explained in~\ref{sec:AppAlfaBeta}.

\begin{figure}
   \centering
   \includegraphics[width=1\textwidth]{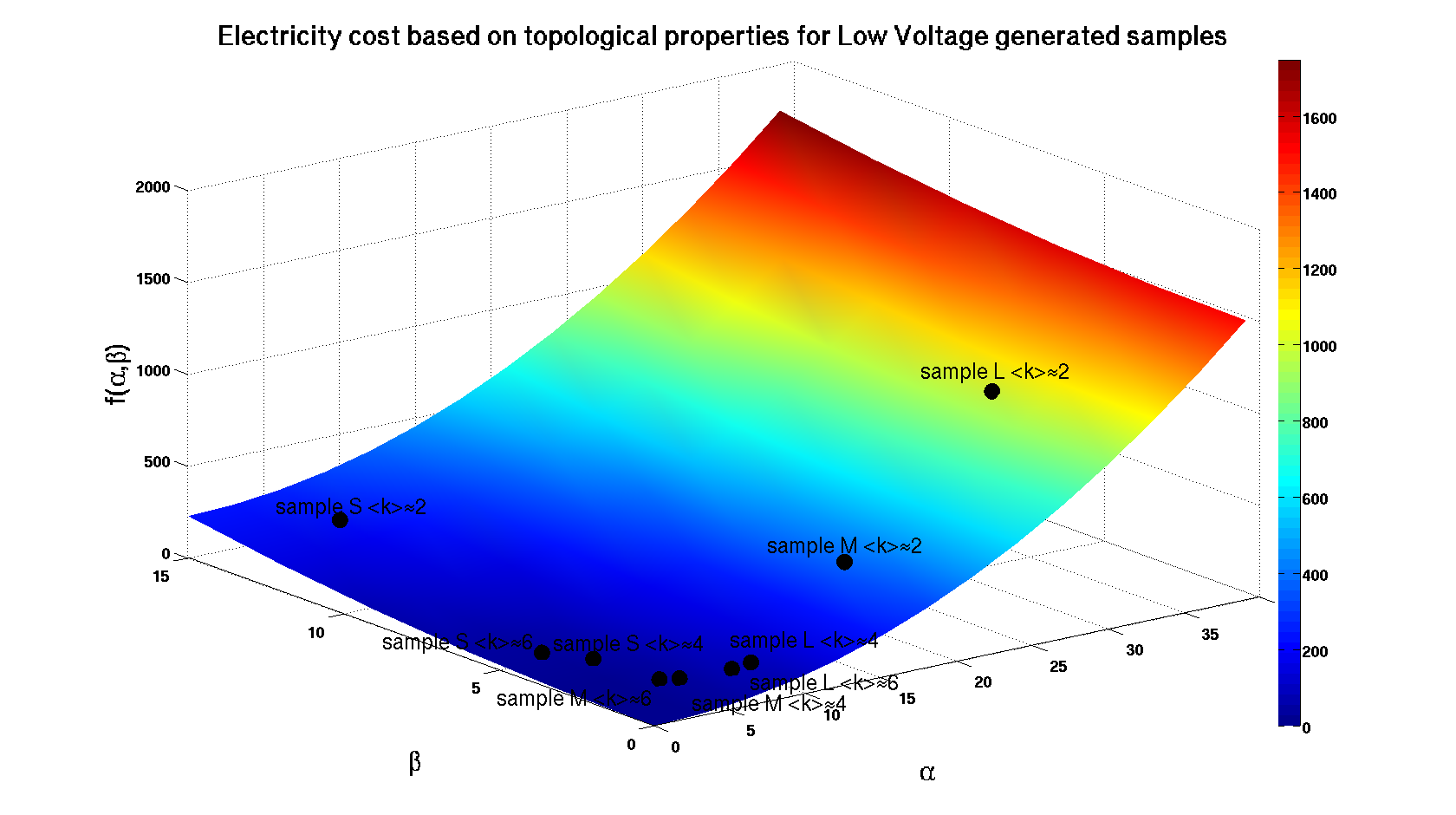}
   \caption{Transport cost of electricity based on the topological properties for synthetic networks based on \sw model for Low Voltage Grid.}
\label{fig:alphaBetaLV}
\end{figure}

\begin{figure}
   \centering
   \includegraphics[width=1\textwidth]{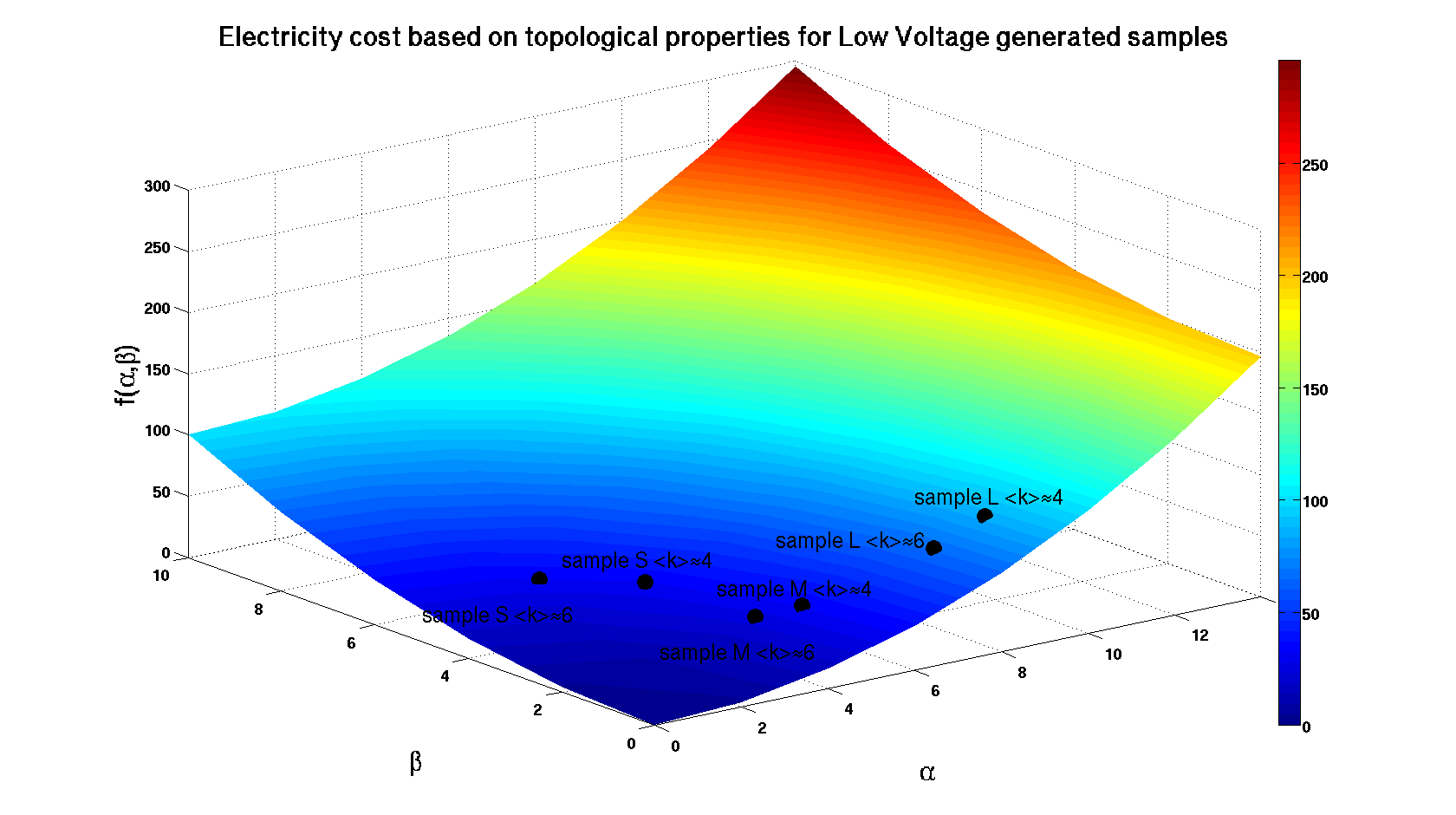}
   \caption{Transport cost of electricity based on the topological properties for synthetic networks based on \sw model for Low Voltage Grid (detail of $<k>\approx4$ and $<k>\approx6$ samples).}
\label{fig:alphaBetaLVML}
\end{figure}

Figure~\ref{fig:alphaBetaLV} shows the values for the $\alpha$ and $\beta$ metrics for the synthetic networks generated following the \sw model with an increasing average node degree ($<k>\approx2$, $<k>\approx4$ and $<k>\approx6$). It is not surprising to see the samples with $<k>\approx2$ score poorer than the other networks. The networks with higher average node degree are better visualized in Figure~\ref{fig:alphaBetaLVML}. One sees how the network with \textit{Medium} size scores best and the difference between the network with $<k>\approx6$ and the network with $<k>\approx4$ is limited. Robustness (i.e., $\beta$ parameter) for the \textit{Medium} and \textit{Large} size networks reaches a high value just with a sufficient connectivity (i.e., $<k>\approx4$) and more connectivity (i.e., $<k>\approx6$) does not improve much this metric. The samples with \textit{Small} size score better in the $\alpha$ metric and this is quite reasonable since the paths, especially in terms of the number of substations traveled in the shortest path, are limited, of course due to the reduced \textit{order} of the network.

\begin{figure}
   \centering
   \includegraphics[width=1\textwidth]{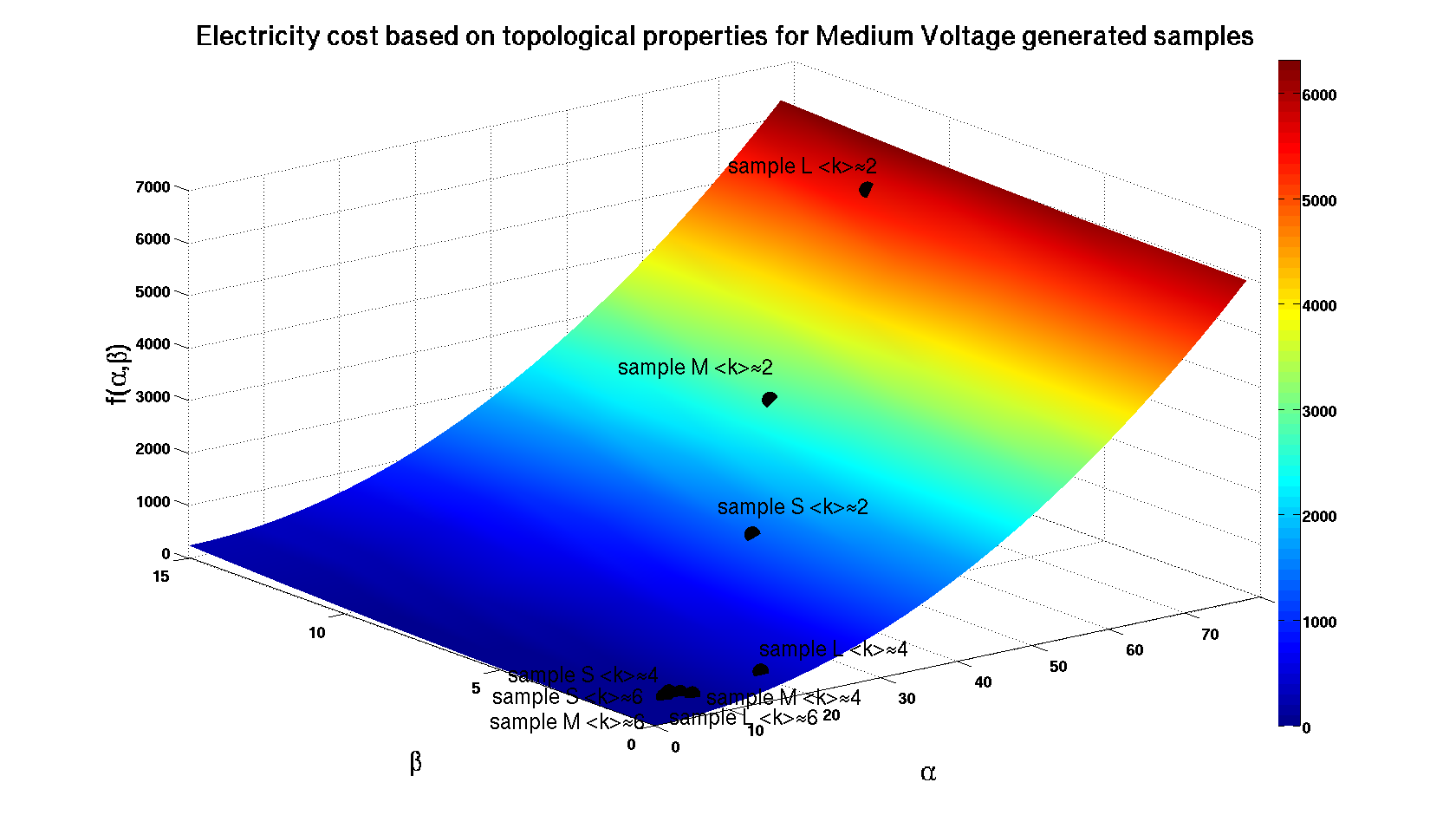}
   \caption{Transport cost of electricity based on the topological properties for synthetic networks based on \sw model for \MV Grid.}
\label{fig:alphaBetaMV}
\end{figure}

\begin{figure}
   \centering
   \includegraphics[width=1\textwidth]{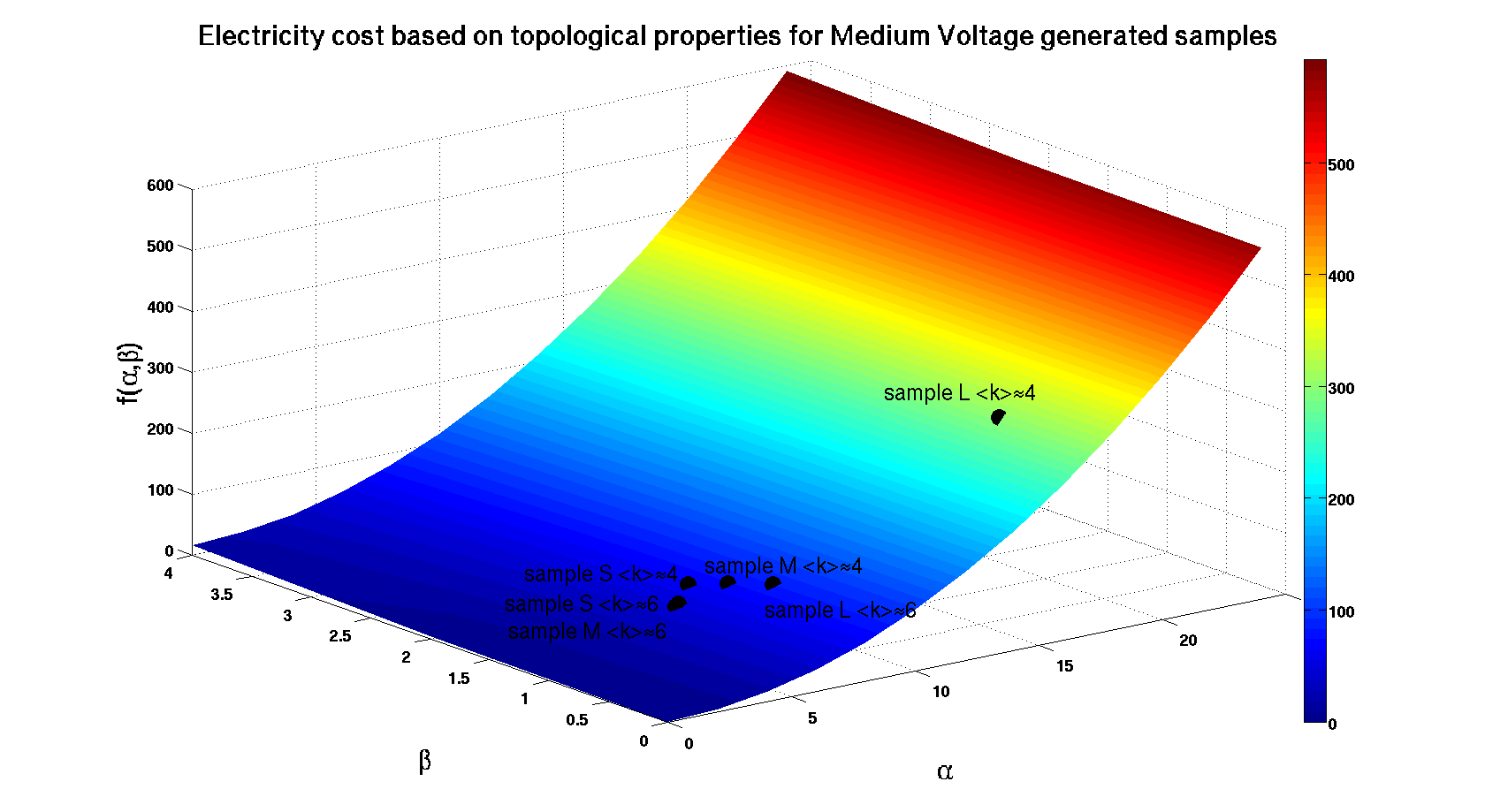}
   \caption{Transport cost of electricity based on the topological properties for synthetic networks based on \sw model for \MV Grid (detail of $<k>\approx4$ and $<k>\approx6$ samples).}
\label{fig:alphaBetaMVML}
\end{figure}

The $\alpha$ and $\beta$ metrics for the networks generated for \MV purposes are shown in Figure~\ref{fig:alphaBetaMV}. The same tendency appears: once the network is sufficiently connected (i.e., $<k>\approx4$) the metrics score definitely better than the $<k>\approx2$ situation. If we dig into the most connected samples (Figure~\ref{fig:alphaBetaMVML}), we see how the values are quite concentrated with the exception of the \textit{Large} sample with $<k>\approx4$. It is interesting to see the change in the $\alpha$ value once there are more links: the value of the metric almost halves with an increase of connectivity i.e., $<k>\approx6$ situation.

\begin{figure}
   \centering
   \includegraphics[width=1\textwidth]{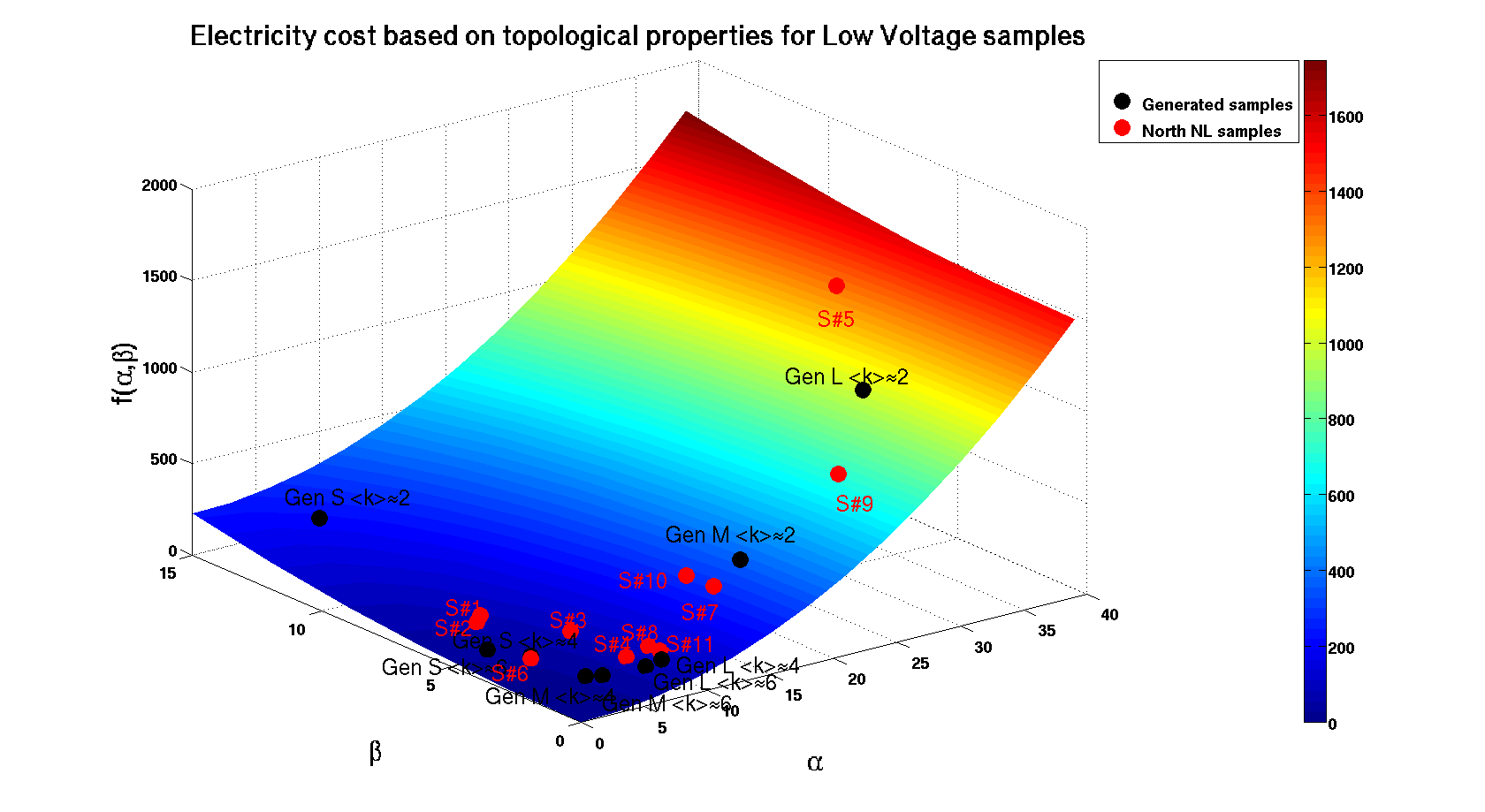}
   \caption{Comparison of the transport cost between synthetic \sw networks (black dots) and Northern Netherlands \LV samples (red dots).}
\label{fig:alphaBetaComplLV}
\end{figure}

Let us compare the $\alpha$ and $\beta$ metrics of the synthetic networks with the values of the real \PG samples of the Northern Netherlands. Considering the \LV samples and the synthetic networks designed for this purpose, we generally see an improvement in the metrics especially in the $\alpha$ values for the $<k>\approx4$ and $<k>\approx6$ networks. In fact, if we do not consider the synthetic networks with $<k>\approx2$, because of the  problems of \sw topology with such small connectivity, there is an improvement on average in the $\alpha$ metric of more than 50\% comparing the Northern Netherlands samples with the $<k>\approx4$ synthetic ones. In fact, for the $\alpha$ metric from an average of about 13 for the physical samples, the $<k>\approx4$ synthetic ones score about 6. The improvement is more than 60\% when considering the $<k>\approx6$ ones where the average for these synthetic networks scores just below 5. There are improvements also in the $\beta$ metric, although limited. From an average around 4 for the physical samples the $<k>\approx4$ on average score just below 2.75; while a better result is obtained by $<k>\approx6$ which on average score 2.30 (about 40\% improvement). The graphical comparison is shown in Figure~\ref{fig:alphaBetaComplLV}.

\begin{figure}
   \centering
   \includegraphics[width=1\textwidth]{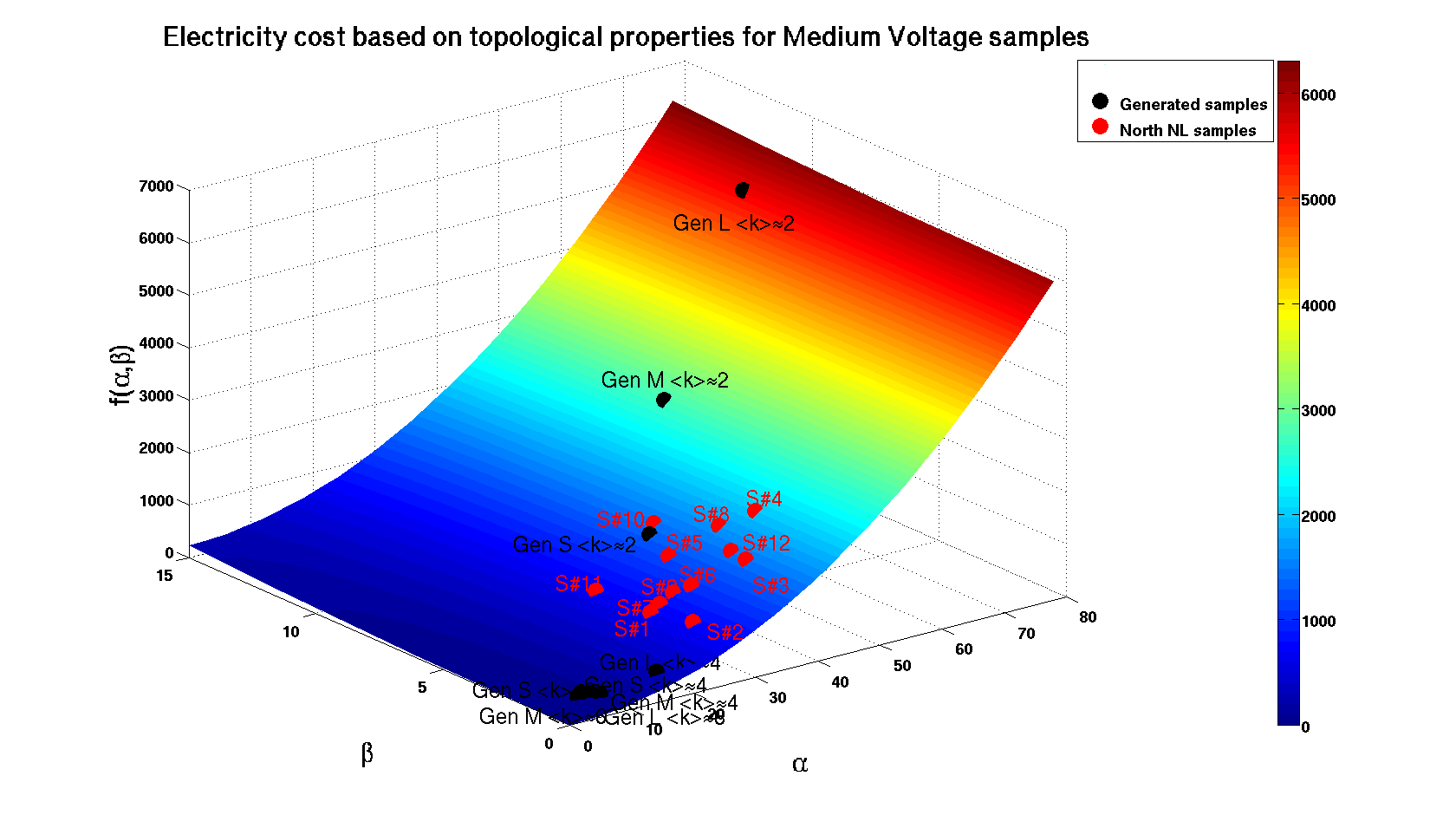}
   \caption{Comparison for transport cost between synthetic \sw networks (black dots) and Northern Netherlands \MV samples (red dots).}
\label{fig:alphaBetaComplMV}
\end{figure}

Taking into account the \MV Netherlands samples and the \sw synthetic networks for this purpose, we see an important improvement in the metrics both in the $\alpha$ and $\beta$ values for the $<k>\approx4$ and $<k>\approx6$ networks. As already mentioned, synthetic networks with $<k>\approx2$ should not be considered.  The improvement on average in $\alpha$ metric is more than 65\% comparing to the $<k>\approx4$ synthetic samples (from an average of $\alpha$ about 33 for the physical samples, the $<k>\approx4$ synthetic ones score about 11), and an improvement of more than 75\% when comparing to the $<k>\approx6$ ones (the average $\alpha$ for $<k>\approx6$ synthetic networks scores around 7.3). There are improvements also in the $\beta$ metric. In particular, from an average around 3.55 for the physical samples the $<k>\approx4$ score on average just below 1.15; a similar result is obtained by $<k>\approx6$ which on average score about 1.2 (more than 65\% improvement). The graphical comparison is shown in Figure \ref{fig:alphaBetaComplMV}.

\subsection*{Discussion}

Watts and Strogatz's \sw model, as shown in Tables~\ref{tab:desiderataSatisf}, \ref{tab:desiderataSatisfIndND} and~\ref{tab:summarytable}, is the model that captures best the requirements for the new Grid compared to the other ones analyzed being these dependent on the average node degree (preferential attachment, R-MAT and Random Graph) or not (Copying Model, Forest Fire, Kronecker and Power Laws). The tighten clustering that this models exhibits provides efficient local distribution with paths that are locally short; at the same time the shortcuts between the local clusters are the elements that keep the average path extremely limited. These two aspects influence the $\alpha$ parameter which then stays relatively limited. At the same time, the \sw model benefits from a general robustness against failures: the absence of big hubs that keep the network together (which are present on the other hand in the power-law-based topologies, for instance) improves the reliability against attacks which help obtaining good scores for the $\beta$ parameter. More quantitatively,
one sees the general improvement in the metrics characterizing both the parameters influencing the losses (i.e., $\alpha$ parameter) and the reliability of the \G (i.e., $\beta$ parameter) while the network becomes more dense, i.e., more edges are added. On average, we see an improvement of at least 50\% when comparing the physical samples of Northern Netherlands with the small-world networks with an average degree $<k>\approx4$, while better results are obtained with more density (i.e., $<k>\approx6$) where the improvement are 60\% compared to the physical samples. This is indeed beneficial to the Power Grid and, according to the relationship with the topology, it should translate into a reduction in the costs for electricity distribution. 

These benefits come literally at a cost. The network needs more connectivity therefore costs for extra cabling need to be considered in addition to the cost for upgrading the substations and end-users electricity gateways. A return on investment analysis on this aspect is beyond the scope of the present study. Nevertheless, it is interesting to see how with the $\alpha$ and $\beta$ metrics it is possible to consider how a certain physical sample belonging to a certain size category (\textit{Small}, \textit{Medium} and \textit{Large}) would improve in its performance if its topology is arranged according to the principles of a synthetic model and more connections are added accordingly.

The benefits reached for $\alpha$ and $\beta$ should translate into a reduction in the cost for electricity transport and distribution since the parameters that influence these metrics are directly connected to aspects related to costs. However the significant investment required to add more connectivity in the network might not immediately enable cheaper electricity costs, but on the contrary make it more expensive.

\section{Related Work}\label{sec:relWks}

Complex Network Analysis works take into account the \PG at the High Voltage level usually to analyze the structure of the network without considering in detail the physical properties of the power lines. In our previous work~\cite{pag:preprint1102}, we have analyzed several works that investigate \PG properties using \CNA approach. There are two main categories: 1) understand the intrinsic property of  Grid topologies and compare them to other types of networks assessing the existence of properties such as small-world or scale-free~\cite{Amaral2000,Watts03,Watts98,Barabasi1999}; 2) better understand the behavior of the network when failures occur (i.e., edge or node removal) and analyze the topological causes that bring to black-out spread and cascading failures of power lines~\cite{Corominas-murtra2008,alb:str04,Crucitti04}. Few studies in the \CNA landscape consider the possibility of using the insight gained through the analysis to help the design. These few cases consider the addition of lines in the network to assess the increase in the reliability of the entire Power Grid. Examples are the study of the Italian \HV \G~\cite{Crucitti2005} and the study of improvement by line addition in Italian, French and Spanish Grids~\cite{Rosato2007}. Also Holmgren~\cite{holmgren06} uses the \CNA to understand which \G improvement strategies are most beneficial showing the different improvement of typical \CNA metrics (e.g., path length, average degree, clustering coefficient, network connectivity) although in a very simple small graph (less than 10 nodes) when different edges and nodes are added to the network. Wider is the work of Mei~\textit{et al.}~\cite{Mei11} where a self-evolution process of the \HV \G is studied with \CNA methodologies. The model for \PG expansion considers an evolution of the network where power plants and substations are connected in a ``local-world'' topology through new transmission lines; overall the \PG reaches in its evolution  the \sw topology after few-steps of the expansion  process. Wang~\textit{et al.}~\cite{Wang2010,wang08} study the \PG to understand the kind of communication system needed to support the decentralized control required by the the new \PG applying \CNA techniques. The analyses aim at generating samples using random topologies based on uniform and Poisson probability distributions and a random topology with small-world network features. The simulation results are compared to the real samples of U.S. \PG and synthetic reference models belonging to the IEEE literature. These works also investigate the property of the physical impedance to assign to the generated Grid samples.
\CNA is not generally used as a design tool to propose new topologies for the future \SG as we use in this paper where we also assess the benefits in terms of economical improvement. 

Traditionally power system engineers adopt techniques which are different from Complex Network Analysis although sometimes exploiting graph theory principles~\cite{wall79,crawford75}. The traditional techniques applied by Power Engineers involve the individuation of an objective function representing the cost of the power flow along a certain line which is then subject to physical and energy balance. This problem translates in an operation research problem. These models are applied both for the \HV planning~\cite{garver70,lee74} and the \MLV~\cite{wall79,crawford75} since long time. Not only operation research, but also expert systems~\cite{machias89} are developed to help in the process of designing grounding stations based on physical requirements as well as heuristic approaches from engineering experience. 
The substation grounding issue is approached as an optimization problem of construction and conductor costs subject to the constraints of technical and safety parameters, its solution is investigated through a random walk search algorithm~\cite{gilbert11}. In~\cite{fratkin96} a pragmatic approach using sensitivity analysis is applied to a linear model of load flow related to various overloading situations and a contingency analysis (N-1 and N-2 redundancy conditions) is performed with different grades of uncertainty in medium and long term scenarios. In the practice the planning and expansion problem is even more complex since it implies power plants, transmission lines, substations and Distribution Grid. In~\cite{grigsby07} all these aspects are assessed separately and several challenges appear. For instance in the planning of a \HV overhead transmission line specific clearance code must be followed and not only load is a key element, but also topography and weather/climatic (above all wind and ice) conditions play an import role in the planning of the infrastructure. For substation planning the authors of~\cite{grigsby07} emphasize, in addition to the need for upgrading the Grid (e.g., load growth, system stability) and budgeting aspects, the multidisciplinary aspects which involve from environmental and civil to electrical and communications engineering. 
A more general approach proposed in~\cite{grigsby07} to deal with power system planning might be regarded as a multi-objective (e.g., economics, environment, feasibility, safety) decision problem thus requiring the tools typical of decision analysis~\cite{keeney82}.

The works mentioned so far take into account mainly the \HV end of the \G while not least important is the Distribution \G especially in the vision of the future electrical system as proposed in this work where the end-user plays a vital role. The integrated planning of \MLV networks is tackled by Paiva \etal~\cite{paiva05} who emphasize the need of considering the two networks together to obtain a sensible optimal planning. The problem is modeled as a mixed integer-linear programming one considering an objective function for investment, maintenance, operation and losses costs that need to be minimized satisfying the constraints of energy balance and equipment physical limits.

Even more challenges to Electrical system planning is posed by the change in the energy landscape with several companies running different aspects of the business (generation, transmission, distribution). In addition, accommodating more players in the wholesale market transmission expansion should follow (as it is already for generation) a market based approach i.e., the demand forces of the market and its forecast should trigger the expansion of the Grid~\cite{bresesti03}. The same consideration regarding the need of a different approach in planning in a deregulated market are expressed in~\cite{shariati08} where an optimization of an objective function in the market environment is applied. 
Another method to evaluate transmission expansion plan takes into account the probability reliability criteria of Loss Of Load Expectation (LOLE); in particular, in~\cite{choi05} an objective function is proposed that takes into account the cost in constructing a transmission line between all buses involved in the line which is then subject to constrains in peak load demand satisfaction and a certain level of LOLE that the line should not outrun.

In the \SG framework the planning techniques might be revised especially for the Distribution Grid which is the segment that is likely to face the greatest changes due to the presence of Advanced Metering Infrastructure (i.e., bidirectional intelligent digital meters at customer location) and Distribution Automation (i.e., feeders can be monitored, controlled in automated way through two-way communication). In addition, the \MLV \G is no longer a layer where only energy is consumed, but Distributed Energy Generation facilities (small-scale photovoltaic systems and small-wind turbines) will be attached to this segment of the Grid; altogether these elements are likely to reshape the way planning for \MLV is realized~\cite{brown08}.



\section{Conclusions}\label{sec:conclusion}

In an evolving electricity sector with end-users able to produce their own energy and sell it on a local-scale market, the \G plays the essential enabling role of supporting infrastructure. Local scale energy exchange is in fact beneficial for several aspects such as the increase in renewable-based energy production, the possibility for the end-user to have an economic contentment by selling surplus energy and, not less important, a step forward to the unbundling of the electricity sector. We studied how different topologies inspired from technological and social network studies have varying properties and can be (or not) adequate for the future \SG networks. We showed that between the various models analyzed, the \sw model appears to have many supporting characteristics, according to a set of topological metrics defined for Power Grids. We  also showed how these topological benefits can be related to economical aspects of electricity distribution through an improvement in the $\alpha$ and $\beta$ parameters. We also performed a statistical investigation related to cables' properties used in \MLV samples to evaluate the cost of cables to be used to realize synthetic networks in order to estimate the investment required for such networks. The benefits reached through topological properties are significant and beneficial to enable a local energy exchange, however the quantification from and economic point of view is not easy due to the high investment in realizing a more connected \MLV Grid. 

The underlying motivation for the present work, is to develop decision support techniques based on \CNA metrics to upgrade the \PG to a \SG and to assess the current infrastructures. In addition, it enables to predict how a change in the topology, according to a certain network model, can be beneficial for the network from an efficiency, resilience and robustness perspective. Finally, the approach enables to quantify how the topology can help in reducing the parameters influencing electricity costs while considering the evolution of the \MLV \PG network into an infrastructure to support Smart Grid.

\newpage

\appendix
\renewcommand\thesection{Appendix \Alph{section}}

\section{Price and Resistance Distribution}~\label{sec:appA}

In Section~\ref{subsec:priceStat} we illustrate
how physical properties of cables and their prices have a correlation. A proper analysis must then follow a bivariate approach. However, one might be interested in studying only one characteristic of cables (e.g., price) separated from others (e.g., resistance). Here we investigate what is the statistical distribution of these separate properties of cables. In particular we look for the presence of power-laws since they appear in several natural phenomena and man-created infrastructures~\cite{Clauset07}.

Fitting the data regarding prices to the most likely distribution obtained from the three different sample sizes of the Dutch \LV \G  gives usually distributions very concentrated towards small price values and only very few cables have very high prices (i.e., more than 10000 euros) which are particularly long, or their technology is extremely expensive. The distribution that best fits the data for the \LV samples is the Log-Normal distribution ($y=f_X(x;\mu, \sigma)= \frac{1}{x\sigma \sqrt{2\pi}}e^{\frac{-(ln x-\mu)^2}{2\sigma^2}}$). Comparing the fitted distribution with the original empirical cumulative distribution function of the data provides significant p-values with the Kolmogorov-Smirnov test. The analytical parameters for the fitted distribution obtained through log-likelihood estimation are shown in Table~\ref{tab:logNormParam}.
\begin{table}[h]
\centering
\begin{tabular}{|l|p{2.5cm}|p{2.5cm}|}
\hline
Sample type & \multicolumn{2}{|c|}{Log-Normal distribution parameters}\\ \hline
  & $\mu$ & $\sigma$ \\ \hline
\LV- Small & \rl 6.104	&\rl 1.513\tn \hline
\LV- Medium  & \rl 6.075	&\rl 1.397\tn \hline
\LV- Large  & \rl 6.939	&\rl 1.258\tn \hline
\end{tabular}
\caption{Log-Normal distribution $\mu$ and $\sigma$ parameters for cable price distribution for \LV samples.}\label{tab:logNormParam}
\end{table}

Information related to prices for \MV cables are only partially available for this study and limited to some technologies and cross-sections of aluminum and copper cables. In order to have an estimate of costs, we fit the prices available to the best interpolating curve. For the aluminum cables we used a cubic polynomial, while for the copper ones a linear relation between price and cross-section. We performed the same probability distribution fitting procedure with the \MV samples. Also in this case, the distributions that best approximate the sample data show a ``fat-tail'' behavior. For the three representative classes of samples, we consider that the best approximation is given by the theoretical distribution of generalized extreme value ($y=f_X(x;k,\mu,\sigma)=\frac{1}{\sigma}(1+k\frac{x-\mu}{\sigma})^{-1-\frac{1}{k}}$exp$\{-(1+k\frac{x-\mu}{\sigma})^{-\frac{1}{k}}\}$). The sample distributions have significant p-values with the Kolmogorov-Smirnov test indicating to accept the hypothesis of this underlined probability law for the \textit{Small} and \textit{Medium} classes of samples. The \textit{Large} class sample poses more problems since the p-value resulting form the test is under the 5\% acceptance threshold. Although the test suggests to reject the hypothesis of the underlying distribution, we consider it anyway a good distribution approximation since this type of distribution is the one that has a p-value closer to significance compared to others distribution tested (e.g., log-normal, exponential, Gaussian). 
The analytical parameters for the fitted distribution obtained through log-likelihood estimation are shown in Table~\ref{tab:extremeVal}.  
\begin{table}
\centering
 \begin{tabular}{|l|p{1.5cm}|p{1.5cm}|p{2cm}|}
  \hline
  Sample type & \multicolumn{3}{|c|}{Extreme value distribution parameters}\\ \hline
  & $k$ &$\sigma$ & $\mu$ \\ \hline
\MV- Small & \rl 0.547& \rl 33082.4&\rl 31988.8\tn \hline
\MV- Medium  &\rl 0.419&\rl 32569.4&\rl 35880.8\tn \hline
\MV- Large  &\rl  0.490	&\rl 16925.2&\rl 16766.9\tn \hline
\end{tabular}
\caption{Extreme values distribution $k$, $\mu$ and $\sigma$ parameters to fit cable price distribution for \MV samples.}\label{tab:extremeVal}
\end{table}


Similar statistical considerations can be applied for fitting the resistance characterizing the cables. The obtained distributions both for \LV and \MV reference networks present once again a ``fat-tail'' characteristic since, although the most of the cables have a small resistance properties, there are some cables with far higher resistance properties. The distributions that best fit the data are either generalized extreme values ($y=f_X(x;k,\mu,\sigma)=\frac{1}{\sigma}(1+k\frac{x-\mu}{\sigma})^{-1-\frac{1}{k}}$exp$\{-(1+k\frac{x-\mu}{\sigma})^{-\frac{1}{k}}\}$) or log-normal ($y=f_X(x;\mu, \sigma)= \frac{1}{x\sigma \sqrt{2\pi}}e^{\frac{-(ln x-\mu)^2}{2\sigma^2}}$). The parameters are shown in Tables~\ref{tab:resiLVfit} and~\ref{tab:resiMVfit} for \LV and \MV respectively.

\begin{table}
\centering
\begin{tabular}{|l|l|p{1.5cm}|p{1.5cm}|p{1.5cm}|}
\hline
Sample type&Distribution type & \multicolumn{3}{|c|}{Distribution parameters}\\ \hline
  & &$k$ &$\mu$ & $\sigma$ \\ \hline
\LV- Small & Log-normal& & \rl-2.27846 & \rl 1.97188\tn \hline
\LV- Medium & Generalized & \rl 0.994657&\rl 0.054877&\rl 0.058296\tn 
&extreme values&&&\\ \hline
\LV- Large & Log-normal& &\rl -0.881168& \rl 1.25617\tn \hline
\end{tabular}
\caption{Distribution parameters to fit cable resistance for \LV samples.}\label{tab:resiLVfit}
\end{table}

\begin{table}
\centering
\begin{tabular}{|l|p{1.5cm}|p{1.5cm}|p{2cm}|}
\hline
Sample type & \multicolumn{3}{|c|}{Extreme value distribution parameters}\\ \hline
  & $k$ &$\sigma$ & $\mu$ \\ \hline
\MV- Small & \rl 1.09862 & \rl 4.1366& \rl 2.96819\tn \hline
\MV- Medium  & \rl 0.613803& \rl 3.35594& \rl 3.41663\tn \hline
\MV- Large  & \rl 0.619069& \rl 3.59693& \rl 3.35337\tn \hline
\end{tabular}
\caption{Extreme values distribution $k$, $\mu$ and $\sigma$ parameters to fit cable resistance for \MV samples.}\label{tab:resiMVfit}
\end{table}

\begin{figure}
   \centering
   \includegraphics[width=0.9\textwidth]{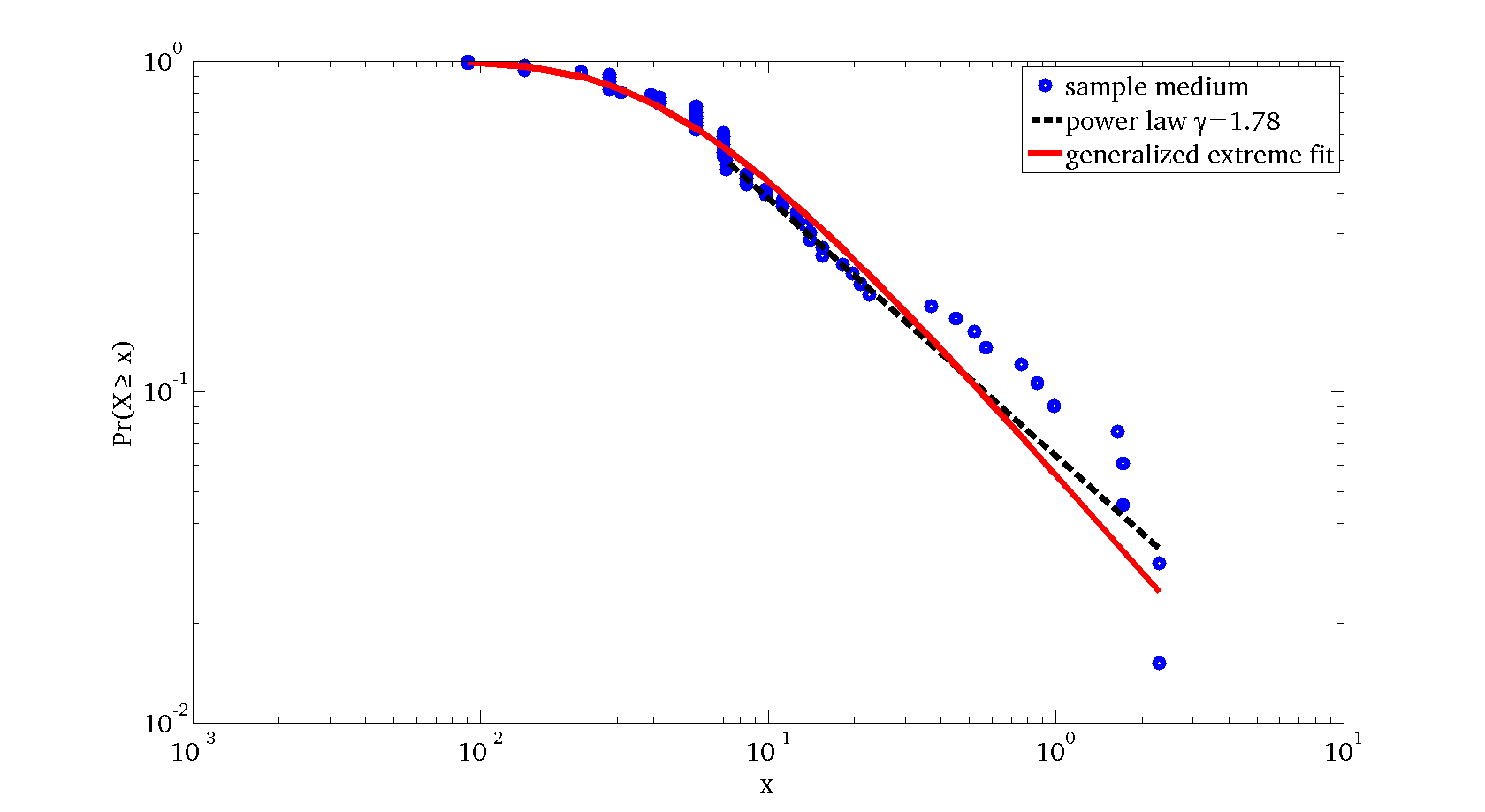}
   \caption{Cumulative probability distribution (complementary) for cable resistance M-size sample \LV network (double logarithmic scale).}
\label{fig:resistanceLogLogMedLV}
\end{figure}

\begin{figure}
   \centering
   \includegraphics[width=0.9\textwidth]{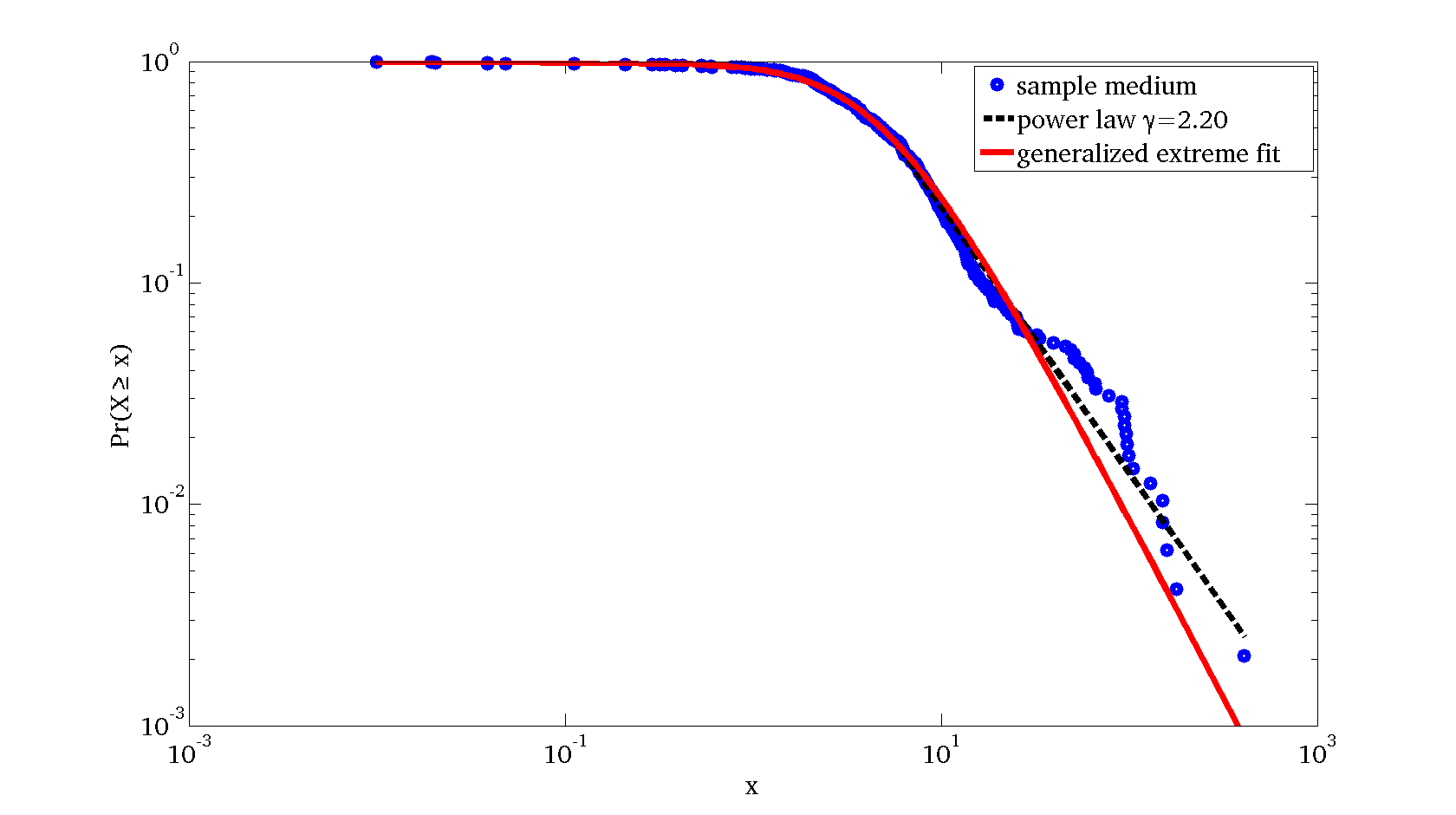}
   \caption{Cumulative probability distribution (complementary) for cable resistance M-size sample \MV network (double logarithmic scale).}
\label{fig:resistanceLogLogMedMV}
\end{figure}

\begin{figure}
   \centering
   \includegraphics[width=0.9\textwidth]{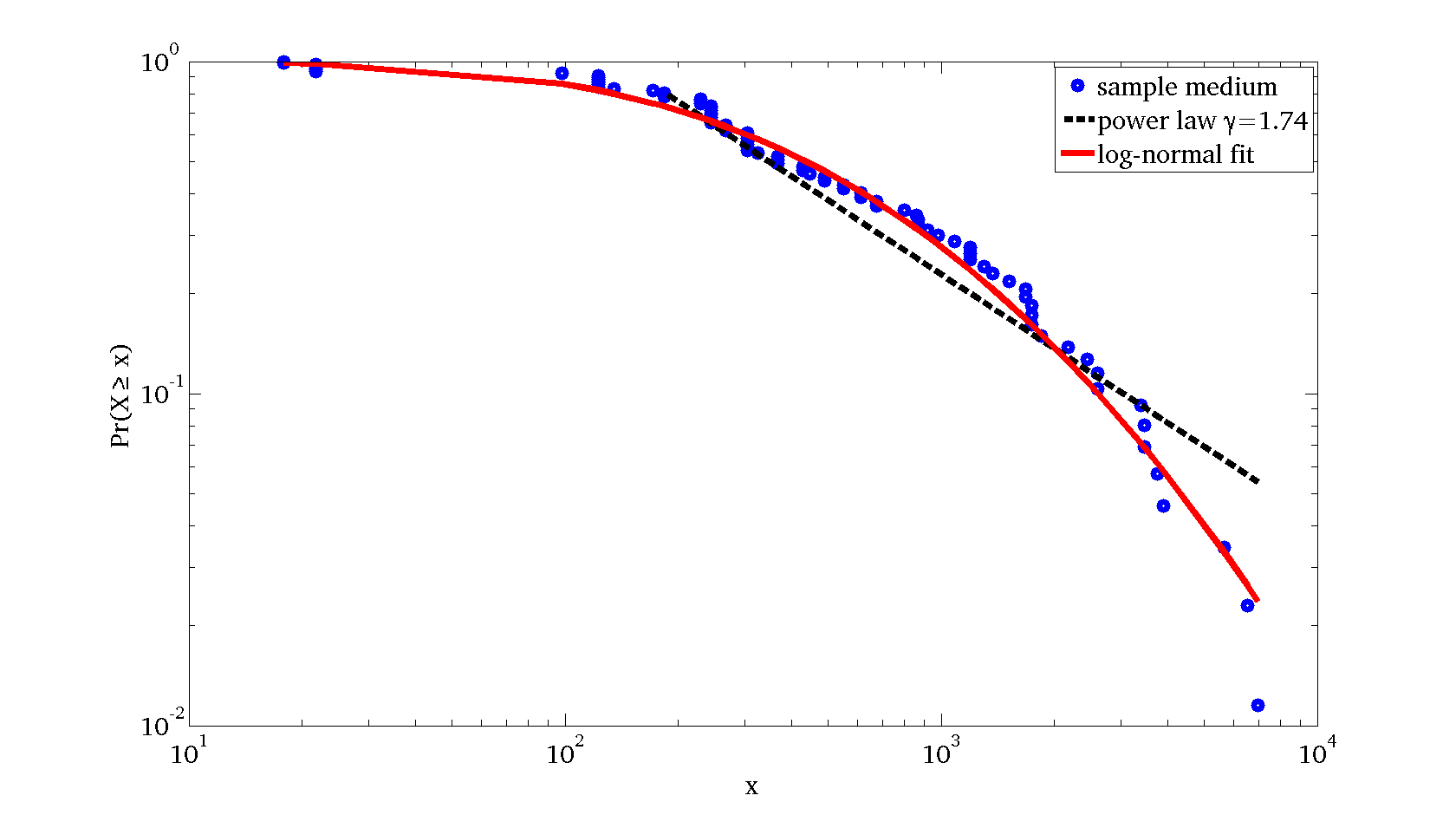}
   \caption{Cumulative probability distribution (complementary) for cable price M-size sample \LV network (double logarithmic scale).}
\label{fig:priceLogLogMedLV}
\end{figure}

\begin{figure}
   \centering
   \includegraphics[width=0.9\textwidth]{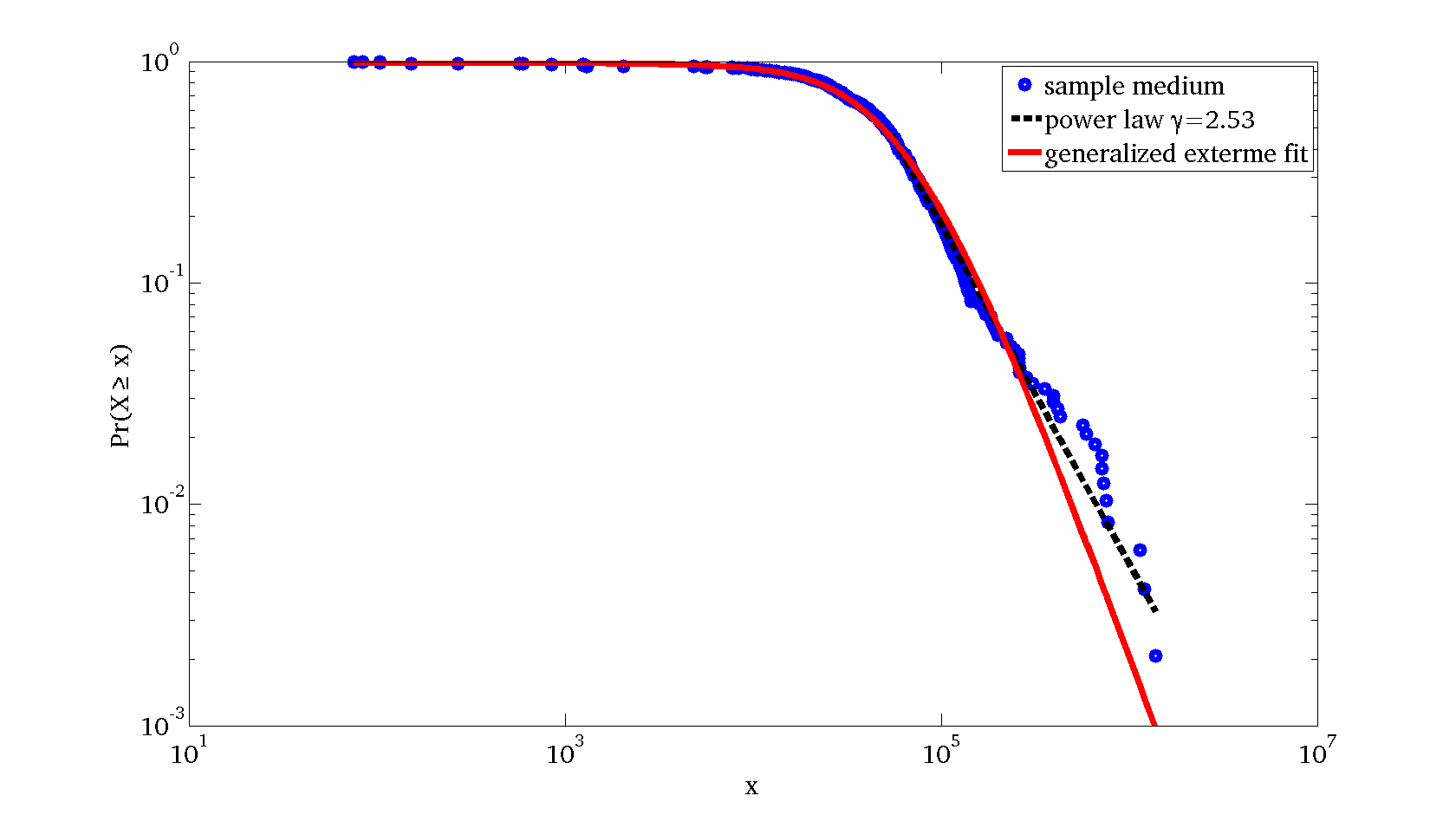}
   \caption{Cumulative probability distribution (complementary) for cable price M-size sample \MV network (double logarithmic scale).}
\label{fig:priceLogLogMedMV}
\end{figure}

Both price and resistance distributions present usually a high probability that is concentrated in the lower values, however there are overall small, but highly significant in terms of their values, contributions in the tail of the distribution. We perform also an investigation considering if this ``heavy-tail'' contribution have power-law properties. We apply the fitting techniques proposed by Clauset \etal~\cite{Clauset07} to understand the presence of significant \pl contributions in these distributions. From this analysis it appears that there are marks of \pl distribution in both the probability of cable prices and cable resistance. These \pl contributions are generally significant in the middle part of the distribution, while the very initial part of the distribution and the final part of the tail tend to deviate from the \pl rule. In fact, the p-value that characterizes the Kolmogorov-Smirnov test is generally higher than the 5\% null hypothesis rejection for the \pl hypothesis in the central part of the distribution. Two examples for the \LV and \MV samples are given in Figures~\ref{fig:resistanceLogLogMedLV} and~\ref{fig:resistanceLogLogMedMV} related to cable resistance, while Figures~\ref{fig:priceLogLogMedLV} and~\ref{fig:priceLogLogMedMV} are related to cable prices. Each figure represents the cumulative probability distribution (complementary) on double logarithmic scale where the blue circles represent the samples data, the red line is the best fitting probability distribution over the whole sample (described above) and the black dashed line represents the best fitting \pl distribution in the interval of the sample closer to power-law. 

\section{Relating Topological Properties to Economical Distribution Benefits}\label{sec:AppAlfaBeta}

In Section~\ref{sec:discussion} we introduce the concept to associate Grid topology and cost of electricity. Here we give a thorough explanation of these concepts based on the findings in our previous work~\cite{PaganiAielloTSG2011}, where we developed a set of metrics to relate topological aspects and electricity cost and applied it to existing Dutch Medium and Low Voltage infrastructure. 
As described in Section~\ref{sec:discussion}, we take advantage of that proposal and apply the same metrics to the generated topologies suitable for the Smart Grid.  The goal is to consider from a topological perspective those measures that are critical in contributing to the cost of electricity as elements in the Transmission and Distribution Networks as described in economic studies such as the one of Harris and Munasinghe~\cite{Harris06,Munasinghe84}:
\begin{itemize}
\item losses both in line and at transformer stations,
\item security and capacity factors,
\item line redundancy, and 
\item power transfer limits.
\end{itemize}

The topological aspects that we consider provide two sorts of measures, the first
one $\alpha$ gives an average of the dissipation in the transmission between
two nodes
\begin{align} 
\alpha & = f(L_{line_N},L_{substation_N})\label{eq:alpha};
 \end{align} the second one $\beta$ is a measure of reliability/redundancy in
the paths among any two nodes
\begin{align}
 \beta & = f( Rob_{N},Red_{N},Cap_{N})\label{eq:beta}.
 \end{align}
The functions to explicitly compute $\alpha$ and $\beta$ parameters can be expressed as follows:
\begin{itemize}
\item Losses on the transmission/distribution line can be expressed by the quotient of the weighted
  characteristic path length and the average weight of a line (a weighted edge in the graph):
\begin{align}L_{line_N} = \frac{WCPL_N}{\overline{w}} \label{eq:lineLoss}\end{align}
\item Losses at substation level are expressed as the number of nodes (on average) that are traversed when computing the weighted shortest path between all the nodes in the network:
\begin{align}L_{substation_N} = \overline{Nodes_{WCPL_N}} \label{eq:stationLoss}\end{align}
\item Robustness is evaluated with random removal strategy and the weighted-node-degree-based removal by computing the average of the order of maximal connected component between the two situations when the 20\% of the nodes of the original graph are removed. It can be written as:
  \begin{align}Rob_N = \frac{|MCC_{Random20\%}|+|MCC_{NodeDegree20\%}|}{2} \label{eq:robust}\end{align} 
\item Redundancy is evaluated by covering a random sample of the nodes
  in the network (40\% of the nodes whose half represents source nodes
  and the other half represents destination nodes) and computing for
  each source and destination pair the first ten shortest paths of
  increasing length. If there are less than ten paths available, the
  worst case path between the two nodes is considered. To have a
  measure of how these resilient paths have an increment in
  transportation cost, a normalization with the weighted
  characteristic path length is performed. We formalized it as:
\begin{align}Red_{N} = \frac{\sum_{i \in Sources, j \in Sinks}{SP_{w_{ij}}}}{WCPL} \label{eq:redund}\end{align}
\item Network capacity is considered as the value of the weighted characteristic path length, whose weights are the maximal operating current supported, normalized by the average weight of the edges in the network (average current supported by a line). That is:
\begin{align}Cap_{N} = \frac{WCPL_{current N}}{\overline{w_{current}}}\end{align}
\end{itemize}
With these instantiations, equations (\ref{eq:alpha}) and (\ref{eq:beta}) become:
\begin{align}
\alpha & = f(L_{line_N},L_{substation_N}) = L_{line_N}+L_{substation_N} \label{eq:alphaEx}\\
\beta & = f(Red_{N},Rob_{N},Cap_{N}) = \frac{Red_N}{Rob_N \cdot ln(Cap_N)}\label{eq:betaEx}
\end{align}

The aspects here considered are just some of the factors (the ones closely coupled to topology) that influence the overall price of electricity.  Naturally, there are other factors that influence the final price, e.g., fuel prices, government policies and taxation, etc., as illustrated for instance in the economic studies of Harris and Munasinghe~\cite{Harris06,Munasinghe84}. 

\section{The Grid Engineering Process based on  Complex Network Analysis}\label{sec:AppEngineering}

In our previous analysis work~\cite{PaganiAielloTSG2011} we considered a topological analysis of the Dutch \MLV Power Grid, while in this work we generate synthetic networks to assess which ones are better to support a \SG where prosumers exchange energy at local scale. Based on both these studies we can define an engineering process to upgrade the existing infrastructures towards a \SG of prosumers. The engineering process is thus based on \CNA metrics and techniques. 

\begin{figure}[htbp]
   \centering   
  \includegraphics[width=1\textwidth]{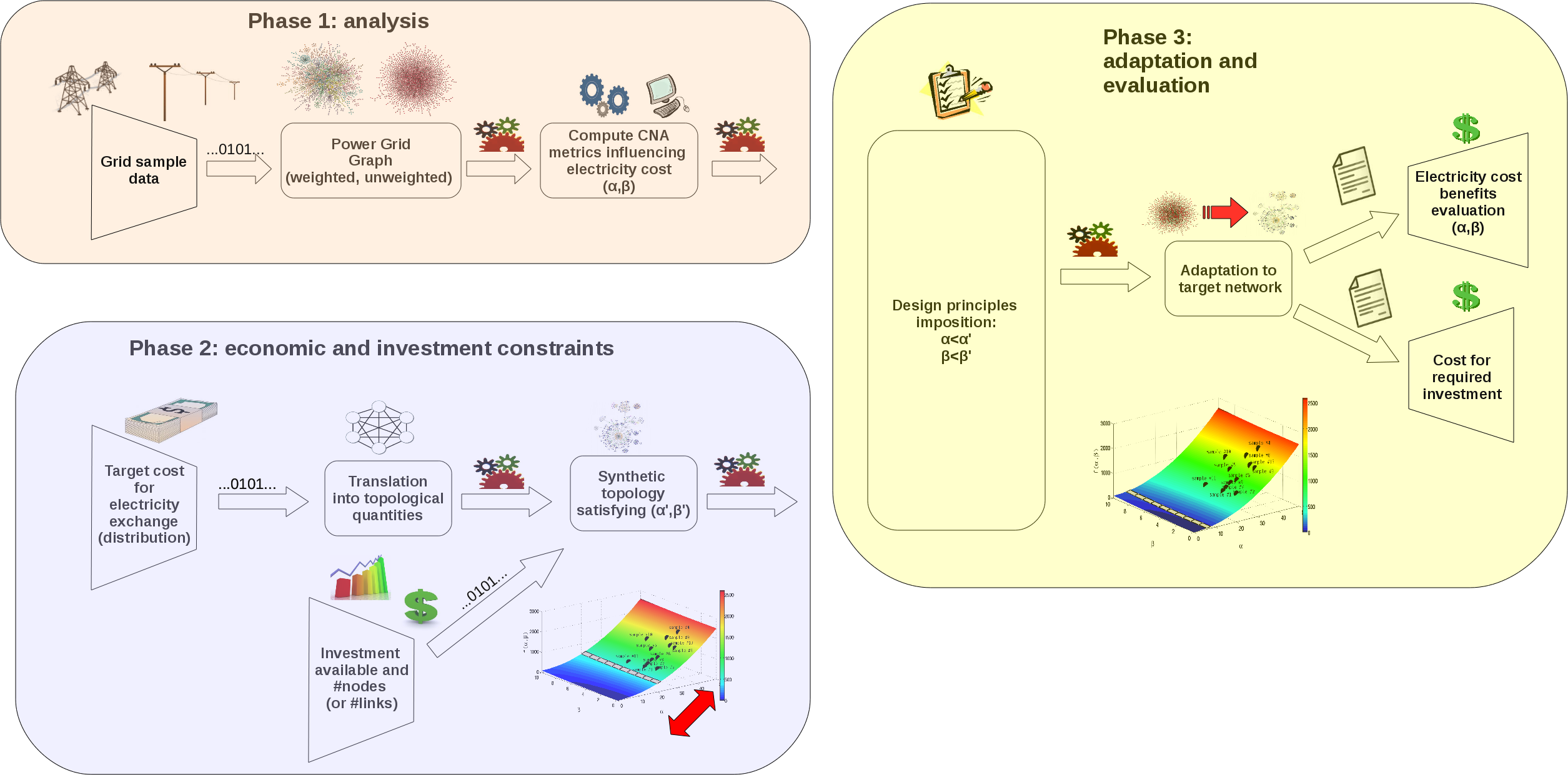}
   \caption{Engineering process for \MLV \G optimized for prosumer-based energy exchange.}
\label{fig:engProc}
\end{figure}
 
This process is intended for energy distribution companies to assess what is the current state of their infrastructures 
considering the influence of the topology on the electricity transport prices. In a totally unbundled scenario for the electricity market the distribution company might be incentivized in providing a better infrastructure closer to prosumer and consumer needs. Distribution company might charge them based on indicators that not only take into account downtime periods, but also topological efficiency which is based on the influence of topology on electricity prices. Figure~\ref{fig:engProc} presents the flow of this process. Each big rectangular box represents a phase of this process and each contains a number of operation in a flow represented by small rectangles. The initial phase is basically an analysis of the existing infrastructure and computation of its topological properties (Phase 1 light orange box) extracted from Grid sample data input (trapezoidal block in the figure). Economic factors (Phase 2 violet box) play a role too which are accounted in considering the desired costs in electricity distribution and the investment available by the electricity distribution company (trapezoidal blocks in Phase 2 box). The match to the actual infrastructure to the desired one is realized (Phase 3 light yellow box) and reports for cost benefits and cost for the investment are provided (trapezoidal blocks in Phase 3 box). In particular, each phase has more articulated processes internally which are detailed below.

The {\bf first phase} of the process starts with the acquisition of the (complete) network topological information, that is, information about the nodes of the Grid (substations, transformers, end-users) and the lines connecting those nodes (cables and links).  Also physical parameters characterizing the cables are necessary such as resistance per unit of length, length of cables and capacity of cables (supported current). Once the information is available it is possible to build a \PG graph. The following step is performing a \CNA extracting the metrics that are essential to assess the influence of topology in electricity prices as shown in~\ref{sec:AppAlfaBeta} (e.g., weighted characteristic path length, average number of nodes involved in the shortest path...). 
From these metrics the summary indicators that relate electricity prices to topology are computed, namely, the $\alpha$ and $\beta$ values are extracted as shown in Section~\ref{sec:discussion} and~\ref{sec:AppAlfaBeta}.

The {\bf second phase} represents the input of the requirements for the evolution of the Grid. These translate into constrains for generation of a network topology which satisfies the desiderata parameters of the electricity distributor provider, or any other actor (e.g., the municipality, a cooperative of users in the neighborhood, a venture capitalist), interested in realizing a Distribution Network that is more prone to the small-scale energy exchange paradigm. The stakeholder in the network defines a cost for the electricity distribution (most likely a range in the cost). This target cost is translated into topological measures ($\alpha$ and $\beta$ parameters) that the target network should satisfy. In addition to the constraint regarding $\alpha$ and $\beta$ parameters, the stakeholder provides additional constraints such as the number of nodes (or transmission lines) the network without improvements should have (it could be the same as the original sample or be different in case of planned increase/decrease in the network assets) and the available budget. The budget quantifies the investment in realizing/upgrading the network to make it more prone to prosumer-based energy exchange (this influences the possibility of increasing the number of substations and power lines).

The {\bf third phase} consists of adapting the physical sample network to the synthetic one, once the two sets of topological measures coming from phase~1 and~2 have been compared. This phase therefore provides a new network that is optimized for the local-scale energy exchange considering the constrains given in Phase 2. Once the network is available it is then possible to compare it with the physical \PG sample in order to evaluate the presumed benefits in terms of topology and its advantages in electricity distribution costs. On the other hand it is possible to evaluate the foreseen cost for the investment to achieve this kind of network.  

We leave as future work to define in details all the steps of the engineering process. Though most of pieces of the puzzles are there: Phase one is covered by~\cite{PaganiAielloTSG2011}; while the current work covers Phase~2 of the picture by offering ways to compare synthetic and physical samples in terms of $\alpha$ and $\beta$. In addition, an assessment of the costs required to build synthetic networks is also given as an aspect of evaluation in the current work.

\bibliographystyle{siam}
\bibliography{andrea.bib}

\begin{thebibliography}{10}

\bibitem{Aiello2000}
{\sc W.~Aiello, F.~Chung, and L.~Lu}, {\em {A random graph model for massive
  graphs}}, Proceedings of the thirty-second annual ACM symposium on Theory of
  computing - STOC '00,  (2000), pp.~171--180.

\bibitem{alb:str04}
{\sc R.~Albert, I.~Albert, and G.~Nakarado}, {\em Structural vulnerability of
  the north american power grid}, Physical Review E, 69 (2004).

\bibitem{albert99}
{\sc R.~Albert, H.~Jeong, and A.-L. Barab\'{a}si}, {\em {Internet: Diameter of
  the World-Wide Web}}, Nature, 401 (1999), pp.~130--131.

\bibitem{Albert2000}
{\sc R.~Albert, H.~Jeong, and A.~L. Barab\'{a}si}, {\em {Error and attack
  tolerance of complex networks}}, Nature, 406 (2000), pp.~378--382.

\bibitem{Amaral2000}
{\sc L.~A.~N. Amaral, A.~Scala, M.~Barth\'{e}l\'{e}my, and H.~E. Stanley}, {\em
  Classes of small-world networks}, Proceedings of the National Academy of
  Sciences of the United States of America, 97 (2000), pp.~11149--11152.

\bibitem{bar:lin03}
{\sc A.~L. Barab\'{a}si}, {\em Linked: The new science of networks}, American
  Journal of Physics, 71 (2004), pp.~409--410.

\bibitem{Barabasi2009}
{\sc A.~L. Barab\'{a}si}, {\em {Scale-free networks: a decade and beyond.}},
  Science, 325 (2009), pp.~412--3.

\bibitem{Barabasi1999}
{\sc A.~L. Barab\'{a}si and R.~Albert}, {\em {Emergence of scaling in random
  networks}}, Science, 286 (1999), p.~509.

\bibitem{Barabasi2000}
{\sc A.~L. Barab\'{a}si, R.~Albert, and H.~Jeong}, {\em {Scale-free
  characteristics of random networks: the topology of the {World Wide Web}}},
  Physica A: Statistical Mechanics and its Applications, 281 (2000),
  pp.~69--77.

\bibitem{belagari75}
{\sc F.~Beglari and M.~Laughton}, {\em The combined costs method for optimal
  economic planning of an electrical power system}, Power Apparatus and
  Systems, IEEE Transactions on, 94 (1975), pp.~1935 -- 1942.

\bibitem{Bollobas79}
{\sc B.~Bollobas}, {\em Graph theory : an introductory course / Bela
  Bollobas.}, Springer Verlag,, New York, 1979.
\newblock Includes index.

\bibitem{Bollobas1998}
{\sc B.~Bollobas}, {\em Modern Graph Theory}, Springer, July 1998.

\bibitem{Nationalbank2003}
{\sc M.~Boss, H.~Elsinger, M.~Summer, and S.~Thurner}, {\em {The network
  topology of the interbank market}}, Quantitative Finance, 4 (2004),
  pp.~677--684.

\bibitem{Brandes2001}
{\sc U.~Brandes}, {\em {A faster algorithm for betweenness centrality•}},
  Journal of Mathematical Sociology, 25 (2001), pp.~163--177.

\bibitem{bresesti03}
{\sc P.~Bresesti, M.~Gallanti, and D.~Lucarella}, {\em Market-based generation
  and transmission expansions in the competitive market}, in Power Engineering
  Society General Meeting, 2003, IEEE, vol.~1, july 2003, p.~4 vol. 2666.

\bibitem{brown09}
{\sc H.~E. Brown and S.~Suryanarayanan}, {\em A survey seeking a definition of
  a smart distribution system}, in North American Power Symposium (NAPS), 2009,
  oct. 2009, pp.~1 --7.

\bibitem{brown08}
{\sc R.~Brown}, {\em Impact of smart grid on distribution system design}, in
  Power and Energy Society General Meeting - Conversion and Delivery of
  Electrical Energy in the 21st Century, 2008 IEEE, july 2008, pp.~1 --4.

\bibitem{Bu11}
{\sc S.~Bu, F.~Yu, P.~Liu, and P.~Zhang}, {\em Distributed scheduling in smart
  grid communications with dynamic power demands and intermittent renewable
  energy resources}, in Communications Workshops (ICC), 2011 IEEE International
  Conference on, june 2011, pp.~1 --5.

\bibitem{cecati11}
{\sc C.~Cecati, C.~Citro, and P.~Siano}, {\em Combined operations of renewable
  energy systems and responsive demand in a smart grid}, Sustainable Energy,
  IEEE Transactions on, PP (2011), p.~1.

\bibitem{Chakrabarti2006}
{\sc D.~Chakrabarti and C.~Faloutsos}, {\em {Graph mining: Laws, generators,
  and algorithms}}, ACM Computing Surveys (CSUR), 38 (2006), p.~2.

\bibitem{Chakrabarti2004}
{\sc D.~Chakrabarti, Y.~Zhan, and C.~Faloutsos}, {\em {R-MAT: A Recursive Model
  for Graph Mining}}, in Fourth SIAM International Conference on Data Mining,
  Apr. 2004.

\bibitem{Chassin05}
{\sc D.~P. {Chassin} and C.~{Posse}}, {\em {Evaluating North American electric
  grid reliability using the Barab{\'a}si Albert network model}}, Physica A
  Statistical Mechanics and its Applications, 355 (2005), pp.~667--677.

\bibitem{choi05}
{\sc J.~Choi, T.~Tran, A.~El-Keib, R.~Thomas, H.~Oh, and R.~Billinton}, {\em A
  method for transmission system expansion planning considering probabilistic
  reliability criteria}, Power Systems, IEEE Transactions on, 20 (2005),
  pp.~1606 -- 1615.

\bibitem{Clauset07}
{\sc A.~Clauset, C.~R. Shalizi, and M.~E.~J. Newman}, {\em {Power-law
  distributions in empirical data}}, SIAM Reviews,  (2007).

\bibitem{Cohen2000}
{\sc R.~Cohen, K.~Erez, D.~Ben-Avraham, and S.~Havlin}, {\em {Resilience of the
  internet to random breakdowns}}, Physical review letters, 85 (2000),
  pp.~4626--8.

\bibitem{Colizza2006}
{\sc V.~Colizza, A.~Barrat, M.~Barth\'{e}lemy, and A.~Vespignani}, {\em {The
  role of the airline transportation network in the prediction and
  predictability of global epidemics.}}, Proceedings of the National Academy of
  Sciences of the United States of America, 103 (2006), pp.~2015--20.

\bibitem{colizza07}
{\sc V.~Colizza, A.~Barrat, M.~Barth\'{e}lemy, and A.~Vespignani}, {\em
  Predictability and epidemic pathways in global outbreaks of infectious
  diseases: the sars case study.}, BMC Med, 5 (2007), p.~34.

\bibitem{cossent09}
{\sc R.~Cossent, T.~G{\'o}mez, and P.~Fr{\'\i}as}, {\em Towards a future with
  large penetration of distributed generation: Is the current regulation of
  electricity distribution ready? regulatory recommendations under a european
  perspective}, Energy Policy, 37 (2009), pp.~1145--1155.

\bibitem{crawford75}
{\sc D.~Crawford and J.~Holt, S.B.}, {\em A mathematical optimization technique
  for locating and sizing distribution substations, and deriving their optimal
  service areas}, Power Apparatus and Systems, IEEE Transactions on, 94 (1975),
  pp.~230 -- 235.

\bibitem{Crucitti2004a}
{\sc P.~Crucitti, V.~Latora, and M.~Marchiori}, {\em {Model for cascading
  failures in complex networks}}, Physical Review E, 69 (2004), pp.~3--6.

\bibitem{Crucitti04}
\leavevmode\vrule height 2pt depth -1.6pt width 23pt, {\em A topological
  analysis of the italian electric power grid}, Physica A: Statistical
  Mechanics and its Applications, 338 (2004), pp.~92 -- 97.
\newblock Proceedings of the conference A Nonlinear World: the Real World, 2nd
  International Conference on Frontier Science.

\bibitem{Crucitti2005}
\leavevmode\vrule height 2pt depth -1.6pt width 23pt, {\em {Locating critical
  lines in high-voltage electrical power grids}}, Fluctuation and Noise
  Letters, 5 (2005), pp.~L201--L208.

\bibitem{Donato2007}
{\sc D.~Donato, L.~Laura, S.~Leonardi, and S.~Millozzi}, {\em {The Web as a
  graph}}, ACM Transactions on Internet Technology, 7 (2007), pp.~4--es.

\bibitem{Doye2002}
{\sc J.~Doye}, {\em {Network Topology of a Potential Energy Landscape: A Static
  Scale-Free Network}}, Physical Review Letters, 88 (2002), pp.~1--4.

\bibitem{dunne08}
{\sc J.~A. Dunne, R.~J. Williams, N.~D. Martinez, R.~A. Wood, and D.~H. Erwin},
  {\em Compilation and network analyses of cambrian food webs}, PLoS Biol, 6
  (2008), p.~e102.

\bibitem{Erdos1959}
{\sc P.~Erd{\H{o}}s and A.~R{\'{e}}nyi}, {\em On random graphs. {I}}, Publ.
  Math. Debrecen, 6 (1959), pp.~290--297.

\bibitem{Erdos1960}
\leavevmode\vrule height 2pt depth -1.6pt width 23pt, {\em On the evolution of
  random graphs}, in publication of the mathematical institute of the Hungarian
  academy of sciences, 1960, pp.~17--61.

\bibitem{ethiraj04}
{\sc S.~K. Ethiraj and D.~Levinthal}, {\em Modularity and innovation in complex
  systems}, Management Science, 50 (2004), pp.~pp. 159--173.

\bibitem{Faloutsos1999}
{\sc M.~Faloutsos, P.~Faloutsos, and C.~Faloutsos}, {\em {On power-law
  relationships of the internet topology}}, in Proceedings of the conference on
  Applications, technologies, archit ectures, and protocols for computer
  communication, ACM, 1999, p.~262.

\bibitem{fratkin96}
{\sc A.~Fratkin and L.~Ostrov}, {\em Ehv transmission grid design in israel
  electric corporation's power system: approaches and experience}, in
  Electrical and Electronics Engineers in Israel, 1996., Nineteenth Convention
  of, nov 1996, pp.~475 --478.

\bibitem{friedman08}
{\sc T.~L. Friedman}, {\em Hot, Flat, and Crowded: Why We Need a Green
  Revolution - and How It Can Renew America}, Picador, 2008.

\bibitem{garver70}
{\sc L.~Garver}, {\em Transmission network estimation using linear
  programming}, Power Apparatus and Systems, IEEE Transactions on, PAS-89
  (1970), pp.~1688 --1697.

\bibitem{Gautreau2008}
{\sc A.~Gautreau, A.~Barrat, and M.~Barth\'{e}lemy}, {\em {Global disease
  spread: statistics and estimation of arrival times.}}, Journal of theoretical
  biology, 251 (2008), pp.~509--22.

\bibitem{gershenson03}
{\sc J.~K. Gershenson, G.~J. Prasad, and Y.~Zhang}, {\em {Product modularity:
  definitions and benefits}}, Journal of Engineering Design, 14 (2003),
  pp.~295--313.

\bibitem{gilbert11}
{\sc G.~Gilbert, Y.~Chow, D.~Bouchard, and M.~Salama}, {\em Optimization of
  high voltage substations using a random walk technique}, in Transmission and
  Distribution Construction, Operation and Live-Line Maintenance (ESMO), 2011
  IEEE PES 12th International Conference on, may 2011, pp.~1 --7.

\bibitem{grigsby07}
{\sc L.~L. Grigsby}, ed., {\em The Electric Power Engineering Handbook, Second
  Edition}, CRC Press, 2007.

\bibitem{Guimer2004}
{\sc R.~Guimer\`{a} and L.~a.~N. Amaral}, {\em {Modeling the world-wide airport
  network}}, The European Physical Journal B - Condensed Matter, 38 (2004),
  pp.~381--385.

\bibitem{Harris06}
{\sc C.~Harris}, {\em Electricity Markets: Pricing, Structures and Economics},
  Wiley, 2006.

\bibitem{hoff97}
{\sc T.~E. Hoff}, {\em Using distributed resources to manage risks caused by
  demand uncertainty}, The Energy Journal, 18 (1997), pp.~63--84.

\bibitem{holmgren06}
{\sc A.~J. Holmgren}, {\em {Using Graph Models to Analyze the Vulnerability of
  Electric Power Networks}}, Risk Analysis, 26 (2006), pp.~955--969.

\bibitem{Jeong2000}
{\sc H.~Jeong, B.~Tombor, R.~Albert, Z.~N. Oltvai, and A.~L. Barab\'{a}si},
  {\em {The large-scale organization of metabolic networks.}}, Nature, 407
  (2000), pp.~651--4.

\bibitem{joskow08}
{\sc P.~L. Joskow}, {\em Lessons learned from electricity market
  liberalization}, The Energy Journal, 29 (2008), pp.~9--42.

\bibitem{keeney82}
{\sc R.~L. Keeney}, {\em Decision analysis: An overview}, Operations Research,
  30 (1982), pp.~pp. 803--838.

\bibitem{kendall48}
{\sc M.~Kendall}, {\em Rank correlation methods}, Griffin, London, 1948.

\bibitem{kephart91}
{\sc J.~Kephart and S.~White}, {\em Directed-graph epidemiological models of
  computer viruses}, in Research in Security and Privacy, 1991. Proceedings.,
  1991 IEEE Computer Society Symposium on, May 1991, pp.~343 --359.

\bibitem{Kleinberg1999}
{\sc J.~Kleinberg, R.~Kumar, P.~Raghavan, S.~Rajagopalan, and A.~Tomkins}, {\em
  {The web as a graph: Measurements, models, and methods}}, in Proceedings of
  the 5th annual international conference on Computing and combinatorics,
  Springer-Verlag, 1999, pp.~1--17.

\bibitem{lasseter11}
{\sc R.~Lasseter}, {\em Smart distribution: Coupled microgrids}, Proceedings of
  the IEEE, 99 (2011), pp.~1074 --1082.

\bibitem{Latora2002}
{\sc V.~Latora and M.~Marchiori}, {\em {Is the Boston subway a small-world
  network?}}, Physica A: Statistical Mechanics and its Applications, 314
  (2002), pp.~109--113.

\bibitem{lee74}
{\sc S.~Lee, K.~Hicks, and E.~Hnyilicza}, {\em Transmission expansion by
  branch-and-bound integer programming with optimal cost - capacity curves},
  Power Apparatus and Systems, IEEE Transactions on, PAS-93 (1974), pp.~1390
  --1400.

\bibitem{Leskovec2010}
{\sc J.~Leskovec, D.~Chakrabarti, J.~Kleinberg, C.~Faloutsos, and
  Z.~Ghahramani}, {\em {Kronecker graphs: An approach to modeling networks}},
  The Journal of Machine Learning Research, 11 (2010), pp.~985--1042.

\bibitem{Leskovec05}
{\sc J.~Leskovec, J.~Kleinberg, and C.~Faloutsos}, {\em Graphs over time:
  densification laws, shrinking diameters and possible explanations}, in
  Proceedings of the eleventh ACM SIGKDD international conference on Knowledge
  discovery in data mining, KDD '05, New York, NY, USA, 2005, ACM,
  pp.~177--187.

\bibitem{lovins02}
{\sc A.~B. Lovins, E.~K. Datta, T.~Feiler, K.~R. Rabago, J.~N. Swisher,
  A.~Lehmann, and K.~Wicker}, {\em Small is profitable: the hidden economic
  benefits of making electrical resources the right size}, Rocky Mountain
  Institute, 2002.

\bibitem{machias89}
{\sc A.~Machias, E.~Dialynas, and C.~Protopapas}, {\em An expert system
  approach to designing and testing substation grounding grids}, Power
  Delivery, IEEE Transactions on, 4 (1989), pp.~234 --240.

\bibitem{mar:mic06}
{\sc C.~Marnay and M.~Venkataramanan}, {\em Microgrids in the evolving
  electricity generation and delivery infrastructure}, in IEEE Power
  Engineering Society General Meeting, 2006.

\bibitem{massey51}
{\sc J.~Massey, Frank~J.}, {\em The kolmogorov-smirnov test for goodness of
  fit}, Journal of the American Statistical Association, 46 (1951), pp.~pp.
  68--78.

\bibitem{Mei11}
{\sc S.~Mei, X.~Zhang, and M.~Cao}, {\em Power Grid Complexity}, Springer,
  2011.

\bibitem{milgram67}
{\sc S.~Milgram}, {\em The small world problem}, Psychology Today, 2 (1967),
  pp.~60--67.

\bibitem{Moreno03}
{\sc Y.~Moreno, R.~Pastor-Satorras, A.~Vazquez, and A.~Vespignani}, {\em
  Critical load and congestion instabilities in scale-free networks},
  Europhysics Letters, 62 (2003).

\bibitem{smartGrid09}
{\sc M.~G. Morgan, J.~Apt, L.~B. Lave, M.~D. Ilic, M.~Sirbu, and J.~M. Peha},
  {\em The many meanings of ``smart grid"}, tech. rep., Carnegie Mellon
  University, 2009.

\bibitem{Munasinghe84}
{\sc M.~Munasinghe}, {\em Engineering-economic analysis of electric power
  systems}, Proceedings of the IEEE, 72 (1984), pp.~424 -- 461.

\bibitem{natGrid}
{\sc {National Grid}}, {\em Undergrounding high voltage electricity
  transmission - the technical issues}, tech. rep., National Grid, 2009.

\bibitem{Newman2003a}
{\sc M.~E.~J. Newman}, {\em The structure and function of complex networks},
  SIAM REVIEW, 45 (2003), pp.~167--256.

\bibitem{pag:preprint1102}
{\sc G.~A. Pagani and M.~Aiello}, {\em The power grid as a complex network: a
  survey}, Tech. Rep. ArXiv preprint arXiv:1105.3338, JBI, University of
  Groningen, 2011.

\bibitem{PaganiAielloTSG2011}
\leavevmode\vrule height 2pt depth -1.6pt width 23pt, {\em Towards
  decentralization: A topological investigation of the medium and low voltage
  grids}, Smart Grid, IEEE Transactions on, 2 (2011), pp.~538 --547.

\bibitem{pag:preprint11}
\leavevmode\vrule height 2pt depth -1.6pt width 23pt, {\em Towards
  decentralized trading: A topological investigation of the dutch medium and
  low voltage grids}, Tech. Rep. ArXiv preprint arXiv:1101.1118, JBI,
  University of Groningen, 2011.

\bibitem{paiva05}
{\sc P.~Paiva, H.~Khodr, J.~Dominguez-Navarro, J.~Yusta, and A.~Urdaneta}, {\em
  Integral planning of primary-secondary distribution systems using mixed
  integer linear programming}, Power Systems, IEEE Transactions on, 20 (2005),
  pp.~1134 -- 1143.

\bibitem{Pareto:1971}
{\sc V.~Pareto}, {\em Manual of political economy (manuale di economia
  politica)}, Kelley, New York, 1971 (1906).
\newblock Translated by Ann S. Schwier and Alfred N. Page.

\bibitem{potter09}
{\sc C.~Potter, A.~Archambault, and K.~Westrick}, {\em Building a smarter smart
  grid through better renewable energy information}, in Power Systems
  Conference and Exposition, 2009. PSCE '09. IEEE/PES, march 2009, pp.~1 --5.

\bibitem{barri10}
{\sc M.~Roman-Barri, I.~Cairo-Molins, A.~Sumper, and A.~Sudria-Andreu}, {\em
  Experience on the implementation of a microgrid project in barcelona}, in
  Innovative Smart Grid Technologies Conference Europe (ISGT Europe), 2010 IEEE
  PES, oct. 2010, pp.~1 --7.

\bibitem{casals07}
{\sc M.~Rosas-Casals, S.~Valverde, and R.~V. Sol\'{e}}, {\em {Topological
  vulnerability of the European power grid under errors and attacks}},
  International Journal of Bifurcation and Chaos, 17 (2007), p.~2465.

\bibitem{Rosato2007}
{\sc V.~Rosato, S.~Bologna, and F.~Tiriticco}, {\em {Topological properties of
  high-voltage electrical transmission networks}}, Electric Power Systems
  Research, 77 (2007), pp.~99--105.

\bibitem{shafiullah10}
{\sc G.~Shafiullah, A.~Oo, D.~Jarvis, A.~Ali, and P.~Wolfs}, {\em Potential
  challenges: Integrating renewable energy with the smart grid}, in
  Universities Power Engineering Conference (AUPEC), 2010 20th Australasian,
  dec. 2010, pp.~1 --6.

\bibitem{shariati08}
{\sc H.~Shariati, H.~Askarian~Abyaneh, M.~Javidi, and F.~Razavi}, {\em
  Transmission expansion planning considering security cost under market
  environment}, in Electric Utility Deregulation and Restructuring and Power
  Technologies, 2008. DRPT 2008. Third International Conference on, april 2008,
  pp.~1430 --1435.

\bibitem{Corominas-murtra2008}
{\sc R.~V. Sol\'{e}, M.~Rosas-Casals, B.~Corominas-Murtra, and S.~Valverde},
  {\em {Robustness of the European power grids under intentional attack}},
  Physical Review E, 77 (2008), pp.~1--7.

\bibitem{Strogatz2001}
{\sc S.~H. Strogatz}, {\em {Exploring complex networks.}}, Nature, 410 (2001),
  pp.~268--76.

\bibitem{teive98}
{\sc R.~Teive, E.~Silva, and L.~Fonseca}, {\em A cooperative expert system for
  transmission expansion planning of electrical power systems}, Power Systems,
  IEEE Transactions on, 13 (1998), pp.~636 --642.

\bibitem{Milgram69}
{\sc J.~Travers and S.~Milgram}, {\em An experimental study of the small world
  problem}, Sociometry, 32 (1969), pp.~425--443.

\bibitem{vai:pow05}
{\sc V.~Vaitheeswaran}, {\em Power to the People}, Earthscan, 2005.

\bibitem{Vazquez2002}
{\sc A.~V\'{a}zquez, R.~Pastor-Satorras, and A.~Vespignani}, {\em {Large-scale
  topological and dynamical properties of the Internet}}, Physical Review E, 65
  (2002), pp.~1--12.

\bibitem{wall79}
{\sc D.~Wall, G.~Thompson, and J.~Northcote-Green}, {\em An optimization model
  for planning radial distribution networks}, Power Apparatus and Systems, IEEE
  Transactions on, PAS-98 (1979), pp.~1061 --1068.

\bibitem{Wang2010}
{\sc Z.~Wang, A.~Scaglione, and R.~J. Thomas}, {\em {Generating Statistically
  Correct Random Topologies for Testing Smart Grid Communication and Control
  Networks}}, IEEE Transactions on Smart Grid, 1 (2010), pp.~28--39.

\bibitem{wang08}
{\sc Z.~Wang, R.~Thomas, and A.~Scaglione}, {\em Generating random topology
  power grids}, in Hawaii International Conference on System Sciences,
  Proceedings of the 41st Annual, jan. 2008, p.~183.

\bibitem{Watts03}
{\sc D.~J. Watts}, {\em Small Worlds: The Dynamics of Networks between Order
  and Randomness}, Princeton University Press, Princeton, NJ, USA, 2003.

\bibitem{Watts98}
{\sc D.~J. Watts and S.~H. Strogatz}, {\em Collective dynamics of \lq
  small-world\rq \space networks}, Nature, 393 (1998), pp.~440--442.

\end{thebibliography}
\end{document}